%% file: main.tex
\documentclass[11pt]{prop_axis} 
\input{headstuff.tex}

\author[1]{Jeremy J. Drake}
\author[2]{Juli\'an Alvarado Gomez}
\author[3]{Costanza Argiroffi}
\author[3]{Ettore Flaccomio} 
\author[4]{Cecilia Garraffo}
\author[5]{Nicolas Grosso}
\author[6,7]{Nazma Islam} 
\author[4]{Margarita Karovska}
\author[4]{Vinay L.\ Kashyap}
\author[4]{Kristina Monsch} 
\author[8]{Jan-Uwe Ness}
\author[3]{Salvatore Sciortino} 
\author[4]{Bradford Wargelin} 

\affil[1]{Lockheed Martin, 3251 Hanover St, Palo Alto, CA 94304}

\affil[2]{Leibniz Institute for Astrophysics (AIP) An der Sternwarte 16, 14482 Potsdam, Germany}

\affil[3]{INAF-Oss. Astronomico di Palermo, Palermo, Italy}

\affil[4]{Center for Astrophysics $|$ Harvard \& Smithsonian, 
Cambridge, MA 02138, USA}

\affil[5]{Aix Marseille Univ, CNRS, CNES, LAM, Marseille, France}

\affil[6]{NASA Goddard Space Flight Center, Astrophysics Science Division, Code 662, Greenbelt, MD 20771, USA}
\affil[7]{Center for Space Sciences and Technology, University of Maryland, Baltimore County, 1000 Hilltop Circle, Baltimore, MD 21250, USA}

\affil[8]{European Space Agency (ESA), Camino Bajo del Castillo s/n, 28692 Villanueva de la
Ca\~nada, Madrid, Spain}

\affil[7]{Department of Astronomy, University of Maryland, College Park, MD 20742-2421, USA}
\affil[z]{Department of Physics and Astronomy, University of Kentucky, 505 Rose Street, Lexington, KY 40506, USA}

\begin{document}

\baselineskip=13.2pt
\sloppy
\pagenumbering{roman}
\thispagestyle{empty}


\title{\textcolor{black}{\sf\huge \textcolor{blue}{\sl LEM GENERAL OBSERVER KEY SCIENCE}\\
\vspace{5mm} Breakthroughs in Cool Star Physics with the Line Emission Mapper X-ray Probe}\footnote{Corresponding author: Jeremy Drake (jeremy.1.drake@lmco.com)}}
\maketitle

\begin{tikzpicture}[remember picture,overlay]
\node[anchor=north west,yshift=2pt,xshift=2pt]%
    at (current page.north west)
    {\includegraphics[height=20mm]{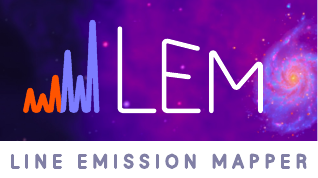}};
\end{tikzpicture}

\vfill

{\footnotesize
\noindent
\\
\phantom{${^52}$}~\textcolor{blue}{\bsf \href{https://lem-observatory.org}{lem-observatory.org}}\\
\phantom{${^52}$}~\textcolor{blue}{\bsf X / twitter: \href{https://www.twitter.com/LEMXray}{LEMXray}}\\
\phantom{${^52}$}~\textcolor{blue}{\bsf facebook: \href{https://www.facebook.com/LEMXrayProbe}{LEMXrayProbe}}\\
}

\centerline{\em White Paper, October 2023}

\clearpage
\twocolumn


\setcounter{page}{1}
\pagenumbering{arabic}

\section*{SUMMARY}
\label{sec:summary}

We outline some of the highlights of the scientific case for the advancement of stellar high energy physics using the Line Emission Mapper X-ray Probe ({\it LEM}). The key to advancements with LEM lie in its large effective area---up to 100 times that of the {\it Chandra} MEG---and 1~eV spectral resolution. The large effective area opens up for the first time the ability to study time-dependent phenomena on their natural timescales at high resolution, such as flares and coronal mass ejections, and also opens the sky to much fainter targets than available to {\it Chandra} or {\it XMM-Newton}.

\section{INTRODUCTION: OUTSTANDING PROBLEMS IN HIGH-ENERGY COOL STAR PHYSICS}

%
%

The prototype for much of the physics involved in high energy processes on cool stars is of course the Sun. The solar X-ray corona comprises bright active regions characterized by plasma trapped within magnetic loops, dark coronal holes, small-scale bright points and diffuse emission components. The Sun also exhibits rapid X-ray variability and flares, resulting from the sudden release of stored magnetic energy in the form of accelerated particles, heat, and ejections of mass. Stars appear to do much of the same thing. The extent to which stellar coronae are similar to or different to the Sun has been an avenue of intense investigation since the {\it Einstein} observatory first prompted the realisation that coronae were magnetically driven  (see, e.g., reviews by Refs.~\citenum{Gudel:04,Gudel2009,Drake2019,sciortino22,Drake2023}). 


Although we can predict the X-ray behaviour of solar-like stars by drawing parallels with our Sun, the majority of stars differ from the Sun in mass, evolutionary phase, in rotation rate, and in metallicity---all aspects of which are expected to influence the underlying interior magnetic dynamo. Some are in close binaries that are forced to rotate at the orbital period through tidal locking. Some are involved in accretion systems, such as T~Tauri stars and cataclysmic variables. In the context of the solar-stellar connection, it is these differences that enable us to use stars as a laboratory to study the general plasma physics and astrophysics of the solar corona under quite different conditions. In the astrophysical context, the solar paradigm provides a reference benchmark.

Space-based studies at X-ray wavelengths that have culminated in {\it Chandra} and {\it XMM-Newton} have provided a fairly comprehensive picture of stellar demographics. X-ray emission from stars is widespread throughout the Hertzsprung-Russell diagram, except for intermediate-mass main-sequence stars (late-B to mid-A spectral types), later-type giants, and possibly stars with spectral types later than mid-M. We now know what their X-ray spectra look like at high spectral resolution, and how their spectra change with rotation period and activity level. However, there are still fundamental problems to solve. We note just a few here that will be touched on in more detail below.

\paragraph{T Tauri Stars, Accretion and Feedback} T~Tauri stars are microcosms of high-energy astrophysics, exhibiting accretion, jets, and giant X-ray flares. However, we have only scratched the surface of how these all tie together. Key questions include how does the accretion proceed, what happens at the accretion shock on the stellar surface, how does this feed back into the corona and into jets?

\paragraph{Coronae on Brown Dwarfs}
Young brown dwarfs are readily detected in nearby star forming regions, but older and cooler objects have only been detected during flares. What their quiescent coronal properties are, how they are heated or are structured and how they build up surface magnetic stresses and flares is not known. 

\paragraph{Magnetic Reconnection Flares}
Nearly all late-type stars are observed to flare, and undertanding flares is important for understanding magnetic energy cycling, space weather, and energetic particles. But there is an enormous dynamic range between solar flares and flares on the most active stars and T~Tauri stars and it is not known if the same processes are at work. It is also not known if giant T~Tauri flares result from star-disk magnetic fields, and chromospheric evaporation has yet to be systematically detected on stars, while large coronal loops hinted at by flare models suggest velocity signatures of plasma sloshing might also be detected.

\paragraph{Coronal Mass Ejections}
Only very scant and inconclusive evidence that CMEs occur on magnetically active stars currently exists. They are important to understand because of their potentially catastrophic impact on exoplanet atmospheres. Sensitivity to detect Doppler-shifted signals from ejected plasma have so far been out of reach of X-ray spectrometers but their signatures  are in principle detectable.

\paragraph{Coronal Chemical Compositions}
The differences in chemical compositions of coronae and photospheres is now well-documented. Through the different patterns of abundances vs stellar activity level and spectral type, this chemical fractionation could provide a powerful means of investigating fundamental plasma physics at work stellar coronae. Abundances can only be reliably determined with high-resolution spectra, and a much wider sample of stars is needed to flesh out abundance patterns. Time-dependent abundance effects are also potentially powerful diagnostics that have so far only been hinted at. 

\paragraph{Atmospheres of Transiting Gas Giants} X-rays are absorbed much higher in a planetary atmosphere than optical or infrared light. In transit, X-rays can probe the scale height and bulk chemical compositions of exoplanet exospheres, probing mass loss and the influence of the ambient stellar wind on erosion and on transit signatures.


\paragraph{Stellar Magnetic Cycles} 
Magnetic cycles have been detected on over 100 stars, but only a handful of cycle signatures have been seen at X-ray wavelengths. Little is known of the structuring of stellar corona other than that of the Sun, and how coronae might change through magnetic cycles in terms of plasma temperature, sizes of coronal structures, and chemical composition.

\paragraph{Symbiotic Stars and Cataclysmic Variables} 
Important for understanding binary stellar evolution and the route toward Type 1a supernovae, symbiotic stars and cataclysmic variables present laboratories for the study mass transfer, accretion, and jets. Many outstanding issues remain concerning the accretion processes, the X-ray generation mechanisms,  and the interpretation of their X-ray spectra.

\paragraph{Novae} Also a member of the cataclysmic variable family, nova explosions provide a panoply of astrophysics including mass transfer and binary star evolution, combined with explosive thermonuclear runaway. Existing observations exhibit a bewildering variety of X-ray spectra and have raised questions regarding the physics of the early explosion and interaction with the circumstellar medium, and the nature of strong variability during the onset of supersoft source phases.

\section{IMPORTANCE OF {\it LEM} FOR COOL STAR SCIENCE}

In the 1990s, the ROSAT All-sky survey and pointed observations fleshed out stellar coronal populations, while {\it ASCA} obtained many low-resolution X-ray spectra that enabled chemical compositions and coronal temperature structure to be investigated. The 2000s ushered in a sea change with {\it Chandra} and {\it XMM-Newton} and the advent of high-resolution X-ray spectroscopy capabilities sensitive enough for studying stellar coronae. This opened up spectroscopic temperature and density diagnostics, velocity resolution and the ability to see non-thermal motion, and resolved lines from different elements to measure chemical abundances. However, only relatively bright stars can be studied using these facilities.

\begin{figure}
    \centering
    \includegraphics[width=1.0\linewidth]{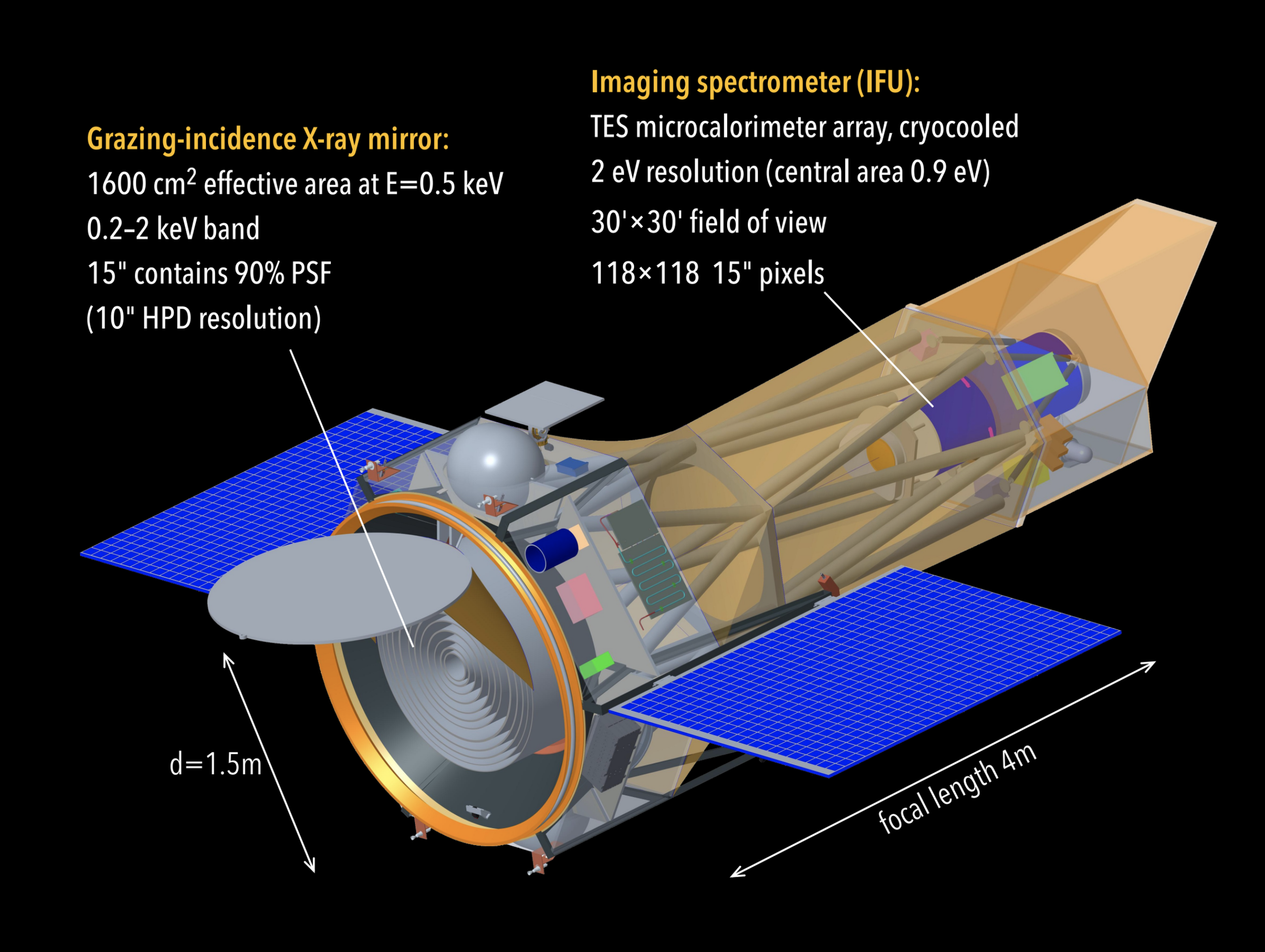}
    \caption{Schematic illustration of the {\it Line Emission Mapper} X-ray Probe showing the key aspects that make it a powerful tool for cool star research: a microcalorimeter array schiving a resolving power of 1~eV in the field center, fed by large area soft X-ray mirror with a field of view of 30'$\times$30' and 10" spatial resolution.}
    \label{f:lem}
\end{figure}

The "Line Emission Mapper" (LEM; Figure~\ref{f:lem}) is a proposed X-ray Probe mission that would launch in the mid-2030s. {\it LEM} combines two major advances into a package that achieves more than order of magnitude performance improvement over existing missions. These are its soft X-ray microcalorimeter array, which achieves a remarkable 1~eV resolution in the central array segment, and its primary mirror utilizing pairs of thin monocrystalline silicon shells (to be coated with either Ir or Pt) that affords a large effective area to mass ratio and with a 10" half power diameter. The resulting instrument achieves an impressive effective area of over 2000 cm$^2$ at 1 keV, with an average of 1600 cm$^2$ over a 0.2-2 keV bandpass. This represents a substantial increase in effective area of up to 100 times greater than the {\it Chandra} MEG, while maintaining similar spectral resolution. 

LEM offers two significant advantages for the study of low mass stars. First, it allows for high-resolution X-ray spectra to be obtained from much fainter objects, greatly expanding the accessible volume of space and range of stars that can be studied at high resolution. Second, it provides the sensitivity needed to conduct time-dependent studies of brighter objects. One further strength of {\it LEM} for studying clusters of stars lies in its large 30'$\times$30' field of view.
Probe Class missions for the present NASA competition are mandated to allot 70\%\ of the available observing time for General Observer access. This opens up extensive opportunities to advance our understanding of the high energy processes of cool stars and how they impact their planetary environments through dedicated pointed observations.



\section{X-RAYS FROM T TAURI STARS}
\label{s:ttauri}


Low-mass pre-main sequence (PMS, T~Tauri) stars are conspicuous X-ray emitters, 
displaying powerful flares, releasing $\sim$$10^{4\mbox{--}8}$ times more energy
than the most X-ray energetic solar flare 
\citep{getman21}.
The analogy with solar activity has been quite successful in
ascribing most of this X-ray emission to an active magnetic corona
\citep{feigelson99}.
In accreting T~Tauri stars, 
the accreted gas is channeled by the stellar magnetosphere 
from the disk to the stellar surface,
where an accretion shock with comparatively high plasma density compared with the ambient corona is generated, 
producing a plasma with a low temperature ($\sim$4~MK) 
compared the coronal one ($\sim$10~MK) in these active stars. 
This source of soft X-rays can be detected 
in addition to the harder X-rays 
emitted by the corona, 
if the emitting source---the accretion shock---is not too embedded into the stellar chromosphere or photosphere.
Several accreting T~Tauri stars are observed to drive strong jets, probably in the directions of their rotation axes. These stars 
exhibit a two-absorber X-ray
spectrum revealing a strongly absorbed, hot and variable plasma component,
and a less absorbed cooler plasma component. The former is
associated with the active corona screened by the
circumstellar disk, whereas the latter originates from
shocks located above the disk midplane, near the base of the jet \citep{guedel05,guedel07d}.

Thanks to its high spectral-resolution an large effective area in the soft X-ray band, {\it LEM} offers breakthrough sensitivity to solve outstanding problems in our knowledge of the physics and physical conditions of accretion shocks, accretion post-shocks, and shocks in jets.  

\subsection{X-rays from accretion shocks at the stellar surface}
\label{s:accretion}

\subsubsection{Accretion feedback on the corona of TW~Hydrae}

\begin{figure}
    \centering
    \includegraphics[width=1.0\linewidth]{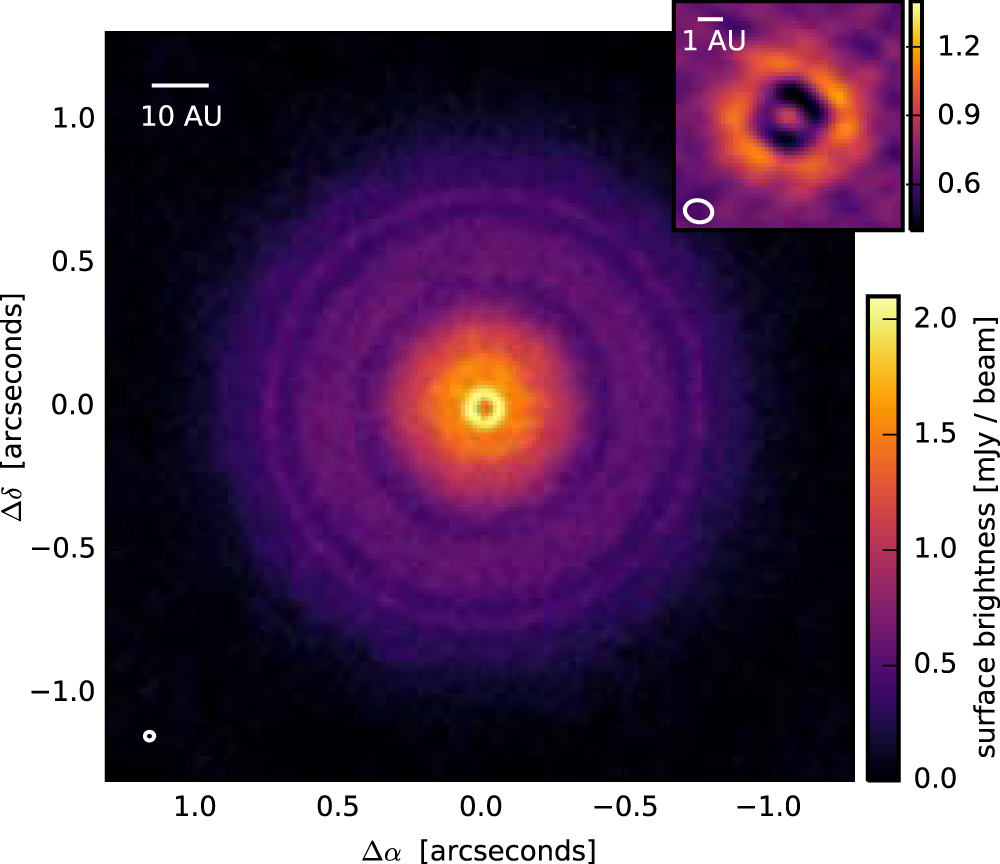}
    \caption{ALMA 870~$\mu m$ continuum image from Ref.~\citenum{Andrews2016} of the protoplanetary disk around the nearest T~Tauri star, TW Hydrae. The inset corresponds to a region 0.2" square and shows the gap near 1~AU from the star that could be due to interactions of the disk with newly-formed planets\cite{Andrews2016}. }
    \label{f:alma_twhya}
\end{figure}

At a distance of only 60~pc, the pre-main sequence star TW~Hydrae (TW~Hya) 
is one of the brightest T~Tauri stars in X-rays. It is accreting from its circumstellar disk, famously seen pole-on in spectacular ALMA images (Figure~\ref{f:alma_twhya}).
The foreground extinction is minimal, allowing the X-ray detection of the accretion shock 
at the stellar surface. In this star, this shock is the main source of the observed X-ray emission
and displays a lower temperature and a higher density  
compared to the stellar corona \citep{kastner02,stelzer2004b,brickhouse10}. 
From the X-ray spectrum obtained with the {\sl Chandra} High Energy Transmission Grating (HETG) 
using a 489.5~ks exposure, the He-like Ne~IX line ratio diagnostics constrain 
the temperature to $2.50\pm0.25$~MK and the electronic density to $3.0\pm0.2\times10^{12}$~cm$^{-3}$ 
in the shock, which is consistent with magnetospheric accretion
\citep{brickhouse10}. However, the He-like O~VII line ratio diagnostics 
indicates a large post-shock coronal region where the electronic density is 5 times lower 
($5.7^{+4.4}_{-1.2}\times10^{11}$~cm$^{-3}$) 
than in the accretion shock and $\sim$7 times lower than the model prediction \citep{brickhouse10}.

By monitoring the total flux of the emission lines formed in the accretion shock 
(N~VII, O~VIII, Ne~IX, Fe~XVII, Mg~XI), Ref.~\citenum{dupree12} 
identify an abrupt increase ($\sim$26\%) in this flux that lasts for 3~ks. 
About 2~h after this X-ray accretion event, the heated photosphere produced 
an increase of the optical veiling. There is a hint for a correlation between 
the optical veiling and the flux of the stellar corona with a delay of $\sim$2.5~h.


\begin{figure}
\centering
\includegraphics[angle=0,width=1.0\columnwidth,trim={0 0 0 1.1cm},clip]{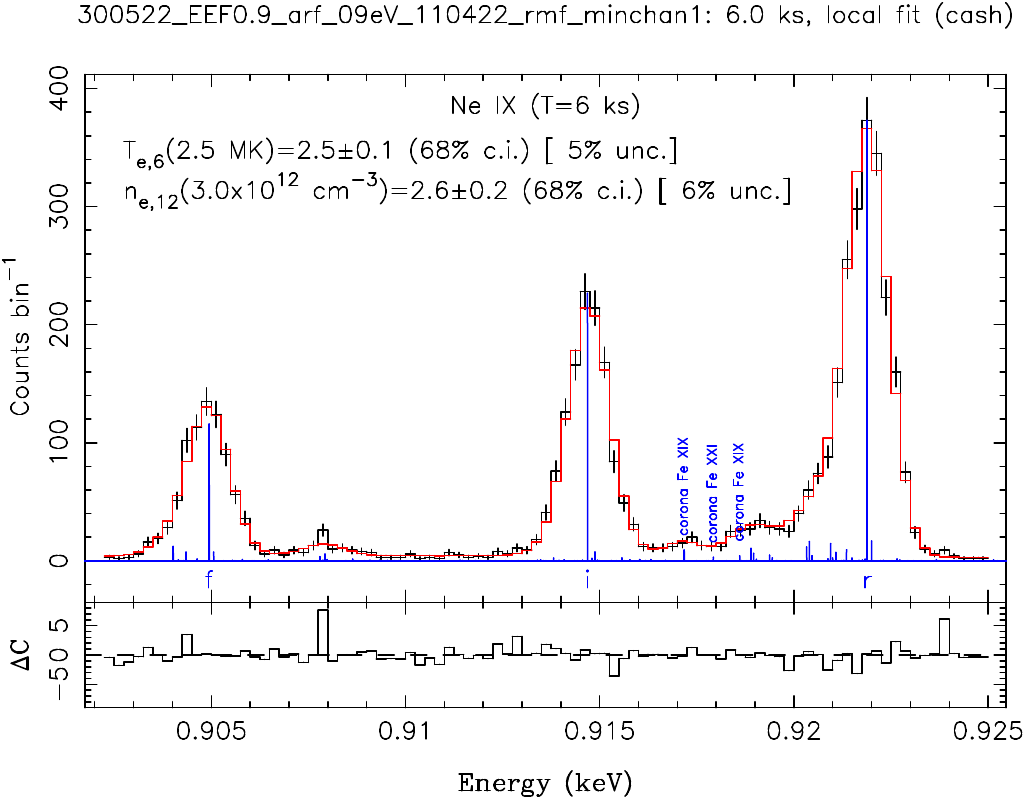} 
\includegraphics[angle=0,width=1.0\columnwidth,trim={0 0 0 1.1cm},clip]{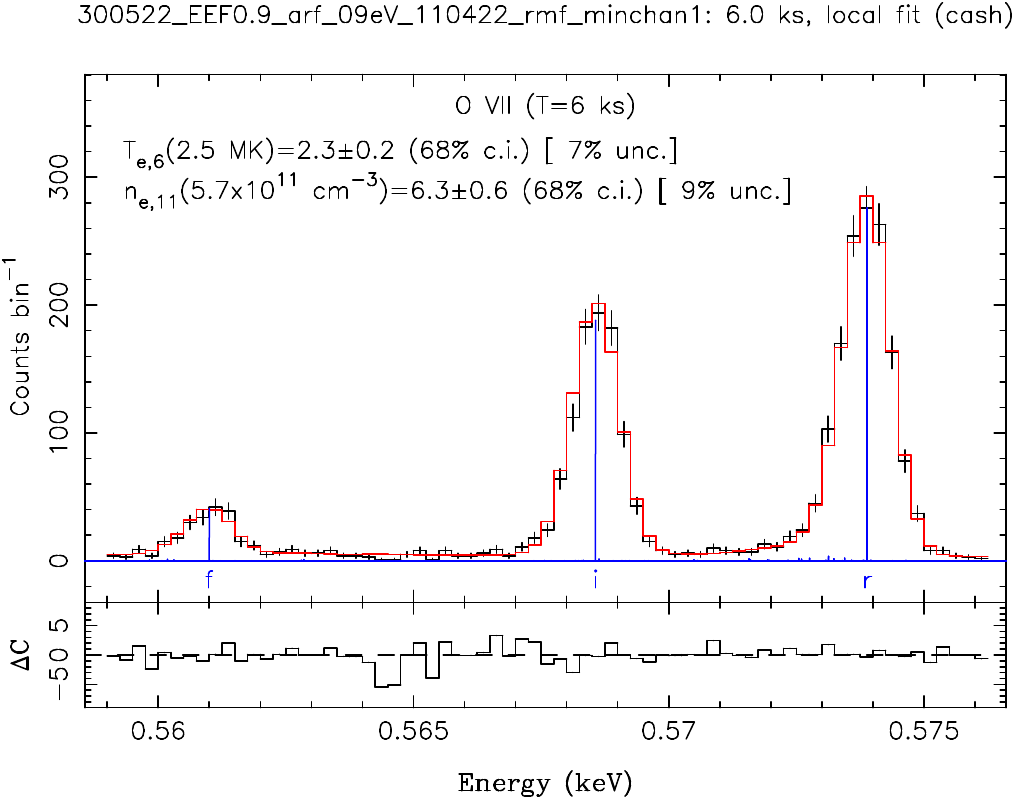}
\vspace{-0.25cm}
\caption{Simulated He-like emission lines 
of the accretion shock and post-shock at the stellar surface of TW~Hya, 
obtained with a single {\it LEM} central-array pixel  
and an exposure of 6~ks.
Top panel: Ne~IX triplet emitted by the accretion shock 
with cool temperature ($T_\mathrm{e}=2.5$~MK) 
and high density 
($n_\mathrm{e}=3\times10^{12}$~cm$^{-3}$).
Bottom panel: O~VII triplet emitted by the accretion post-shock with the same temperature but lower density 
($n_\mathrm{e}=5.7\times10^{11}$~cm$^{-3}$).
In each case, the blue solid lines are the TW~Hya spectra
with labels for the forbidden (f), intercombination (i), and resonance (r) lines 
and highly-ionized iron lines from the hot corona; 
the black histograms with the error bars are the simulated {\it LEM} spectra 
with instrumental-background 
subtraction for visualisation only; 
the red histograms are the best-fit models 
obtained on this energy range using modified Cash-statistics.
The black histograms in each narrow bottom panel show the fit residuals.
LEM will determine the plasma temperature and density 
of the accretion shock and post-shock from plasma diagnostics using very short exposures, opening the window to time-dependent accretion studies.    
} 
\label{fig:TW_Hya}
\end{figure}

\begin{figure}
\centering
\includegraphics[angle=0,width=1.0\columnwidth]{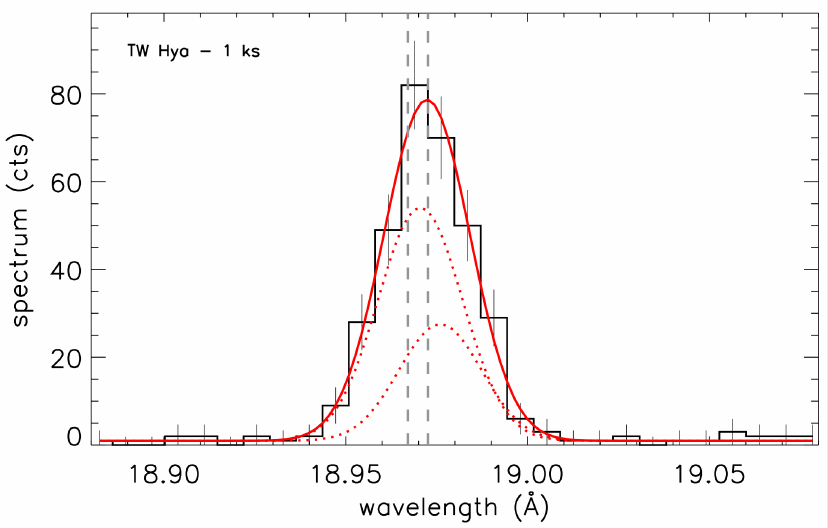}
\caption{Simulated O~VIII Ly$\alpha$ doublet profile of TW~Hya, collected in a simulated exposure of just 1\,ks. The black solid histogram represents the expected TW~Hya spectrum. Gray dashed lines indicate the rest wavelengths of the two contributions to the O~VIII doublet. Red solid line marks the best-fit profile, corresponding to the superposition of two Gaussian functions (red dotted lines) that mimic the two doublet contributions. The best-fit redshift, $39\pm15$\,km\,s$^{-1}$, is consistent with the assumed input value.} 
\label{fig:TW_Hya_dopplershift}
\end{figure}

We computed {\it LEM} simulations of TW~Hya assuming an emission model composed of   
cool (accretion shock and post-shock regions) 
and hot (corona) plasma components, 
and a single absorber 
(Table~6 model B parameters and turbulent velocity of 165~km~s$^{-1}$ from Ref.~\citenum{brickhouse10}).
For the cool component, we adopt a line of sight velocity of 38.3~km~s$^{-1}$ 
\citep{argiroffi17}. 
In {\tt ISIS} \citep{houck00}, we use the modifier of the He-like ion line emissivities \citep{huenemoerder11b}
to include the electronic densities obtained from the He-like triplets (Table 3 of Ref.~\citenum{brickhouse10}).
We assume signal is in a single {\it LEM} pixel of the central array with a spectral 
resolution of 0.9~eV, collecting 90\% of the {\it LEM} point spread function.
Figure~\ref{fig:TW_Hya} shows the predicted {\it LEM} spectra 
centered on the Ne~IX and O~VII lines 
for an exposure of 6~ks. 
We model simultaneously and consistently the triplet lines 
and their satellite lines in these small energy bands 
by fitting the physical parameters of a single plasma component 
(temperature, density, emission measure, element abundance, 
turbulent velocity, radial velocity). 
For the Ne~IX complex, several iron lines emitted by the hot corona 
are blended with the satellite lines, 
therefore, additional gaussian components at fixed energy 
are also fitted.

From the Ne~IX triplet, (1$\sigma$) uncertainties of 4\% and 6\% 
are obtained for the temperature and the density, 
respectively, of the accretion shock; for comparison, (1$\sigma$) uncertainties of 10\% and 7\% are obtained with HETG based on a 489.5~ks-exposure.  From the O~VII triplet, (1$\sigma$) uncertainties of 7\% and 9\% 
are obtained for the temperature and the density, 
respectively, of the post-shock; for comparison, (1$\sigma$) uncertainties of $\sim$38\% and $\sim$50\% are obtained with HETG using 489.5~ks-exposure.

Hence, using diagnostics such as the Ne~IX and O~VII triplets, {\it LEM} will be able to probe with for the first time on short timescales, the temperature and the density of the accretion shock, 
and also the hydrogen column density using, e.g., 
the Ne~IX resonance  
He$\alpha$(1.022~keV)/He$\beta$(1.074~keV) ratio, 
allowing the monitoring of the mass-accretion rate\citep{brickhouse12} and the stability of the accretion shock\cite{Drake2009}.

In addition, we stress that the {\it LEM} spectral resolution will allow us to measure the radial velocity of the accreting plasma component. In fact, in the TW~Hya predicted spectrum, several cool lines display significant Doppler shifts in agreement with the assumed input value. As an example, in Fig.~\ref{fig:TW_Hya_dopplershift} we show the TW~Hya expected profile of the O~VIII Ly$\alpha$ doublet, collected in an exposure time of just 1 ks, with superimposed the best fit profile, whose redshift is $39\pm15$\,km\,s$^{-1}$. Since the free-fall velocity of the accreting material is known, the measurement of the radial velocity component provides us with important constraints on the inclination (and hence geometry) of the accretion stream, and how it changes on short time scales.

Therefore, {\it LEM} will investigate, again for the first time on short timescales, 
any correlation between the physical parameters 
of the accretion shock and the post-shock, 
and the coronal variability, thereby probing for any accretion feedback on the corona.

\subsubsection{Episodic accretion in EX~Lupi}

The PMS star EX~Lupi is located at a distance of 155~pc 
and displays repetitive but irregular outbursts in the optical 
with decay timescales ranging from a few months to years,
produced by episodic accretion \citep{audard14}.
EX~Lupi is the prototype of "EXor" eruptive variable stars identified by Herbig\cite{Herbig1989} as ``Stars exhibiting large-range outbursts, but having T Tauri-like spectra when bright.''
In 2008, it exhibited its brightest optical outburst in 50 years that lasted 8 months, peaking $\sim$30 times above its pre-outburst optical flux. X-ray observations with {\sl Chandra} and {\sl XMM-Newton}
show a strong correlation between the decreasing optical and X-ray fluxes following the optical peak.
The cool ($\sim$4.7~MK) plasma component faded as EX~Lupi returned to its pre-outburst optical level, 
which is consistent with a decrease in the overall emission measure 
of accretion shock-generated plasma \citep{teets12}.

\begin{figure*}
\centering
\includegraphics[angle=0,width=1.0\columnwidth,trim={0 0 0 1.25cm},clip]{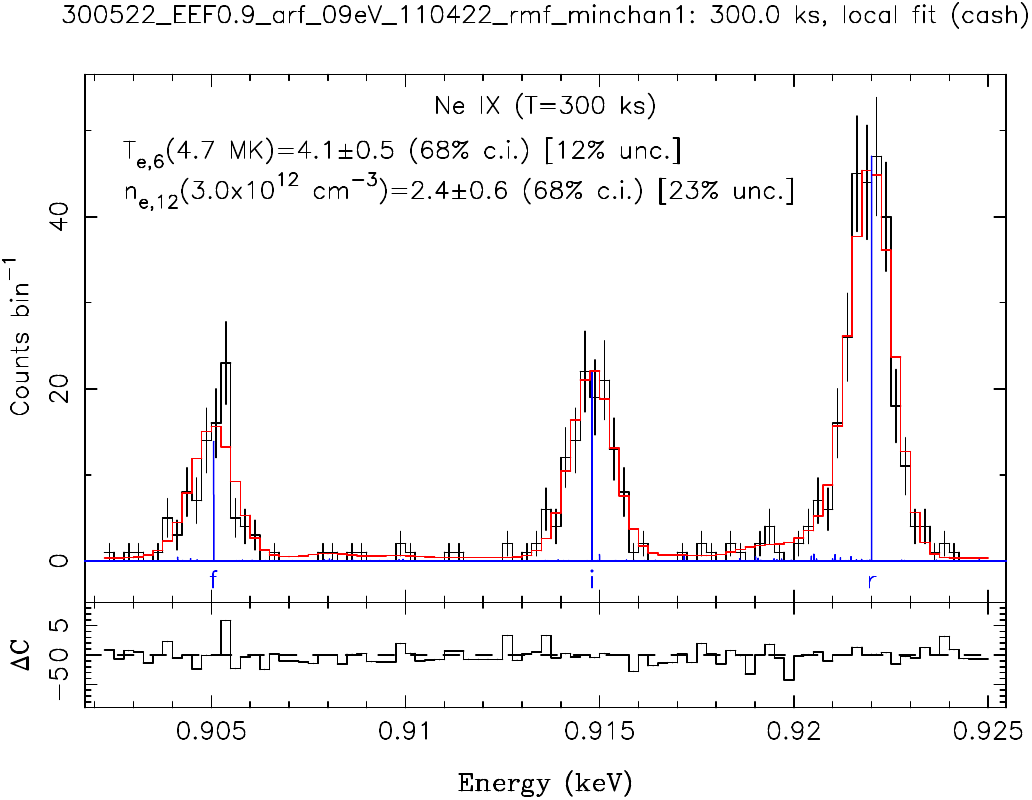} 
\includegraphics[angle=0,width=1.0\columnwidth,trim={0 0 0 1.25cm},clip]{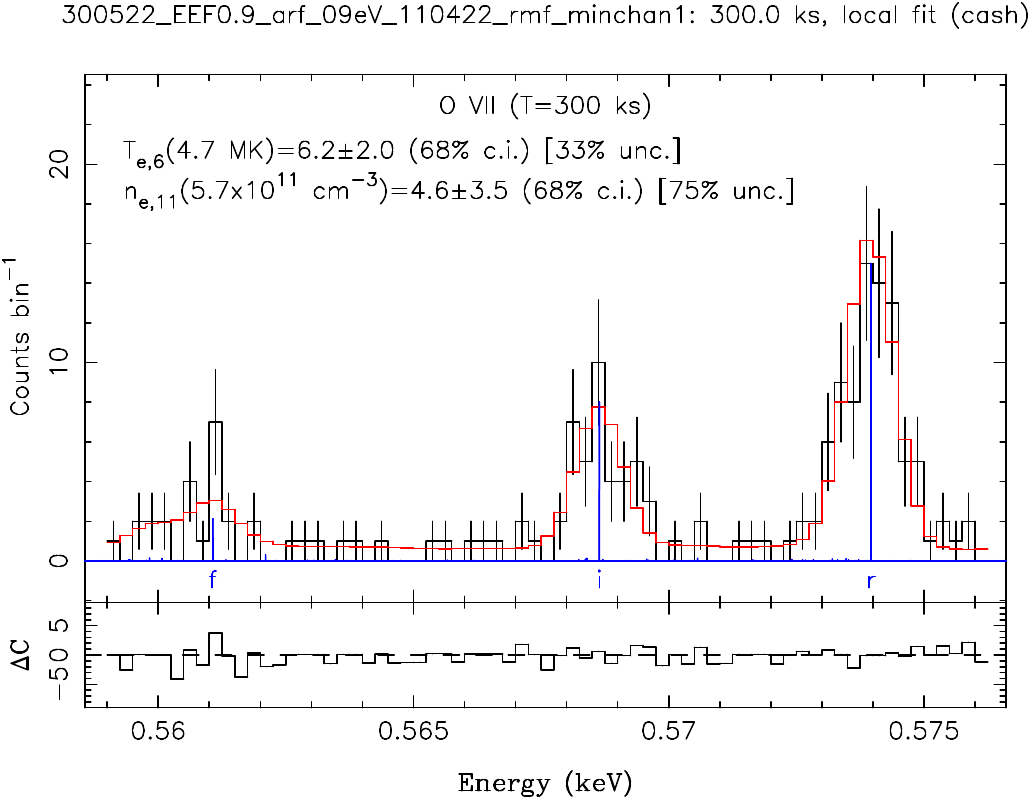}
\vspace{-0.25cm}
\caption{Simulated He-like emission lines 
of accretion shock and post-shock at the stellar surface of EX~Lupi, obtained with a single {\it LEM} central-array pixel  and an exposure of 300~ks.
Left panel: Ne~IX triplet emitted by the accretion shock 
with cool temperature ($T_\mathrm{e}=4.7$~MK) 
and high density 
($n_\mathrm{e}=3\times10^{12}$~cm$^{-3}$ was assumed).
Right panel: O~VII triplet emitted by the accretion post-shock with same temperature but lower density 
($n_\mathrm{e}=5.7\times10^{11}$~cm$^{-3}$).
The color coding is the same as in Fig.~\ref{fig:TW_Hya}.
Using  He-like ions, {\it LEM} will be able to probe the physics of an episodic accretion burst of EX~Lupi, including
the temperature and the density of the accretion shock
and post-shock.
} 
\label{fig:EX_Lup}
\end{figure*}

We computed {\it LEM} simulations of EX~Lupi using the X-ray flux that was observed
a few days before the end of its 2008 outburst, 
when its optical flux was only 4 times above its pre-outburst level 
\citep{grosso10},
this optical flux value being typical of the peaks of more regular outbursts from EX~Lupi.
Figure~\ref{fig:EX_Lup} shows the predicted {\it LEM} spectra 
centered on the Ne~IX and O~VII lines for an exposure of 300~ks. 

From the Ne~IX triplet, (1$\sigma$) uncertainties of 
12\% and 23\% are obtained for 
the temperature and the density, 
respectively, of the accretion shock.
From the O~VII triplet, (1$\sigma$) uncertainties of 
33\% and 75\% 
are obtained for the temperature and the density, 
respectively, of the accretion post-shock.
This again demonstrates that {\it LEM} can probe, 
for the first time during an episodic accretion 
from He-line ions, 
the temperature and the density of the accretion shock, 
constraining the mass-accretion rate\citep{brickhouse12}, 
and the temperature and density of the post-shock during these remarkable enhanced accretion episodes.

\subsection{X-rays from a shock in the blueshifted jet of DG~Tau~A}
\label{s:dgtau}


\begin{figure*}
\centering
\includegraphics[angle=0,width=1.0\columnwidth,trim={0 0 0 1.25cm},clip]{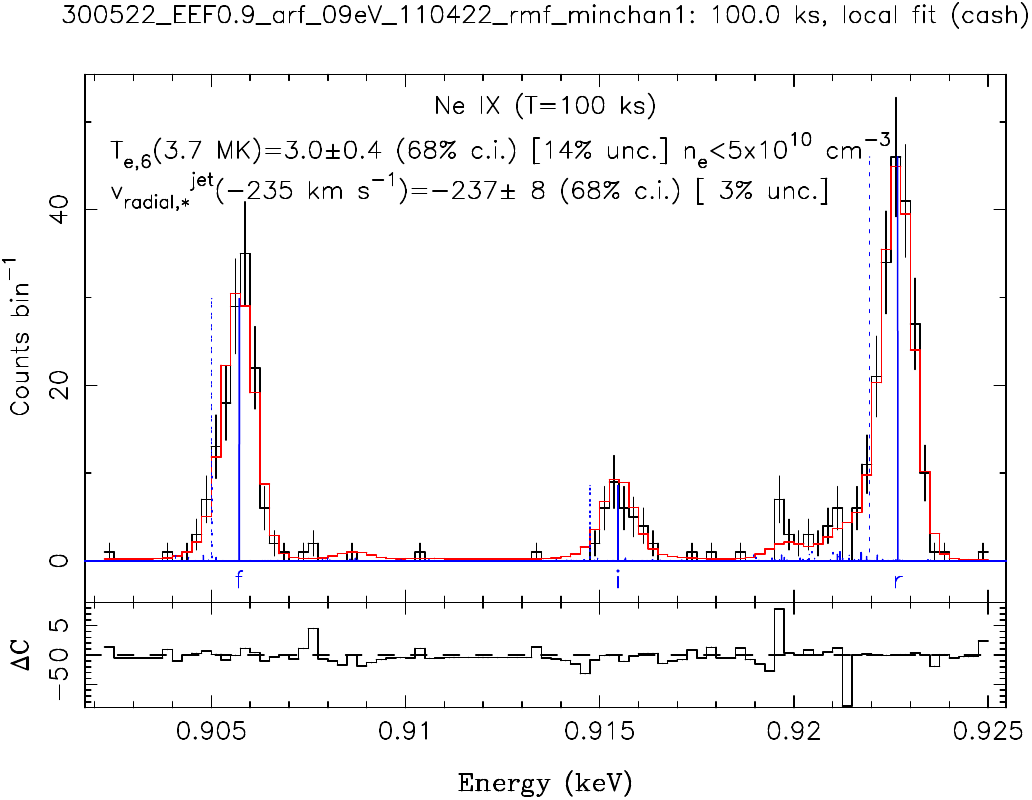} 
\includegraphics[angle=0,width=1.0\columnwidth,trim={0 0 0 1.25cm},clip]{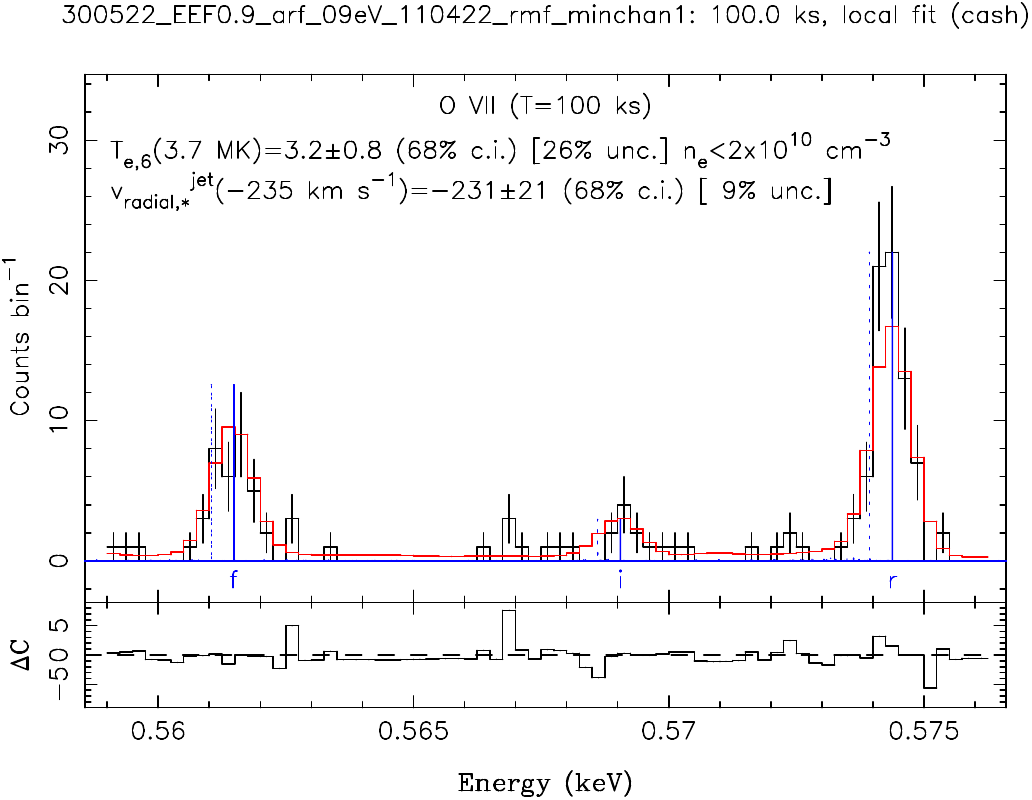}
\vspace{-0.25cm}
\caption{Simulated He-like emission lines 
of a shock in the blueshifted jet of DG~Tau~A, 
obtained with a single {\it LEM} central-array pixel  
and an exposure of 100~ks.
Left panel: Ne~IX triplet. 
Right panel: O~VII triplet.
The color coding is the same as in Fig.~\ref{fig:TW_Hya}.
In the top panels, the solid lines in blue are the theoretical spectra
for a radial velocity of -235~km~s$^{-1}$ between the X-ray emitting shock and the more embedded star;
for comparison purposes, the dashed lines in blue indicate the positions of the He-like emission lines
for the radial velocity of the star.
LEM will determine the plasma temperature 
of the X-ray emitting shock from plasma diagnostics, 
and for the first time its radial velocity.
} 
\label{fig:DG_Tau}
\end{figure*}

DG~Tau~A is a T~Tauri star located at a distance of 140~pc and has one of the highest mass accretion 
and mass outflow rates known. It drives optical jets 
with characteristic flow velocity of 300~km~s$^{-1}$ 
relative to the central star, with an inclination of 38$^\circ$ 
\citep{eisloffel98,dougados00}.
DG~Tau~A displays a two-absorber spectrum: 
the hard component is
associated with the active stellar corona screened by the
circumstellar disk, whereas the soft component originates 
from shocks located above the disk midplane, 
near the base of the jet \citep{guedel05,guedel07,guedel07d}.
A prominent X-ray (blueshifted) jet and a fainter and absorbed 
counter X-ray jet, 
both coincident with the optical jet axis, are 
resolved with {\sl Chandra}, and detected out to a distance of
$\sim$5$^{\prime\prime}$ from the central source \citep{guedel08b}. 
The bulk of the X-ray emission of DG~Tau~A is from 
the unresolved central source. 
The central soft and hard components are separated 
spatially by only $\sim$0.2$^{\prime\prime}$ 
\citep{schneider08}.
The soft X-ray emission from the shock in the jet 
is located above the circumstellar disk midplane, 
and it therefore contributes to X-ray heating and ionization of the gaseous disk surface \citep{glassgold04}.
However, {\it ``the temperature} [of the soft component] 
{\it is only poorly constrained and plasma models 
with a wide range of temperatures can reproduce the observations''}
\citep{guenther09b}.

We computed {\it LEM} simulations of the two-absorber spectrum
of DG~Tau~A \citep{guedel07}, applying 
radial velocities of +16.3~km~s$^{-1}$ and -235~km~s$^{-1}$ 
for the hard (embedded star) and soft (unresolved shock in the blueshifted jet) 
components, respectively. From the optical emission lines, 
the jet density close 
to the star is $\ge10^4$~cm$^{-3}$ \citep{bacciotti00}, 
therefore, we use the smallest available value 
in the electron density grid ($n_\mathrm{e}$=$10^{7}$~cm$^{-3}$). 
As this electron density is well below the critical densities of the He-like Ne~IX and O~VII ions, 
the forbidden line and intercombination line ratios 
will be similar for lower values still
\citep[see Fig.~8 of][]{porquet10}.
Figure~\ref{fig:DG_Tau} shows the predicted {\it LEM} spectra 
centered on the Ne~IX and O~VII lines for an exposure of 100~ks.
As a result of the low-density regime of the He-like Ne~IX and O~VII ions, 
only upper limits can be obtained for the electron density using these ions \citep[see Fig.~9 of][]{porquet10}.
The accretion shock and post-shock emissions are 
too absorbed by the circumstellar disk to be detectable, 
and in any case their radial velocities would 
be redshifted compared to the radial velocity 
of the star and spectrally well-resolved 
from the emission of the blueshifted jet.

From the Ne~IX and the O~VII triplets, 
uncertainties (1$\sigma$) of: 
14\% and 26\%, respectively, 
are obtained for the temperature;
and 3\% and 9\%, respectively, for the radial velocity.
LEM will therefore determine, for the first time,  
the radial velocity and (unambiguously) the plasma temperature 
of the shock in the X-ray jet of DG~Tau~A 
that is pointing towards us.  This would provide important and unique constraints for a T~Tauri jet.

\subsection{The Orion Nebula Cluster}
\label{s:onc}

The {\em Chandra} Orion Ultradeep Project (COUP)\cite{Getman2005,Feigelson2005}, an 850 ksec long continuous observation of the Orion Nebula Cluster (ONC), was arguably one of the most successful {\em Chandra} observations of star forming regions and young stars. More than 1600 sources were detected in the {\em Chandra} ACIS FoV. Although many sources are concentrated around the center of the ONC and will not be accessible at the spatial resolution of {\it LEM} due to confusion, the outer regions are much less dense and individual stars will be resolved. Thanks to the larger field FoV of {\it LEM} with respect to {\em Chandra} ACIS, a much larger region will be observable in a single exposure, including more sparse regions outside of the cluster center. For studies such as that of bright flares, discussed below, source confusion will not be an issue since these events are typically two dex higher in count rate than the quiescent level. Hence they will be easily recognized above the "background" level due to the unresolved emission from the weak and spatially unresolved sources. 

\begin{figure*}
    \centering
    \includegraphics[width=0.8\textwidth]{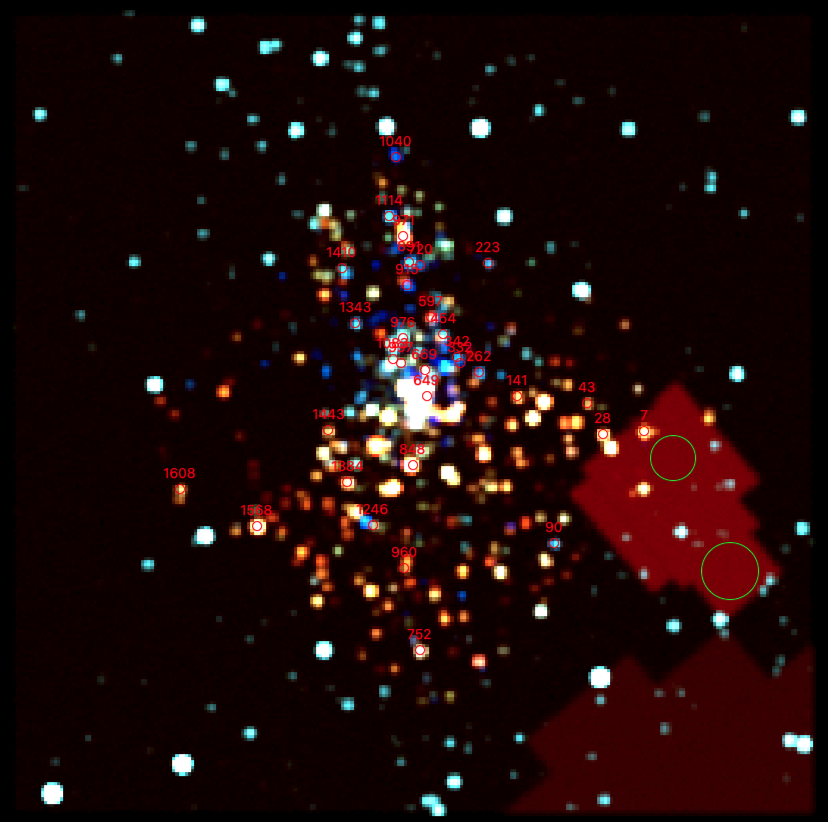}
    \caption{Simulated {\it LEM} observation of the Orion Nebula Cluster. Point sources in the central parts of the field are simulated with the same Chandra-determined position, spectrum and flux of those detected in the COUP observation~\cite{Getman2005}. Outer sources are simulated from a list of candidate members \cite{Prisinzano+2022AA,He2022} using the average spectrum of ONC members and a flux determined from the mean J magnitude vs.\ L$_X$ relation. Two diffuse emission regions, detected in XMM-Newton data (Ref.~\citenum{guedel08b}; see their Fig.\,S1), are also simulated with the intensity and spectral shape reported in that work. Their spatial extent is schematically represented by the two reddish areas in the bottom-right part of the image. Three X-ray spectral bands, 200-900\,eV, 900-1500\,eV, and 1500-3000\,eV are used for the red, green, and blue components of the false color image, respectively. Red circles, with source IDs above them, indicate the positions of a sample of sources studied by \cite{Favata2005}, which experienced strong flares during the COUP observation. The circles are also used as photon extraction regions for the spectra shown below. The two green circle in the northern diffuse emission regions was also used to extract events for spectral analysis.} 
    \label{fig:COUPsimu}
\end{figure*}

We have performed a simulation of a 1\,Ms {\it LEM} observation of the ONC---comparable to the COUP {\it Chandra} exposure---in order to estimate the number of PMS stars that will be observable (i.e. not confused) and the signal-to-noise ratio expected for the resulting X-ray spectra. Figure\,\ref{fig:COUPsimu} shows a false color image of the central part of the {\it LEM} field as obtained with the SOXS simulator. Source position and spectral models (for 1- or 2-T absorbed thermal plasma) for the central part of the field, observed by the COUP project, were taken from Ref.~\citenum{Getman2005}. Outside of this area we adopted a list of candidate cluster members from Refs.~\cite{Prisinzano+2022AA,He2022}. For these we adopted an average spectrum and estimated the flux from the mean J magnitude vs. L$_X$ relation.


The simulated dataset also included a crude representation of the extended emission detected by {\em XMM-Newton}\cite{guedel08b}. The two reddish areas in the bottom-right of the image in Figure~\ref{fig:COUPsimu} represent these two soft emission regions (one of which extends beyond the FoV), each simulated with its published spectrum, and with an approximate (boxy) geometrical shape. 

A sample of 32 stars in which bright X-ray flares occurred\cite{Favata2005} during the COUP observation is marked with red circles, with the respective COUP source numbers plotted above them. The circles also correspond to the areas from which photons were extracted for spectral analysis. The larger green circle in the northern diffuse emission region was also used for spectral extraction. 

\begin{figure*}
    \centering
    \includegraphics[width=0.49\textwidth]{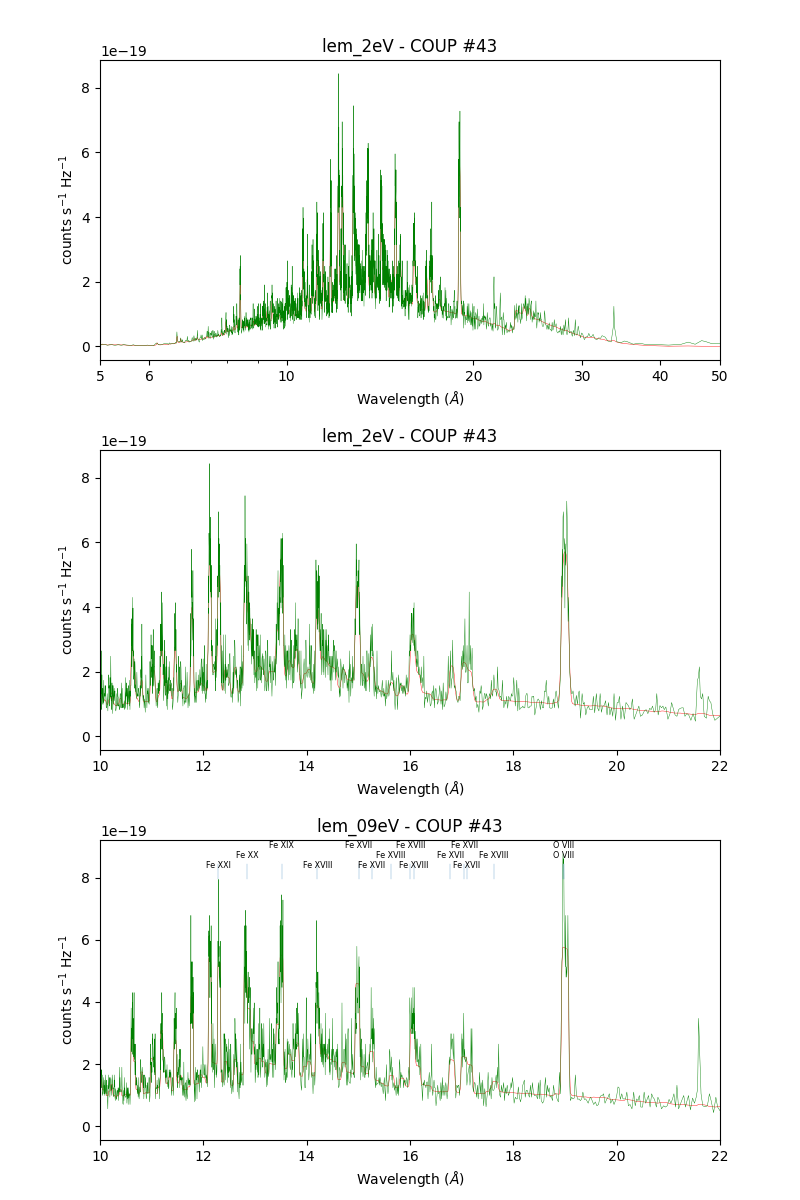}
    \includegraphics[width=0.49\textwidth]{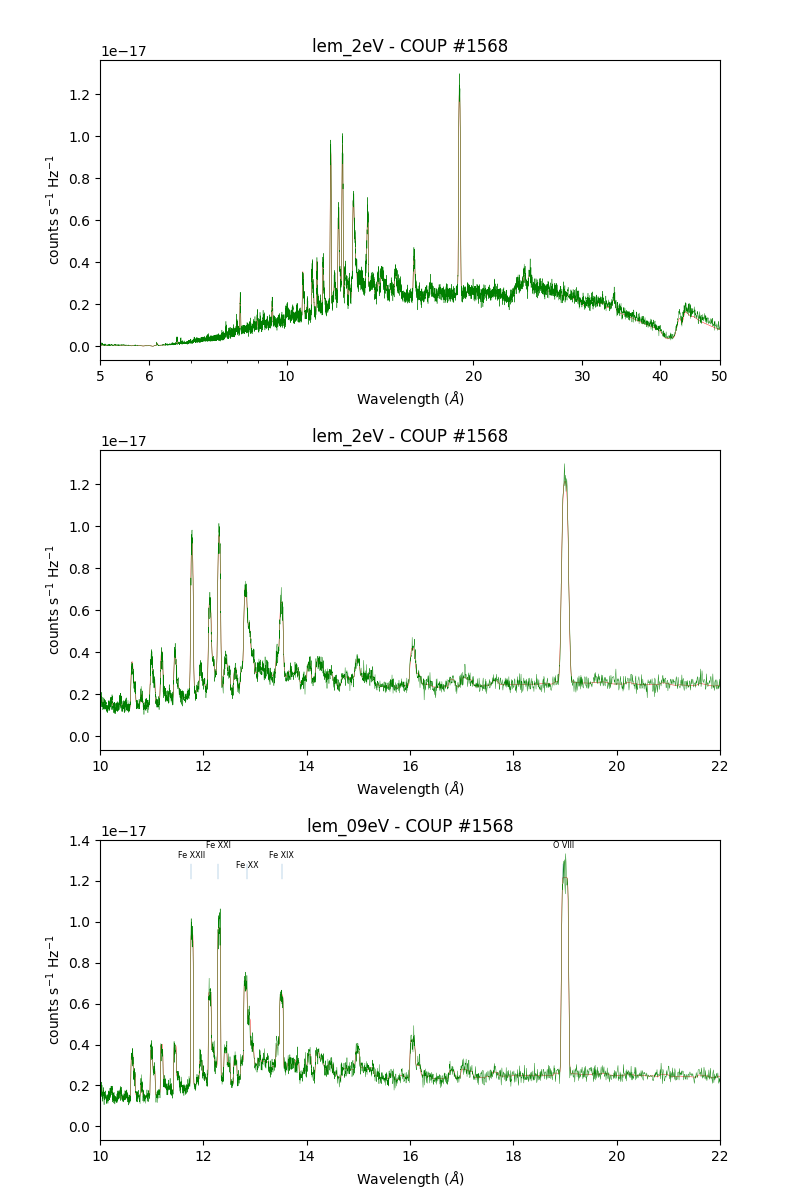}
    \caption{Simulated {\it LEM} spectra for two COUP sources, COUP 43 on the left (7021 Chandra counts 42000 simulated {\it LEM} counts) and COUP 1568 on the right (89400 Chandra/730000 {\it LEM}). Both sources are isolated and uncontaminated by neighbouring sources, with the closest neighbor at $\sim$24$^{\prime\prime}$ and $\sim$20$^{\prime\prime}$, respectively. For each source we show, from top to bottom: the full spectrum simulated with the 2\,\AA-resolution response matrix appropriate for the outer part of the microcalorimeter array, in the 5-50\AA\, wavelength range; the same spectrum in the 10-22\,\AA\, range; the spectrum simulated with the 0.9\,\AA-resolution response matrix corresponding to the central 7~arcmin region of the array, in the same 10-22\,\AA\, range. In all cases, the simulated spectrum is plotted in green and the best-fit model in red. In the bottom panels some prominent lines are marked with the ion that produces them.}
    \label{fig:COUPsimu_spectra}
\end{figure*}

A significant fraction of the stars studied by Ref.~\citenum{Favata2005}, which we take  as a representative sample of ``interesting'' sources, is isolated enough to allow the analysis of their spectra without contamination from any surrounding sources. Two such spectra, simulated with the models derived from COUP data\cite{Getman2005}, are shown in Fig.\,\ref{fig:COUPsimu_spectra} to illustrate the foreseen range of quality of spectra. Approximately 100 COUP sources have more photon events than COUP 43, the faintest of the two.

\begin{figure}
\centering
\includegraphics[angle=0,width=1.0\columnwidth,trim={0 0 0 1.1cm},clip]{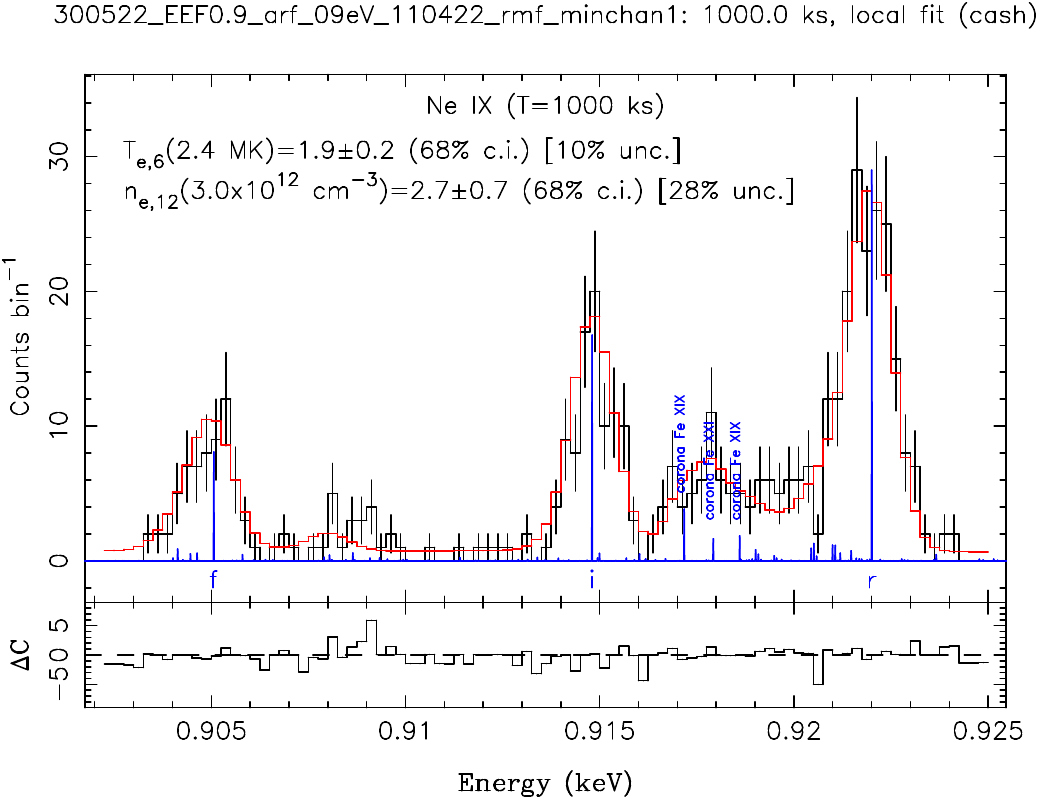}
\caption{Simulated {\it LEM} spectrum of COUP~1150 at 0.9~eV resolution, focusing 
on the Ne~IX triplet. 
The color coding is the same as in Fig.~\ref{fig:TW_Hya}.
A cool plasma component was detected 
from CCD spectroscopy with {\it Chandra}
in this isolated (closest neighbor at $\sim$15.1$^{\prime\prime}$) accreting young star. 
LEM will be able to diagnose the density 
of the accretion shock.}
\label{fig:COUP1150}
\end{figure}

In addition, {\it LEM} will be able to diagnose the densities of accretion shocks in these Orion stars, building up a large sample of objects to allow systematic study of accretion properties and how they correlate with disk and stellar properties. An example is shown in Figure~\ref{fig:COUP1150} illustrating the Ne~IX He-like complex in COUP~1150 in which a cool plasma component likely associated with accretion was detected.

\begin{figure}
    \centering
    \includegraphics[width=0.49\textwidth]{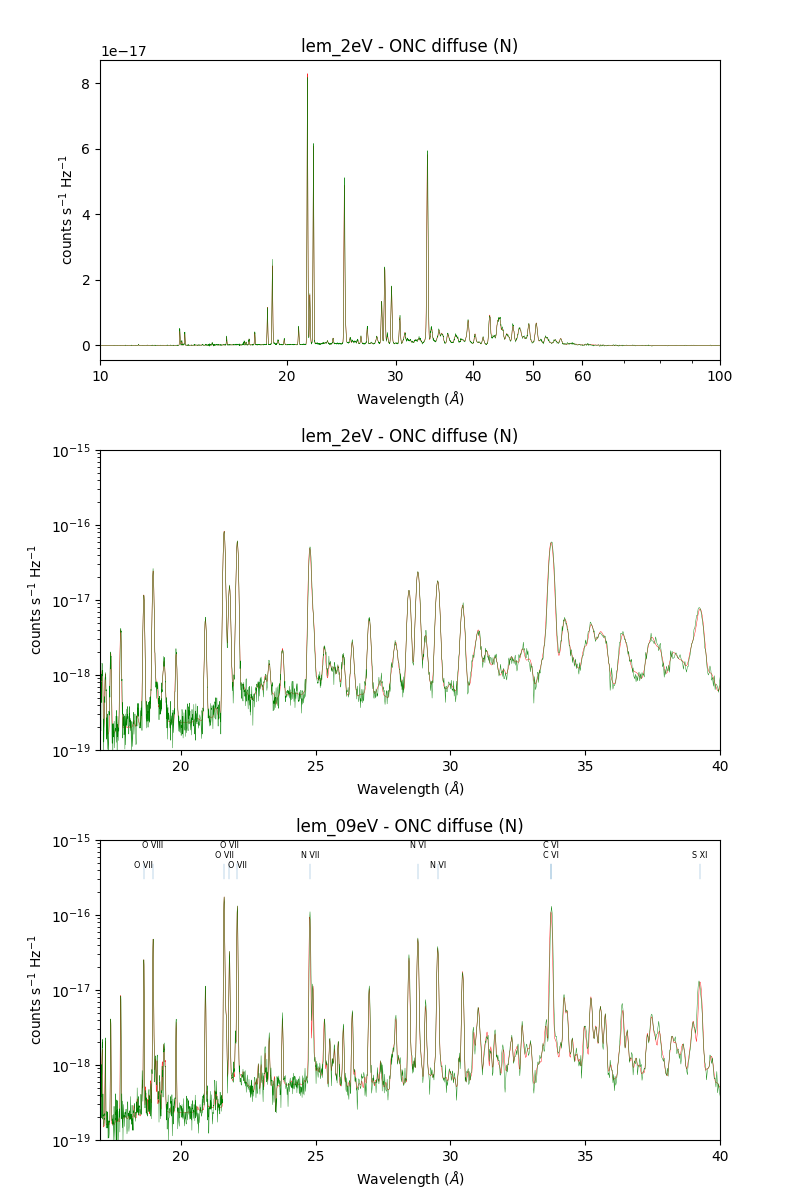}
    \caption{Simulated {\it LEM} spectra of the diffuse emission, as extracted from the two green circles in Fig.\,\ref{fig:COUPsimu}. For a description of the panels and color coding see  Figure~\ref{fig:COUPsimu_spectra}. Note that the flux scale in the two bottom panels is logarithmic.}
    \label{fig:COUPsimu_diffuse_spec}
    \label{fig:spec_ONC_diffuse}
\end{figure}

Figure~\ref{fig:COUPsimu_diffuse_spec} shows spectra extracted from the two green circles in Figure~\ref{fig:COUPsimu}  background-subtracted using a nearby region of similar area (not shown). The input spectrum was based on the {\it XMM-Newton} detection\cite{guedel08b}. 
The X-ray emitting plasma fills a stellar-wind bubble, nicknamed the Veil bubble,
which has a 2.7~pc radius, centered at a projected distance 1.5~pc South-West 
from the Trapezium cluster\citep{pabst19,pabst20}.
The adiabatic expansion of this ionized bubble inside the surrounding neutral gas produces a symmetrical shell, expanding at a bulk velocity 
of 13~km~s$^{-1}$ (see Fig.~3 of Ref.~\citenum{pabst22}).

Note the high signal-to-noise ratio of the simulated spectra in Figure~\ref{fig:COUPsimu_diffuse_spec} and the wealth of detail. The low energy sensitivity of {\it LEM} will enable a detailed spatial analysis of the soft diffuse emission. 
The best fit to the simulated {\it LEM} spectrum in Fig.\,\ref{fig:spec_ONC_diffuse} indicates that the uncertainty on the radial velocity, left as a free parameter along with temperature and abundances, is of the order of 1~km~s$^{-1}$, which is easily sufficient for investigating the plasma dynamics.
LEM will obtain the plasma temperature 
of the collisional plasma inside the Veil bubble
from two complementary methods: 
the fit of the full spectrum 
and the He-like line ratio, $G=(f+i)/r$.
Inconsistent results between the former and the latter 
will point to the contribution of charge exchange 
\citep{yang20b}.

In a collision-dominated plasma, the brightest line of the He-like triplet 
is the resonance line, whereas charge exchange 
produces no continuum emission 
and a He-like triplet where the brightest line is the forbidden line
(see Fig.~6 of Ref.~\citenum{porquet10}).
Therefore, the $G$ ratio is sensitive to any contribution 
of charge exchange, since here photoionization that could otherwise skew the line ratios can be excluded.
Indeed, a contribution of charge exchange 
to the X-ray diffuse emission that is observed 
in massive star-forming regions has been proposed\citep{townsley11b,montmerle12}.
Based on the strong O~VII triplet of the ONC diffuse X-ray emission (top panel of Fig.~\ref{fig:spec_ONC_diffuse}), 
LEM will diagnose any contribution of non-thermal X-ray emission
that could be produced by charge exchange between 
the recombining O~VIII ions and neutral elements.


\section{X-RAYS FROM YOUNG BROWN DWARFS}

In contrast to young stars, which are massive enough to trigger at their center the fusion of hydrogen 
as they evolve towards the main sequence, brown dwarfs are of lower mass than the H-burning minimum mass of 0.075~$M_\odot$/78.5~$M_\mathrm{Jup}$ \citep{chabrier23}. Since they cannot
burn hydrogen, they slowly cool down as they evolve.
Not surprisingly, field brown-dwarfs were first identified in the near-infrared \citep{nakajima95,rebolo95}.
But a few young brown dwarfs were also detected in X-rays by {\sl ROSAT} in star-forming regions 
\citep{neuhaeuser98,mokler02}. 

With {\sl XMM-Newton} and {\sl Chandra}, the detected number of 
young brown dwarfs in clusters in X-rays exceeds now the one of field brown dwarfs, which are older and hence cooler, 
and, therefore, mainly detected during X-ray flares \citep{rutledge00,stelzer04}. 
In the Orion nebula cluster, only 9 of the 34 young brown dwarfs observed by {\sl Chandra} were detected, those 
mainly being the ones having an extinction lower than 5~mag \citep{preibisch05}.
In the Taurus molecular cloud (TMC), half of the sample (9 out of 17 brown dwarfs) was detected by {\sl XMM-Newton} 
(7 of these brown dwarfs were detected for the first time in X-rays), including an X-ray flare \citep{grosso07}.
The atmospheres of the young brown dwarfs are likely warm enough 
to be mildly ionized and, therefore, to sustain coronal activity.

\begin{figure*}
\centering
\includegraphics[angle=0,width=0.49\textwidth,trim={0 0 0 1.1cm},clip]{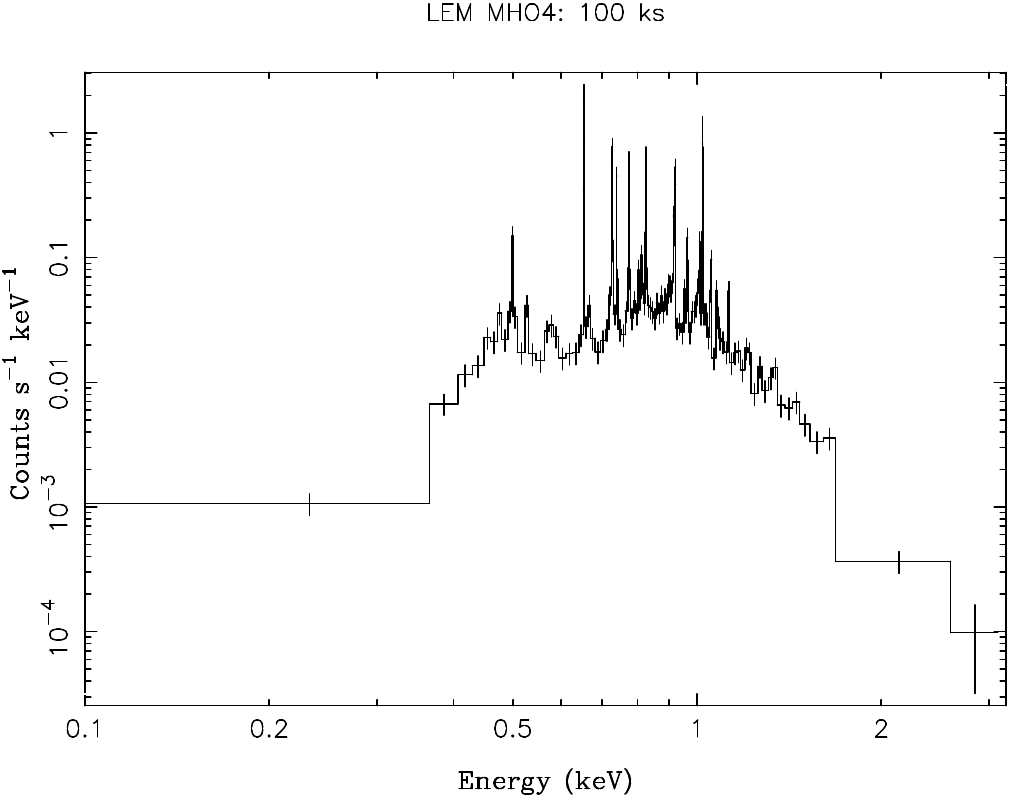} 
\includegraphics[angle=0,width=0.49\textwidth,trim={0 0 0 1.1cm},clip]{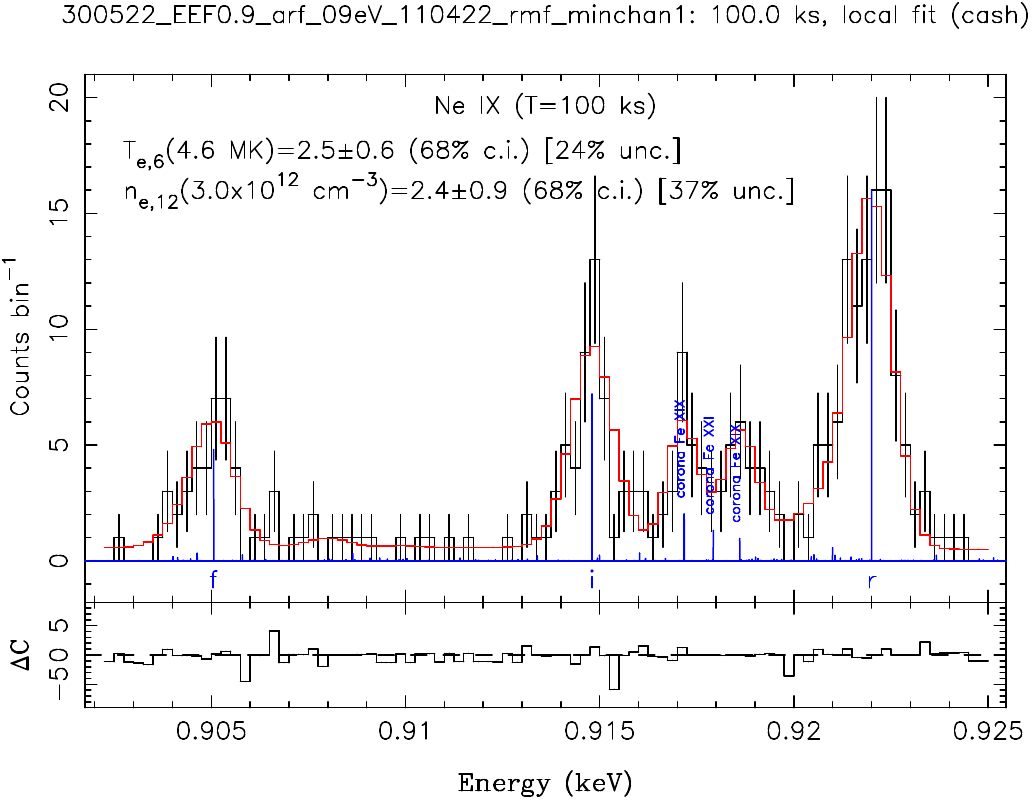}
\includegraphics[angle=0,width=0.49\textwidth,trim={0 0 0 1.1cm},clip]{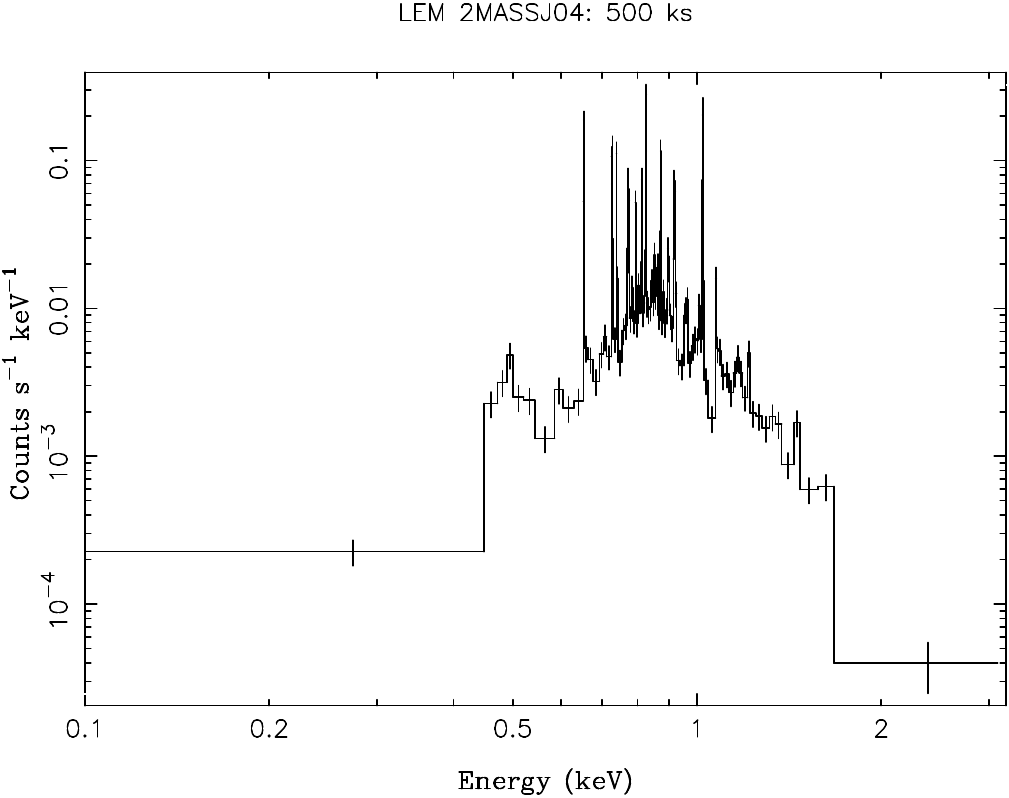} 
\includegraphics[angle=0,width=0.49\textwidth,trim={0 0 0 1.1cm},clip]{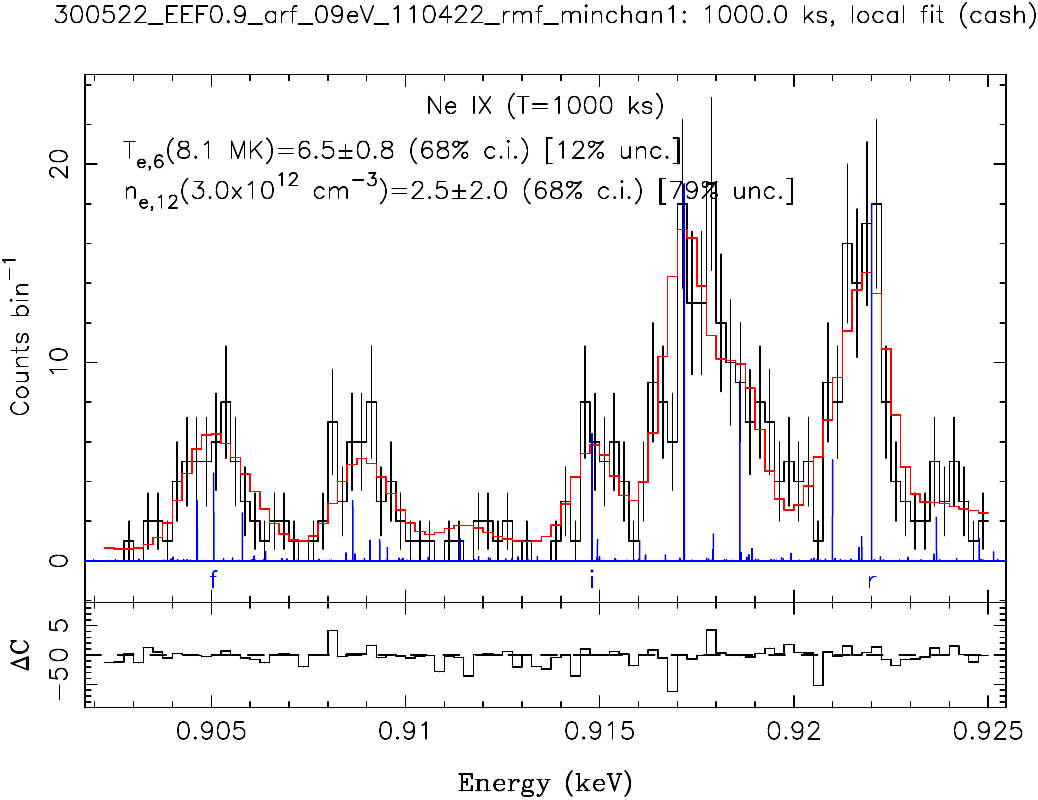}
\caption{Simulated spectra of the young brown dwarfs MHO~4 (top row) and 2MASS~J0422 (bottom row)
in the Taurus molecular cloud, 
obtained with a single {\it LEM} central-array pixel  and an exposure of 100~ks.
Left panels show the normalized X-ray spectrum of the absorbed, two-component plasma \citep{grosso07}.
Right panels illustrate the Ne~IX triplet emission of the cool plasma ($T_\mathrm{e}=4.6$~MK) 
assuming a (putative) high-density 
($n_\mathrm{e}=3\times10^{12}$~cm$^{-3}$)
to mimic an accretion shock. The color coding is the same as in Fig.~\ref{fig:TW_Hya}.
} 
\label{fig:MHO4}
\end{figure*}

We focus here on the young brown dwarfs of the TMC, which are closer (located at 140~pc) and less extinguished than the young brown dwarfs of the Orion nebula cluster, and use for our {\it LEM} simulations the best-fit X-ray parameters \citep{grosso07}.
We select two representative young brown dwarfs of this TMC sample:
MHO~4 (M7 spectral type, corresponding to $T_\mathrm{eff}$$\sim$2850~K), which is the brightest in X-rays, 
and 2MASS~J0422, which has the later spectral-type 
(M8 spectral-type, corresponding to $T_\mathrm{eff}$$\sim$2700~K) of the X-ray detected, young brown dwarfs of this sample
(Fig.~\ref{fig:MHO4}).

The comparison between the {\it LEM} simulated normalized spectra (left panels of Figure~\ref{fig:MHO4} 
and the previously obtained {\sl XMM-Newton EPIC} spectra 
(Fig.~4, top rows of Ref.~\citenum{grosso07}) 
shows the giant leap in spectral resolution that will be achieved with {\it LEM} 
in 100~ks and 500~ks, respectively, for MHO~4 and 2MASS~J0422, 
which will allow, for the first time, the detailed study of young brown dwarf coronae.

The Ne~IX triplet emission will be obtained with an exposure of only 100~ks for MHO~4; 
a 10 times deeper exposure will be need for the fainter 2MASS~J0422.
LEM will probe from the Ne~IX triplet,
for the first time in young brown dwarfs, 
the temperature and the density of the cool plasma-component, providing insights into its nature and structuring.


\section{STELLAR FLARE PHYSICS}
\label{s:flares}



\subsection{Flares in Stellar Clusters}

The large {\it LEM} fov and angular resolution will allow, for the first time, to obtain with a single pointing 2eV (or better) resolution spectra of several tens or even hundreds members of young clusters (e.g., the Pleiades) and star forming regions (e.g., Cha-II, Orion, $\rho$ Oph) allowing to systematically investigate possible relations between
the stellar and/or the circumstellar disk parameters and/or the evolutionary stage of the systems and 
spectroscopically-revealed features in X-ray spectra: coronal chemical composition, density of emitting plasma and its possible variation during a flare, bulk velocity of emitting plasma etc. 

Available observations (e.g. the COUP or DROXO) that a fraction of 1-3 Myr old YSOs show evidence of intense long-decay flares that require the existence of confined structures (loop-like) 10 times or more longer than the one associated with "normal" solar and stellar flares. It is still matter of debate\cite{getman21,Reale2018} 
if such structure footpoints are both anchored on the stellar surface, like solar coronal loops, or if one is on the star and other on the circumstellar disk near the corotation radius\cite{Favata2005,Reale2018}, or even both on the disk\cite{Galeev1979}.

The Chandra ONC Ultradeep Project (nicknamed COUP) has allowed to detect about 30 such intense flares during the 850 ksec long continuous observation. The flare peak X-ray luminosities in the 0.5-8.0 keV bandpass are in the range $L_{X} \sim 10^{31} - 8 10^{32}$ erg/sec, $f_X \sim 4~10^{-13} - 3~10^{-11}$ erg/sec/cm$^2$ \cite{Favata2005}, the emitting plasma has a rather high temperature in the 50--100 MK range. The predicted {\it LEM} count rate should be sufficient to perform spectroscopy of flare emission with a time resolution of about 1 ksec. Considering big flare frequency (30 out of 1600 sources during 850 ksec of Chandra observation), the predicted {\it LEM} rates and the extents of the ACIS and {\it LEM} fov, one can predict than in a 300 ksec long {\it LEM} observation at least 10 big flares will occur and at least 3-4 of them with {\it LEM} rates adequate for a detailed investigation. This number rises to more than 10 events in a 1~Ms legacy observation, analogous to COUP.
Time-resolved spectroscopy of one of such large flare with a time resolution of about 1ks should be sufficiently sensitive to detect the sloshing ("up and down") velocities at flare onset  (cf. Fig. 3 of Ref.~\citenum{Argiroffi2019} and constrain the nature of the structure where the flare occurs.

While COUP has shown the existence of more than 1600 sources in the Chandra FoV (mostly concentrated around the center of the Orion Nebula Cluster), for investigating the nature of large intense flares source confusion will not be a problem since such flare emission is typically two dex higher that the quiescent level, and hence it easily recognized above the pedestal "background" level due to the unresolved emission from the weak, and possibly spatially unresolved, sources.  


\subsection{Flares on Active Field Stars}

\begin{figure*}
    \centering
    \includegraphics[width=0.9\textwidth,angle=0]{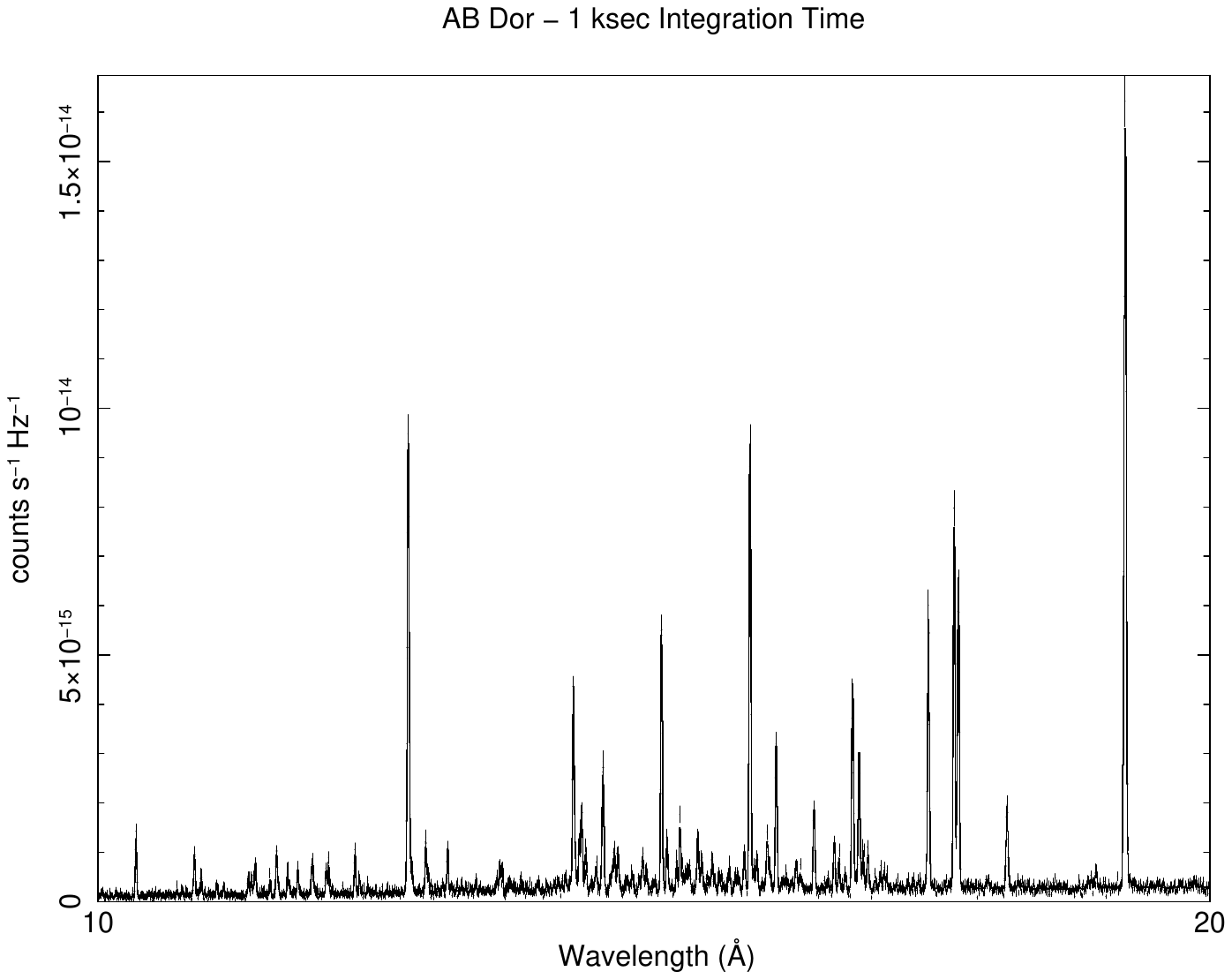}
    \vspace{-0.2in}
    \caption{The simulated spectrum of AB Dor in the 10-20~\AA\ range obtained with a 1 ksec {\it LEM} observation with a resolution of 0.9eV. The model spectrum includes the quiescent and the flare emission components. All emission lines are thermally-broadened and the flare component is Doppler-shifted with a velocity of 100 km/sec.  The two most prominent emission lines are the OVIII and FeXVII with rest wavelengths of 18.967 Ang and 15.014 Ang, respectively.}
    \label{fig:ABDORspectrum}
\end{figure*}

AB Dor (K0V+M8) is a well known young stellar system in the constellation Dorado. The primary is a flare star that shows periodic increases in activity. AB Dor has been extensively studied with the {\it XMM-Newton} EPIC/PN and RGS and detailed best fit to its coronal emission have been published\cite{Gudel2001} 
both for the quiescent and flare emission. Adopting AB Dor as a prototype of a possible source for a {\it LEM} study one can start from those published results. 

By assuming 
that during a flare rising phase (with a typical duration of $\sim$ 1 ksec)  some bulk motion of the emitting plasma is thought to occur as a result of chromosheric evaporation of plasma into the flaring loop. One can simulate the predicted source emission and investigate the capability of {\it LEM} to recover a Doppler shift velocity (more precisely its projection along the line of sight) of 100 km/s, with an integration time of 1 ksec.  Such a value is consistent with several solar studies as well as with a few stellar studies\cite{Argiroffi2019}. Since the plasma bulk motion should be mainly concentrated during the flare onset, the strategy of improving the line signal-to-noise ratio by increasing the integration time can hardly work since it will imply a smearing of observable velocity shift.

For all simulated spectra it has been assumed that the flare occurs "on top" of the quiescent emission (that is modelled according to published best fits) and that the X-ray flux, $f_X$, during flare (indeed during the fraction of flare light-curve considered) is 2$\times$ the quiescent $f_X$.  The simulations have been performed assuming $f_X$ equal, 10 and 100 times weaker than  the AB Dor case observed (heretofore the nominal) one. The quiescent AB Dor $f_X$ is $4~10^{-11}$ erg/sec/cm$^2$ in the 0.3-10 keV band pass. The quiescent emission is modeled as a 3 Temperature Vapec and the (additional) flare as an absorbed Doppler-shifted 4 Temperature Vapec. The {\it LEM} predicted quiescent and flare nominal $f_X$ in the 0.2-2.0 keV bandpass are $\sim 3.8~10^{-11}$ and $\sim 7.2~10^{-11}$ erg/sec/cm$^2$, respectively,  corresponding to count rates of $\sim 5.8~10^{1}$ and $1.1~10^{2}$ count/sec. 

The simulated spectra allow us to investigate for the 3 f$_X$ values: a) the capability to recover the input Doppler shift velocity by studying a single isolated rather strong emission line; b) the capability  to recover various input parameters (Doppler shift velocity, abundances, kTs, normalizations) by fitting the model to the entire spectrum, i.e. to {\bf all} spectral lines (a procedure usually adopted for optical spectra).
In the simulated spectra all the strong lines are in the 10-20 Ang range (Figure. \ref{fig:ABDORspectrum}), which is exactly the "sweet spot" of the {\it LEM} bandpass. The Fe~XVII line at 15.014~\AA\ (E= 825.763 eV), is the second strongest line and is almost isolated (according to the adopted APEC model).  Energy resolution at the FeXVII rest energy is $\sim$ 25\% better than for the strongest line, OVIII 18.967 \AA.

\begin{figure*}
    \centering
    \includegraphics[width=0.32\textwidth,angle=0]{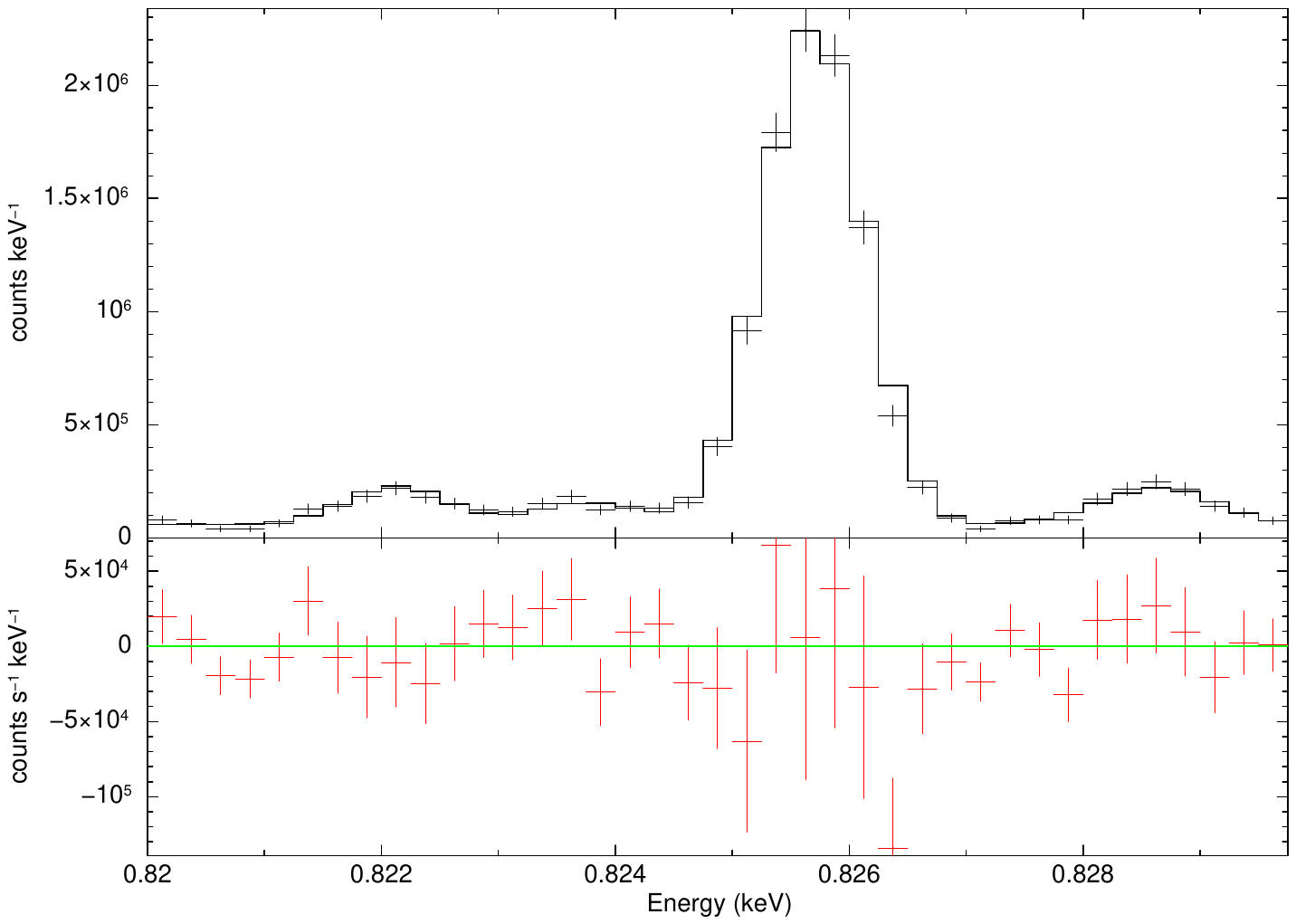}
     \includegraphics[width=0.32\textwidth,angle=0]{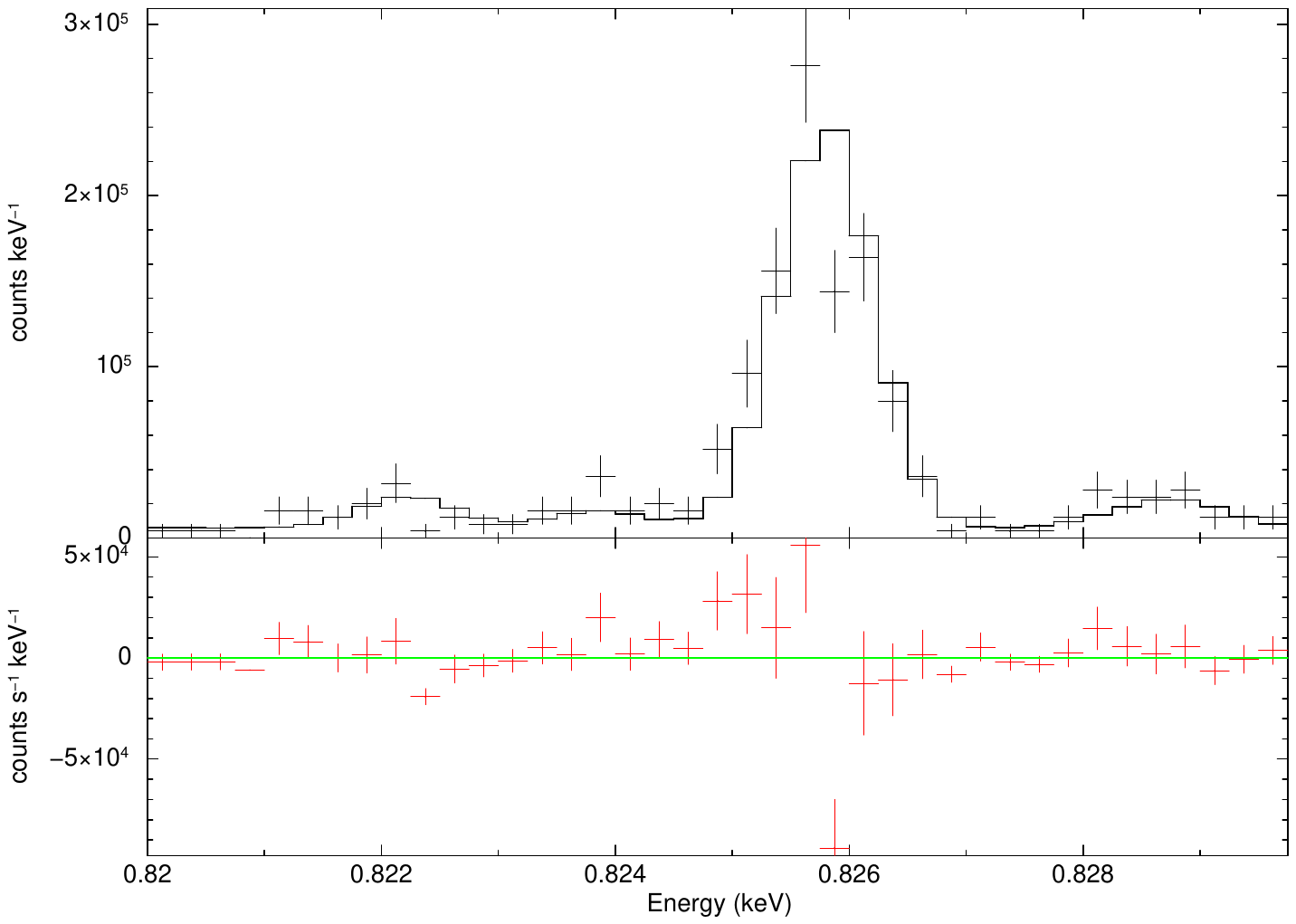}
     \includegraphics[width=0.32\textwidth,angle=0]{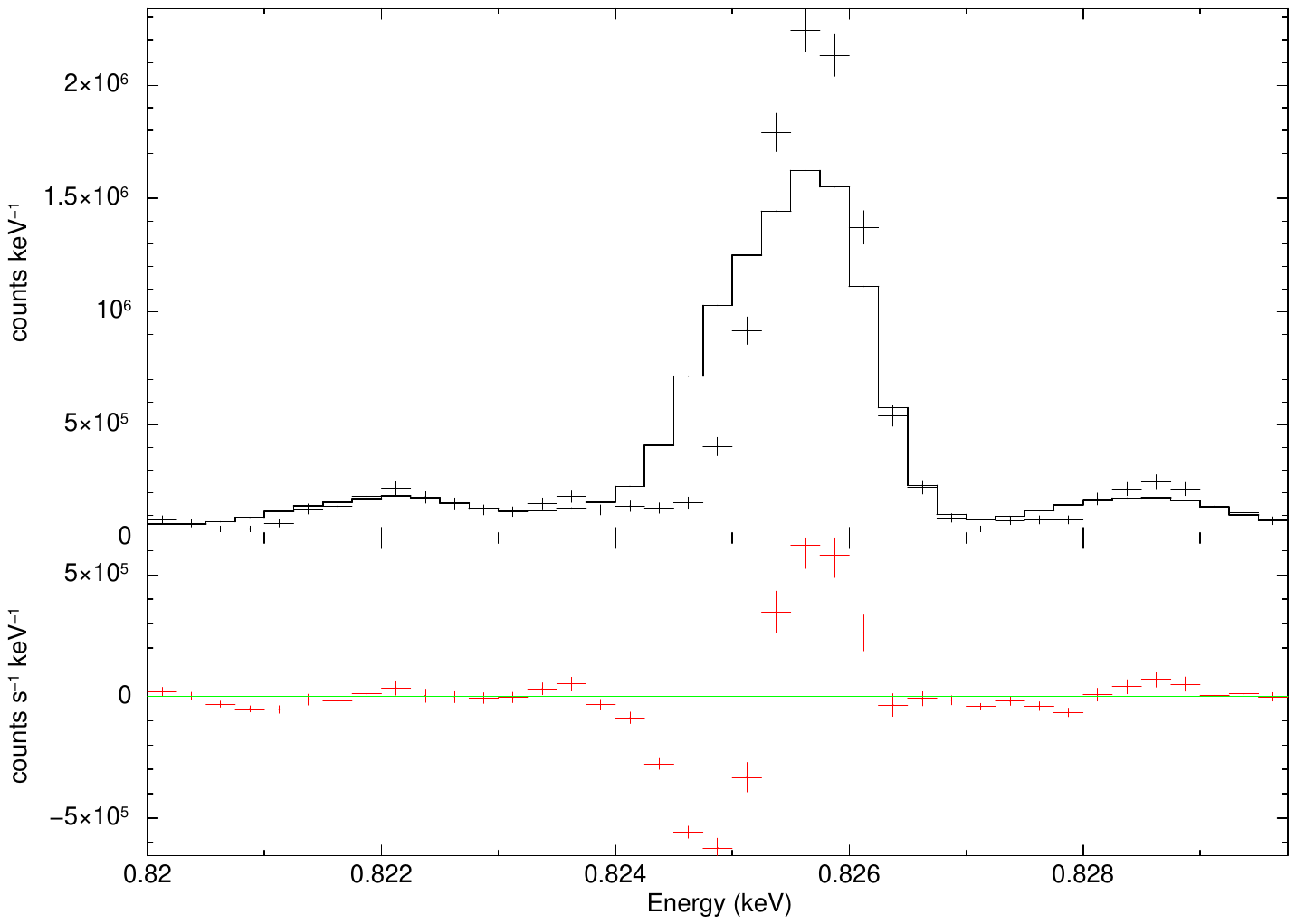}
\includegraphics[width=0.32\textwidth,angle=0]{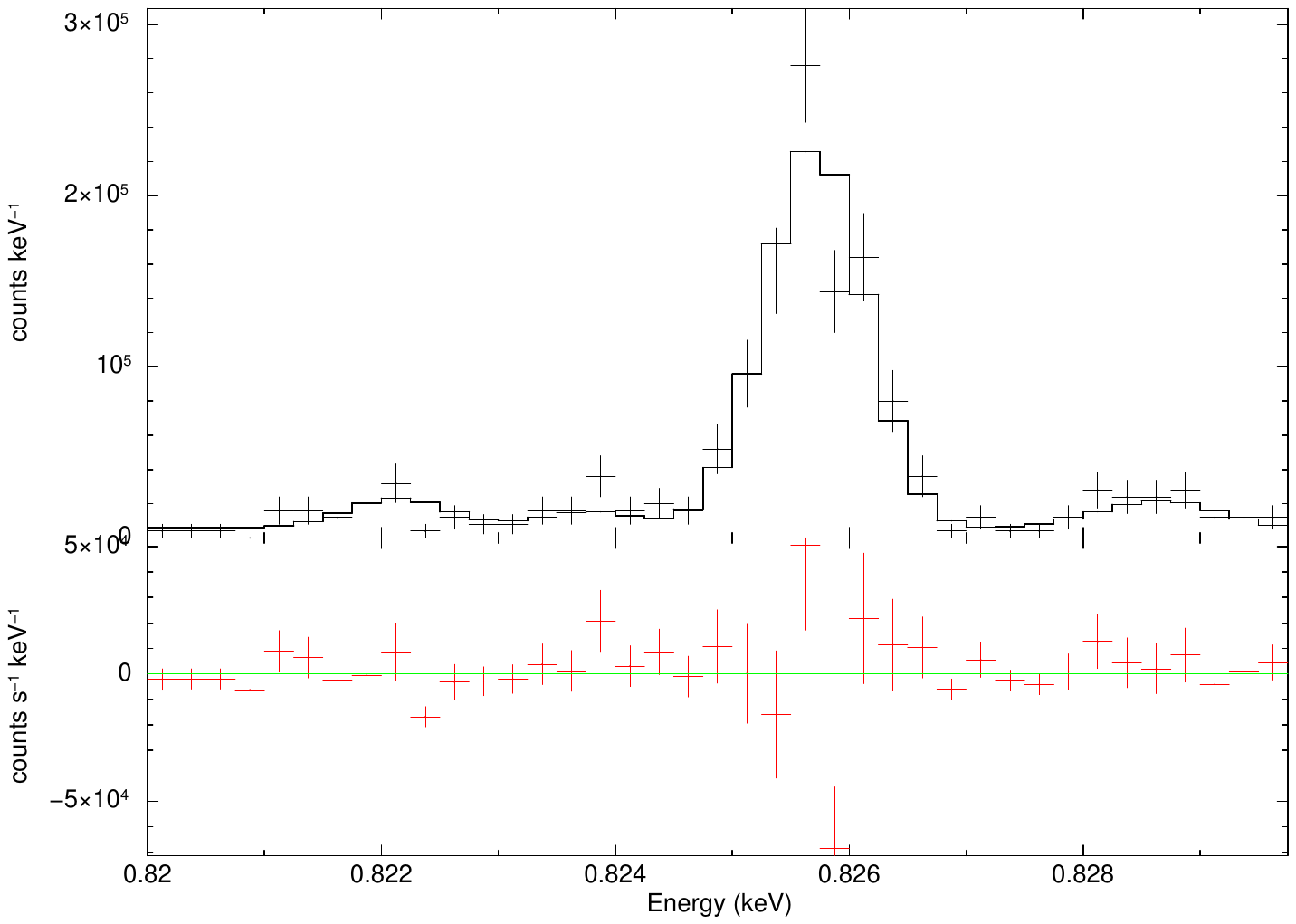}
     \includegraphics[width=0.32\textwidth,angle=0]{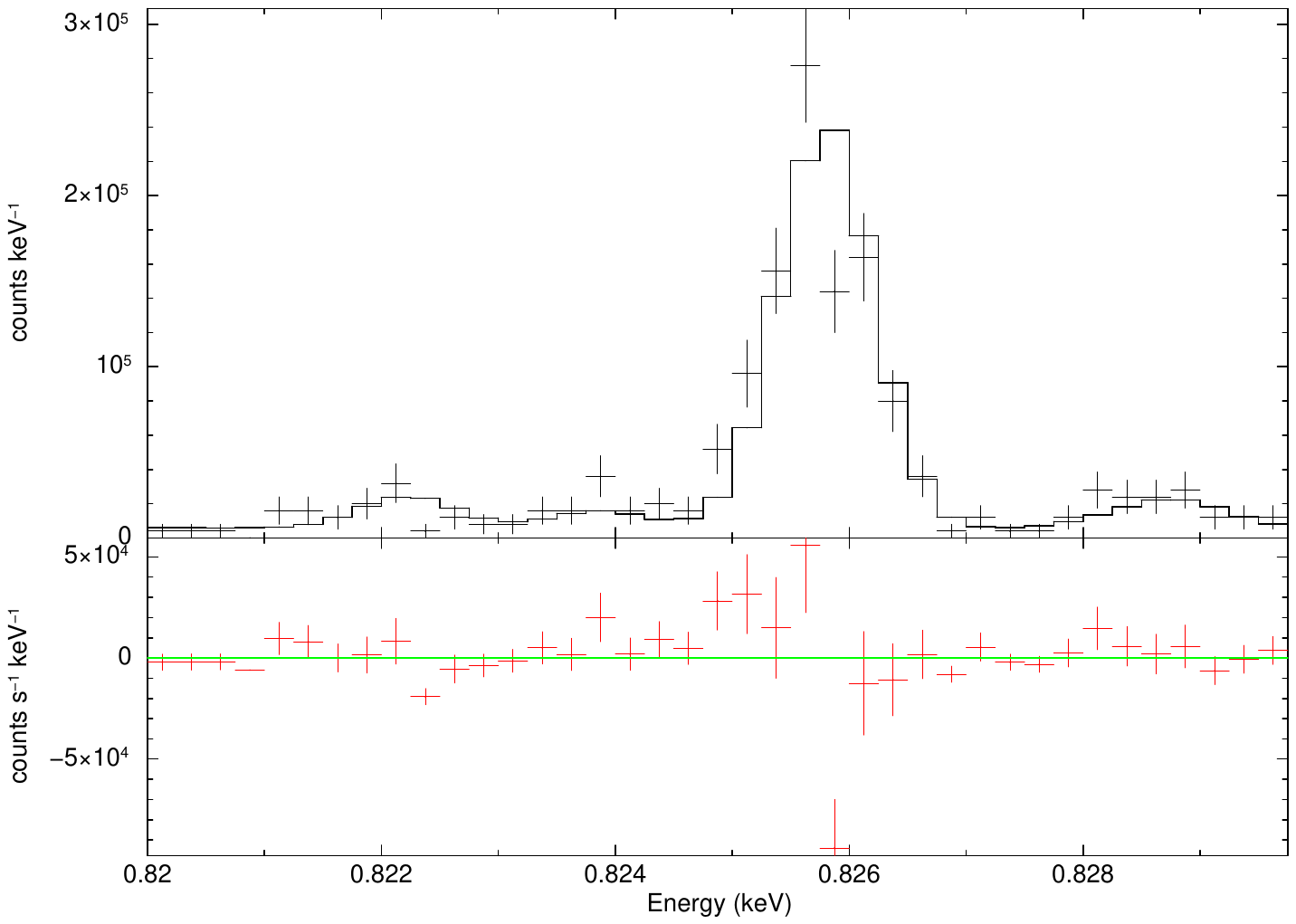}
     \includegraphics[width=0.32\textwidth,angle=0]{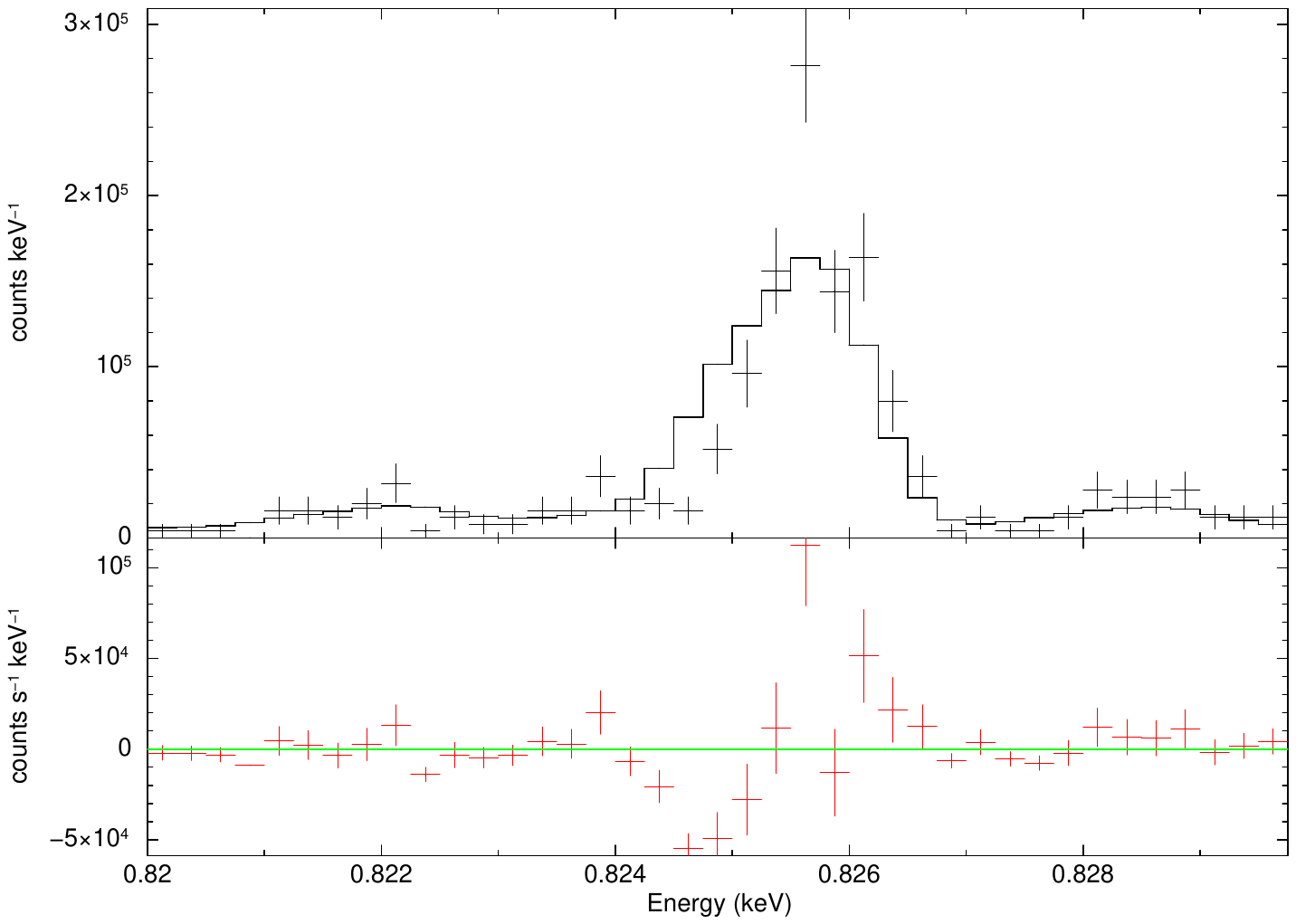}
     
    \caption{AB Dor simulated spectra of the isolated FeXVII line at 15.014 Ang (points with error bars) for an assumed Doppler shift velocity of 100 km/s compared with model predictions (solid histogram) for the resulting best fit (left panels), no Doppler shift (centre panels) and Doppler shift of 300 km/s, respectively. In the upper plots the source X-ray flux in the {\it LEM} 0.2-2.0 keV bandpass is fx $\sim$ 7 10$^{-11}$ erg/sec/cm$^2$, in the bottom plots fx is 10 times lower. For all simulated spectra the integration time is 1 ksec; in all plots the lower sub-panel shows the data residuals with respect to the model.}
    \label{fig:ABDORsingleline09eV}
\end{figure*}

The capability to recover the Doppler shift velocity by analyzing a single line of a source with $f_X$ $\sim$ 5-8 10$^{-11}$ erg/sec/cm$^2$ in the 0.2-2.0 keV bandpass is shown in Figure~\ref{fig:ABDORsingleline09eV}. The best fit Doppler velocity is 101.8$\pm$ 2.5 km/sec, to be compared  with the input value of 100 km/sec. Note the characteristic features that appear in the residual plots (center and right panels) when the model Doppler velocity if far off the best-fit value.

The recovery capability is reduced for a 10 times lower source $f_X$; in this case the best fit Doppler velocity is 96.3 $\pm$ 8.1 km/sec. In this or even 10 times lower flux (i.e. down to $\sim$ 7 10$^{-13}$ erg/s/cm$^2$) regime one can resort to a global fit that accounts for the contributions of all emission lines. With such an approach the Doppler velocity can be recovered within a factor 2 even if some other fitted parameters are much less well constrained. An analysis of the best fit parameters obtained with a global fit shows that for the nominal flux most of them can be recovered within a factor 2 with the exception of some element abundances that are poorly constrained for the lack of strong emission lines of those elements in the {\it LEM} bandpass.  

Based on hardness ratio variations, it has been suggested \citep{2023ApJ...951..152A} 
that abundance upwellings have been observed to occur in the corona of HD\,179949 during the course of a \textit{Chandra} CCD observation.  {\it LEM}, with its higher effective area and higher resolution, will be able to detect such changes directly by monitoring the emission in individual lines, even during quiescence.

\section{CORONAL MASS EJECTIONS}
\label{s:cmes}

Coronal mass ejections (CMEs) are among the most powerful magnetic phenomena occurring in the atmospheres of late type stars. In late type stars CMEs are expected to produce substantial effects on both stellar evolution and exoplanetary systems \citep[e.g.][]{drake2013,osten2015}. Nonetheless, because of current instrumental limitations, observations of stellar CMEs are very few and uncertain \citep[e.g.][]{moschou2019}. The {\it LEM} combination of large effective area and high spectral resolution in the soft X-ray band will boost our current capabilities to detect and identify stellar CMEs.

Solar observations show that CMEs usually occur simultaneously with flares. Each flare-CME pair is thought to originate from the same magnetic energy release. The mean CME mass ($M_{CME}$) and kinetic energy ($K_{CME}$) scale with the energy radiated by the flare in the X-ray band ($E_X$). Solar flares emit at most $E_X$ of $\sim10^{31}$\,erg. The few observations of candidate stellar CMEs are indeed associated with much more energetic flares, with $E_{X}$ ranging from $10^{31}$ to $10^{38}$\,erg. The properties of these candidate stellar CMEs suggest that the extrapolation of the solar relation between $M_{CME}$ and $E_X$ approximately holds up to the highest energy regimes. Conversely, for very high flare energy, the kinetic energy $K_{CME}$ of the associated stellar CMEs appears to be lower by two orders of magnitude the extrapolation of the solar case, being approximately of the same order of $E_X$.

Since both the CME and the flare emit in X-rays and involve substantial plasma motions, we simultaneously synthesized the X-ray emission of both these components. We took as a reference the X-ray flare observed in detail on Prox~Cen \citep{guedel2004}. This flare represents a well suited case because: its moderate $E_X$ (i.e. $\sim1.5\times10^{32}$\,erg) makes plausible the assumption that the solar flare-CMEs relations hold; the detailed HD model developed for this flare \citep{reale2004} allows an accurate synthesis of the flaring loop X-ray emission, including Doppler shifts due to chromospheric evaporation.

Starting from the observed flare $E_X$, we inferred the mass of the associated CME ($3\times10^{17}$\,g) from the solar case relation. Since solar CMEs are usually composed of plasma at different temperatures (from $10^4$ to a few $10^6$\,K), we assumed that only 10\% of the CME mass is hot enough to contribute to the CME X-ray emission. To be conservative, we set the CME terminal kinetic energy assuming it is 10 times higher than $E_X$ (i.e. $1.5\times10^{33}$\,erg, i.e. more than one order of magnitude lower than that corresponding to the solar case extrapolation (i.e. $4\times10^{34}$\,erg). To synthesize the X-ray emission of the CME we made assumptions on how it moves, expands, and cools down. In particular, using as a reference the evolution of the solar CME studied by Ref.~\citenum{rivera2019},
we assumed that: the temperature evolution with respect to the stellar radius is exactly the same; the CME velocity evolution with respect to the stellar radius is equivalent with only a higher terminal value (i.e. $v\approx10^{3}$\,km\,s$^{-1}$, in agreement with CME mass and terminal kinetic energy); the CME expansion rate with respect to the stellar radius is equivalent with only a different initial value (i.e. at $r=1.5\,R_{\star}$ we assumed that CME volume is 10 times the volume of the flaring loop).

\begin{figure*}
\centering
\includegraphics[]{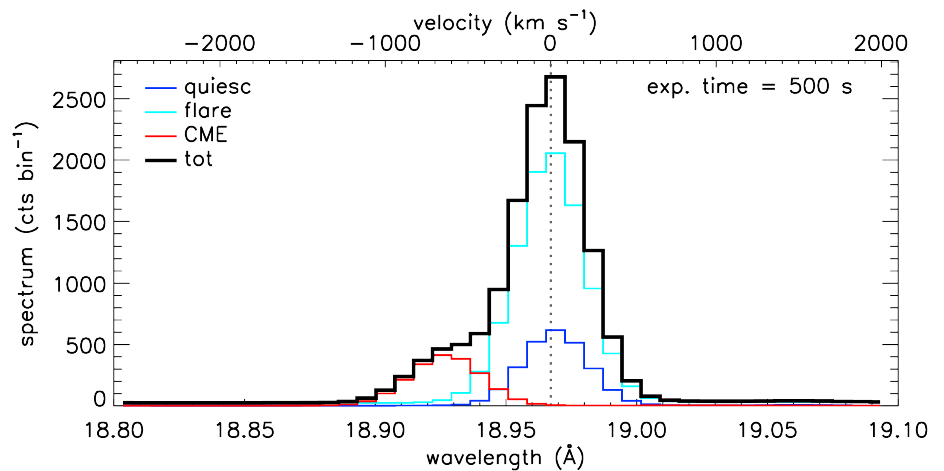}
\caption{Simulated profile of the O\,VIII line at 18.97\,\AA\, of a flare-CME event from Prox~Cen, assuming that the CME ejection occurs along the line of sight. The dotted line marks the position of the strongest contribution of the doublet, used as a reference for the velocity axis.}
\label{fig:CMEspec}
\end{figure*}

We synthesized the predicted X-ray emission of the CME, including also the contribution of the flaring loop (assumed to have half-torus shape) and quiescent corona. Assuming we are observing the flaring loop from above and that the CME is ejected in that direction, the predicted X-ray spectrum. As an example we report the profile of the O\,VIII lyman alpha line, integrated over an exposure time of 500\,s (corresponding to the initial phases of the CME motion, i.e. from 2 to $6\,R_{\star}$) in fig.~\ref{fig:CMEspec}, assuming the Prox~Cen distance and the 0.9\,eV spectral resolution. We find that: the blue-shifted CME contribution is clearly detectable in addition to the flaring and quiescent emission; the velocity difference of the CME and flaring plasma at a few MK allows us to resolve the two contributions even assuming CME lower radial velocities; several X-ray lines, produced by plasma a few MK, display profiles similar to that shown in fig.~\ref{fig:CMEspec}, indicating that multi-line analysis would allow us to detect the CME-associated blue-shifted emission also in sources at larger distances.

\section{CHEMICAL ABUNDANCES AS PROBES OF CORONAL PLASMA PHYSICS}
\label{s:abuns}

Observations of the solar corona at EUV and X-ray wavelengths beginning in the 1960s found evidence that the chemical composition of coronal plasma is significantly different to that of the solar photosphere (see, e.g., the review by Ref.~\citenum{Laming:15}). The abundance differences are strongly correlated with the element first ionization potential (FIP) and the abundance anomaly is now commonly referred to as the ``FIP Effect''.
Elements with low FIP (FIP$ \leq 10$~eV, e.g., Si, Mg, Fe) are enhanced on average in the solar corona by factors of 2--4 relative to elements with high first ionization potentials (FIP$ \geq 10$~eV, e.g., N, Ne, Ar). However, different regions of the solar corona are now known to exhibit different magnitudes of FIP effect, with quiet Sun and coronal holes showing very little, if any, FIP-based abundance bias, and active regions generally showing the largest effect \cite{Laming:15}.

In the 1990's, the {\it ASCA} and {\it EUVE} satellites found that stellar coronal abundances also differed from photospheric values \cite{Drake:96,Drake:96b,White:96,Drake2003}. {\it Chandra} and {\it XMM-Newton} confirmed the results in much greater detail with high-resolution X-ray grating spectroscopy. Stars of ``low'' magnetic activity---like the Sun---showed similar abundance anomalies to the solar FIP Effect. However, magnetically active stars tended to show the opposite: an ``{\em inverse FIP}'' (iFIP) effect where low-FIP abundances are depleted relative to high-FIP elements \cite{Brinkman.etal:01,Drake.etal:01}.

\begin{figure}
\begin{center}
\includegraphics[width=\linewidth]{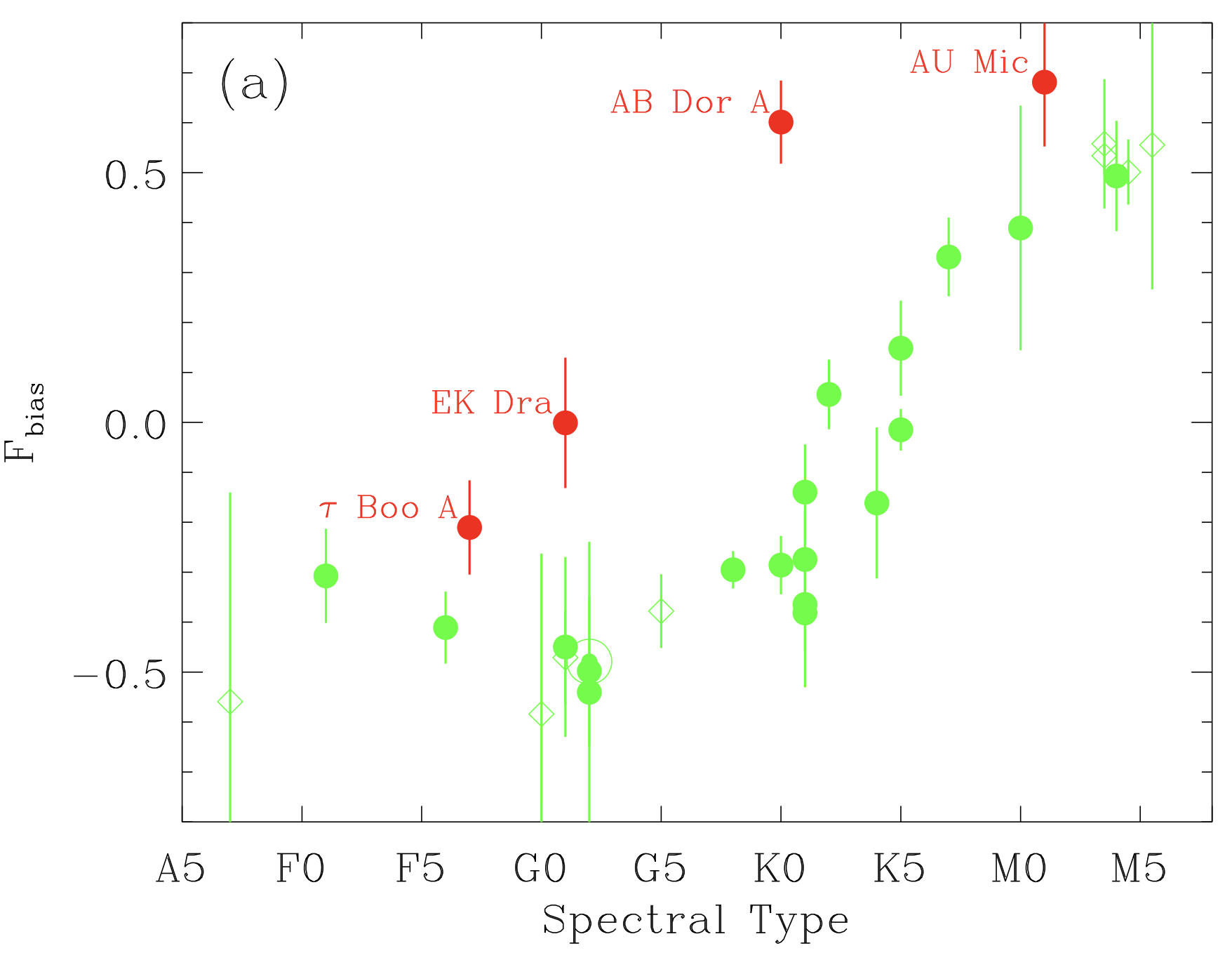}
\end{center}
\caption{The ``FIP bias'' as a function of spectral type from  Ref.~\citenum{Wood.etal:18}. Active stars are shown as red points. The FIP bias is the average ratio of high FIP element abundances to that of Fe relative to the solar mixture. The results indicate both a spectral type and activity dependence of the FIP effect.}
\label{f:fipeffect}
\end{figure}

As more stars were observed, an additional dependence of the FIP and iFIP effects on stellar spectral type emerged \cite{Wood.etal:18}. The ``FIP bias''---an average of high FIP element abundances relative to the low FIP element Fe compared with a solar photospheric mixture---based on an analysis of a sample {\it Chandra}/LETG+HRC-S spectra \cite{Wood.etal:18} is illustrated in Figure~\ref{f:fipeffect}.  The FIP bias is negative for the solar FIP effect in which low FIP elements are relatively enhanced, and positive for the inverse FIP effect. The data show a trend of increasing FIP bias with later spectral type, together with an increase in FIP bias with increasing activity level reminiscent of the early stellar results.

The FIP boundary that differentiates low and high FIP behaviour is approximately 10~eV, corresponding to a temperature of ($\sim 10,000$~K) and pointing to the chromosphere as the fractionation site. Mechanisms that might explain the fractionation remained elusive until it was realised that the ponderomotive force exerted by Alfv\'en waves excited by convection and turbulence and propagating into the chromosphere could provide the selective acceleration of ions, enhancing their abundances \citep{Laming:15}. 
The variation in FIP effect with spectral type indicates that the Alfv\'en wave intensity and spectrum probably change. This is not surprising since the depth and characteristics of the convection zone change with effective temperature ($T_{\rm eff}$).  Alfv\'en waves are thought to be important as sources of coronal heating and for driving the winds of solar-like stars. 
Coronal abundances then offer the tantalizing possibility of providing unique diagnostics of Alfv\'en waves in stellar coronae.

The mechanism that drives the FIP effect is still unresolved.  Monitoring of the high-cadence variations in individual spectral lines during the onset of a flare and searching for time lags will constrain these mechanisms.  {\it LEM} will directly observe whether FIP or inverse-FIP effects are preceded by flare-induced chromospheric evaporation effects by detecting time lags in flare onset in lines arising from different species.

There is evidence that given the uncertainties in abundance measurements, the FIP bias is consistent with 0, or is at least increased from negative values, for hot Jupiter hosting stars \citep{2023ApJ...951..152A}.
The high spectral resolution and effective area of {\it LEM} will definitively answer this question when coronal and photospheric abundances of planet-bearing stars are compared.

Alfv\'en waves are also likely to be initiated from above in the corona through perturbations driven by magnetic reconnection, especially in more active stars. 


Differences in chemical composition of the corona will also impact the radiative loss function  
and thus the cooling rates and energy balance of coronal plasma. A metal-poor corona would naively be expected to be hotter because of the absence of metal lines that dominate the radiative loss at temperatures $T\lesssim 10^7$~K. This is indeed what was found for the Pop~II interacting close binary HD89499, whose coronal temperature based on {\it ASCA} spectra was $2.5\times 10^7$~K \cite{Fleming1996}, or approximately $10^7$~K hotter than typical for the Pop I interacting binaries \cite{Dempsey1997}.

\section{EXOPLANET SCIENCE}
\label{s:exo}

The high effective area of {\it LEM} will make it possible for the exospheres of several transiting systems to be detectable, as was found for HD\,189733 with \textit{Chandra} observations \cite{2013ApJ...773...62P}.
A new method to model transit light curves \cite{2022HEAD...1911063H} 
using a realistic model of the absorption of coronal X-rays by the intervening planetary exosphere holds promise that {\it LEM} data can be analyzed to obtain the size and composition of the atmospheres of transiting exoplanets.  The model parameterizes the exosphere by the base density, scale height, and composition of the exosphere, and computes the absorption numerically at different orbital phases.  Realistic coronal spectra are used to determine the absorption effects in different passbands and even in individual lines.  Systematic effects such as varying numbers of spots and active regions on the stellar surface are included by considering variations modeled on observed solar active regions \citep[see][]{2015ApJ...802...41L}. 

\begin{figure}
    \centering
    \includegraphics[width=\linewidth]{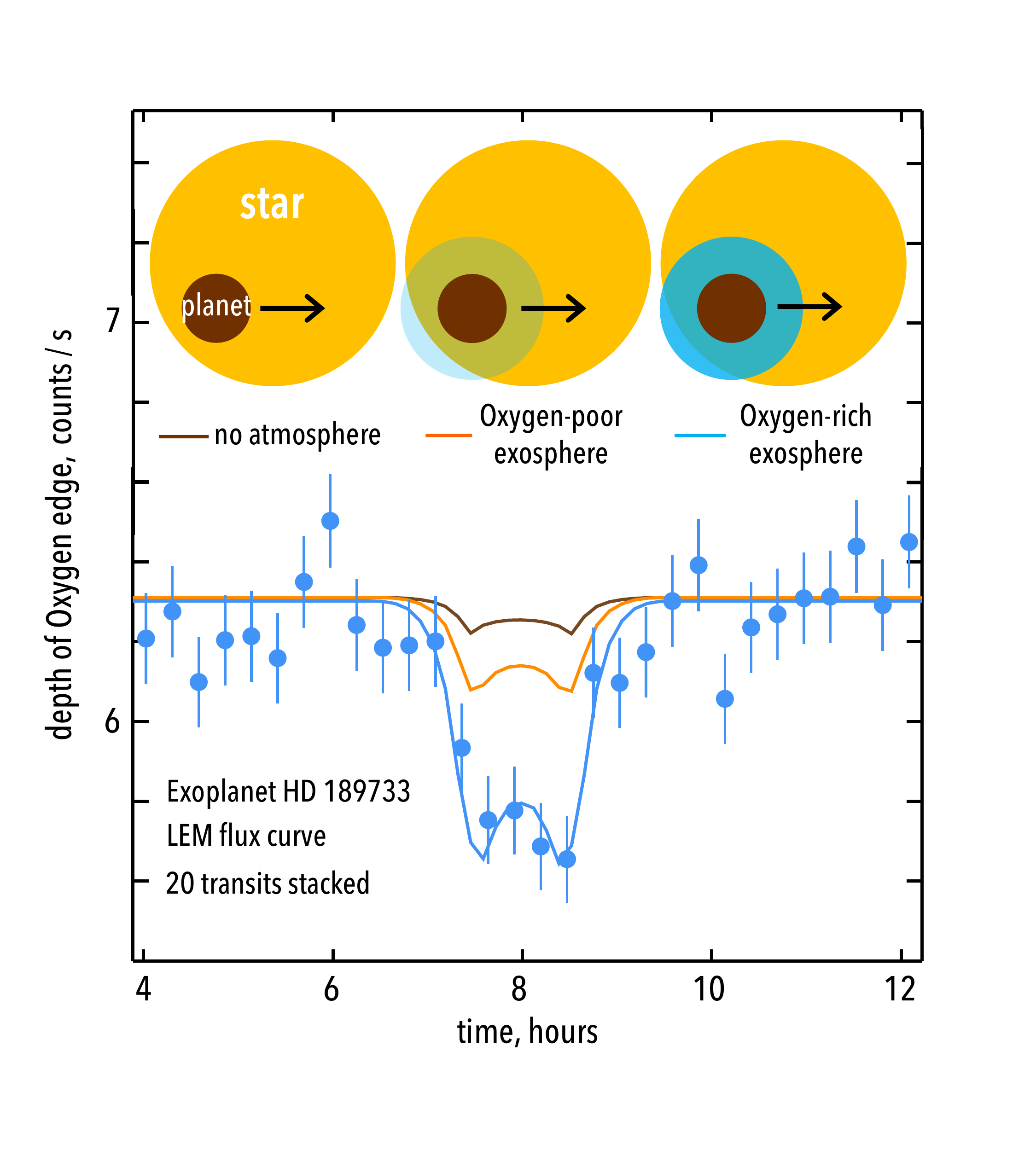}
    \caption{Demonstration of the suitability of {\it LEM} to estimate abundances in exospheres.  The plot shows the depth of the O edge for three separate scenarios: one where the planet has had its atmosphere entirely ablated (brown curve), one with a relatively oxygen-poor atmosphere (solar composition; orange curve), and one with planetary atmospheric abundance enhanced 10$\times$ relative to solar (blue curve with synthetic data points overplotted).  The three scenarios are sketched for illustration in the upper part.}
    \label{fig:OedgeTransit}
\end{figure}

As a practical strategy to minimize the effects of active region inhomogeneities, {\it LEM} will adopt a strategy of observing a transit several times so that the effects of variations during any single pass are averaged out over multiple passes, bringing out the signature of the transit clearly.  We demonstrate this with simulations (see Figure~\ref{fig:OedgeTransit}).  In particular, the high spectral resolution of {\it LEM} will allow us to measure the depth of the absorption edges with high precision, yielding accurate and direct measurements of the C, N, and O abundances in the exosphere.

Soft X-rays also play an important role in the heating and thus dispersal of planet-forming disks and consequently can have strong impact on giant planet migration and where these planets ultimately end up within a planetary system. Knowing the X-ray emission of planet-hosting stars can therefore be important in understanding the observed demographics of exoplanetary systems.


Soft X-rays also heat exoplanetary atmospheres of sub-Neptunes and are an important ingredient in the model for the "photoevaporation valley" that is observed for super Earths/sub-Neptunes.




Superflares can be extraordinary events with measurable effects on the ablation of exospheres.  While an observational program that waits for superflares to occur is not practical, in synergy with synoptic surveys like Rubin LSST, it is possible for {\it LEM} to follow up rapidly if a superflare were to occur on a planet hosting star.  If the planet happens to have been in the path of the blast, {\it LEM} can target the transits that occur immediately after.  If exosphere gets ablated, it will leave a comet-like tail that will alter the transit profile.


\section{STELLAR CYCLES}
\label{s:cycles}


Of order 100 nearby stars are known to have stellar cycles, but fewer than ten have had regular X-ray monitoring. Of those, X-ray intensities over a cycle typically vary by factors of a few, as opposed to roughly 10-20\% in CaII HK or a few percent in optical photometry.  Although the current sample is very small, X-ray cycles also often appear to be more regular than in other bands, for example in $\iota$ Hor \citep{SanzForcada2019} and $\epsilon$ Eri \citep{Coffaro2020}, perhaps because of X-rays' greater emission volume and correspondingly less sensitivity to viewing geometry effects.  With {\it LEM}'s large collecting area, spectral analysis of emission from cycle maxima and minima \citep{Orlando2017} would not only determine the temperatures and surface coverage fractions of quiet emission, active regions, and AR cores, but also easily compare coronal abundances over a cycle.  Cycles have recently been reported in F stars \citep{Mittag2019}, which have very thin convective zones, and, very surprisingly, in fully convective M stars \citep{SuarezMascareno2016}, both subjects of great relevance to the study of magnetic field generation and stellar evolution. Currently, however, only one fully convective star, Proxima Cen, has a stellar cycle amenable to detailed study in any band because of the faintness of such small stars.

Observations of the Pleiades cluster (136 pc, age 100 Myr) would provide an efficient means to monitor cycles in many young stars at once over a wide range of spectral type.  To date, the youngest star with X-ray monitoring is $\iota$ Hor (F8, age $\sim$600 Myr). Young, i.e.~active, stars have cycle periods approximately proportional to their convective turnover time scales \citep{Irving2023};
for F,G, and K stars in the Pleiades, $P_{\rm cycle}$ is generally under 5 years, and two or three observations of 20 ks per year would allow accurate measurements of cycle amplitudes.

\section{SYMBIOTIC STARS}
\label{s:symbiotic}

Symbiotic stars, which are interacting binaries where an accreting compact object (e.g., White Dwarf) closely orbits a red giant, are crucial for the understanding of the origin of asymmetric Planetary Nebulae and some SN Type Ia. Understanding how mass builds up in the compact object is a key to determining the potential  of symbiotics as progenitors of a fraction of Type Ia supernovae - key cosmological distance indicators.  Several symbiotics \citep[see e.g.,][]{2013A&A...559A...6L}  have been identified and observed with existing observatories.  {\it LEM}, with its significantly higher effective area, will expand the reach to increase the symbiotics' sample size by a factor of at least an order of magnitude.

The morphology and X-ray spectra of some Symbiotic Systems (e.g., CH\,Cyg, R\,Aqr, RT\,Cru) show great similarity with the morphology and spectral distribution of AGN (e.g., Seyferts), and are therefore sometimes called nano-quasars \cite{2006MNRAS.372.1602W,2023MNRAS.522.6102T}.
LEM spectroscopy will provide information needed for understanding the underlying accretion and jet powering mechanisms of these systems and their large "cousins",  which otherwise differ by orders of magnitude in spatial, velocity, and mass scales.

\begin{figure*}
    \centering
    \includegraphics[width=0.48\linewidth]{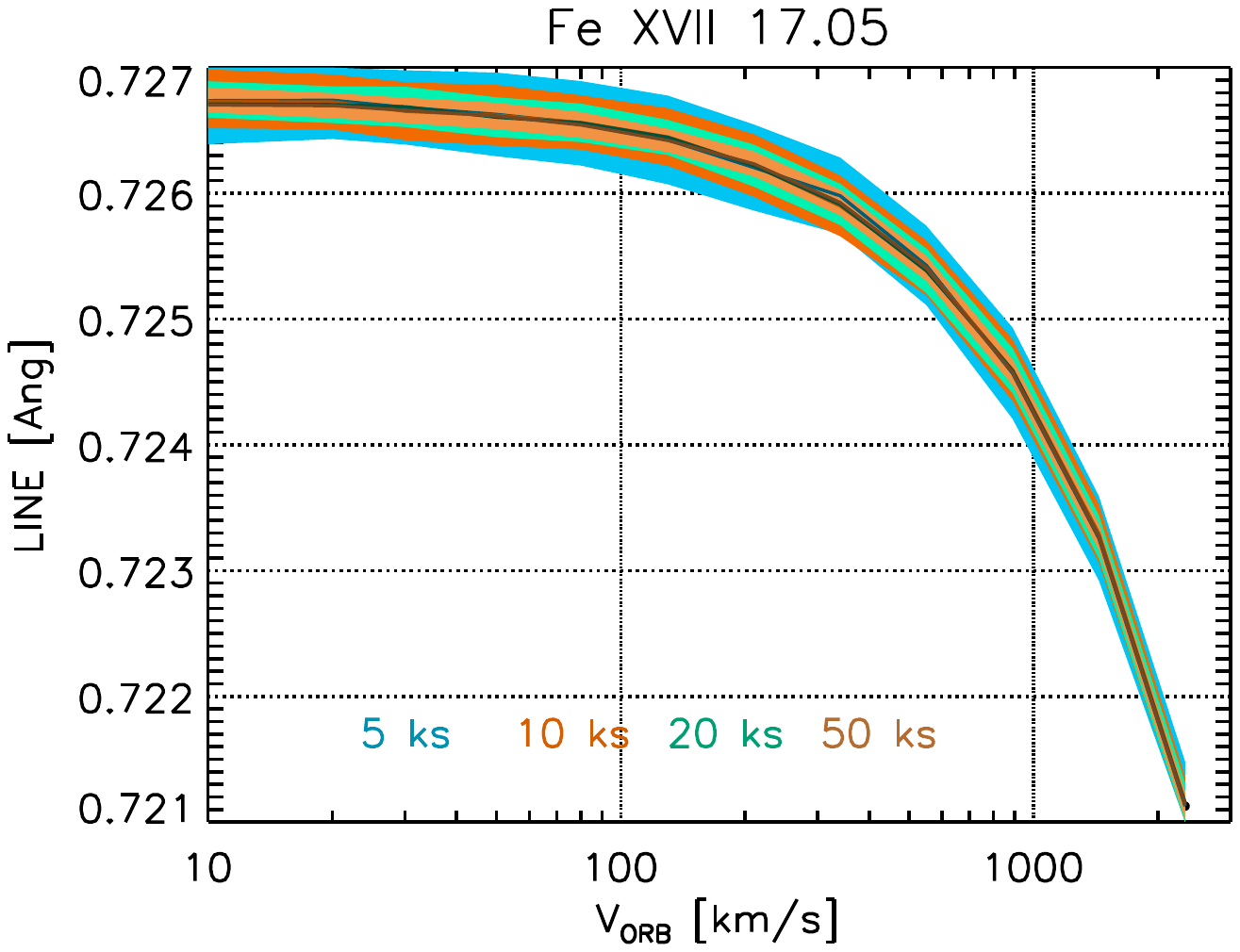}
    \includegraphics[width=0.48\linewidth]{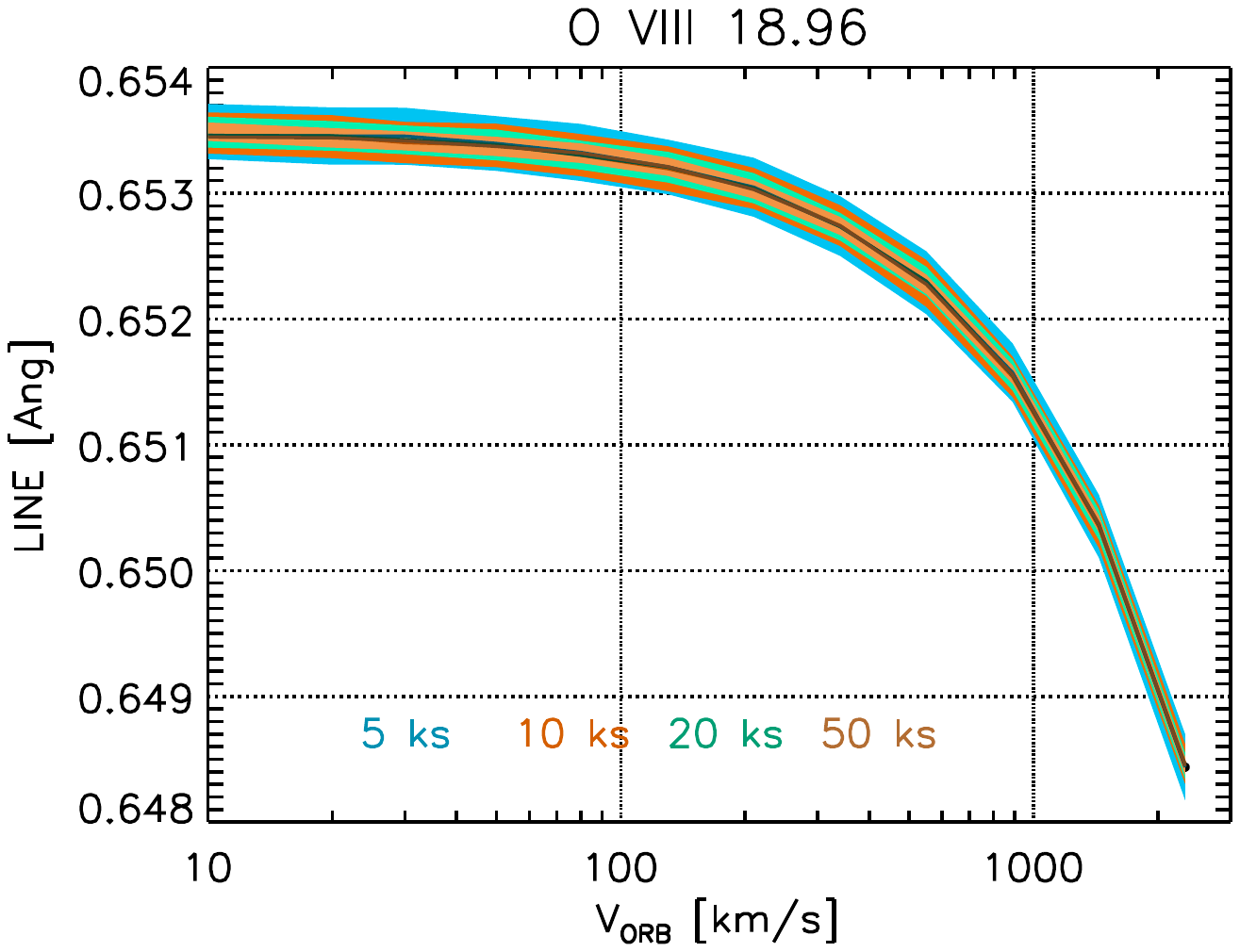}
    \includegraphics[width=0.48\linewidth]{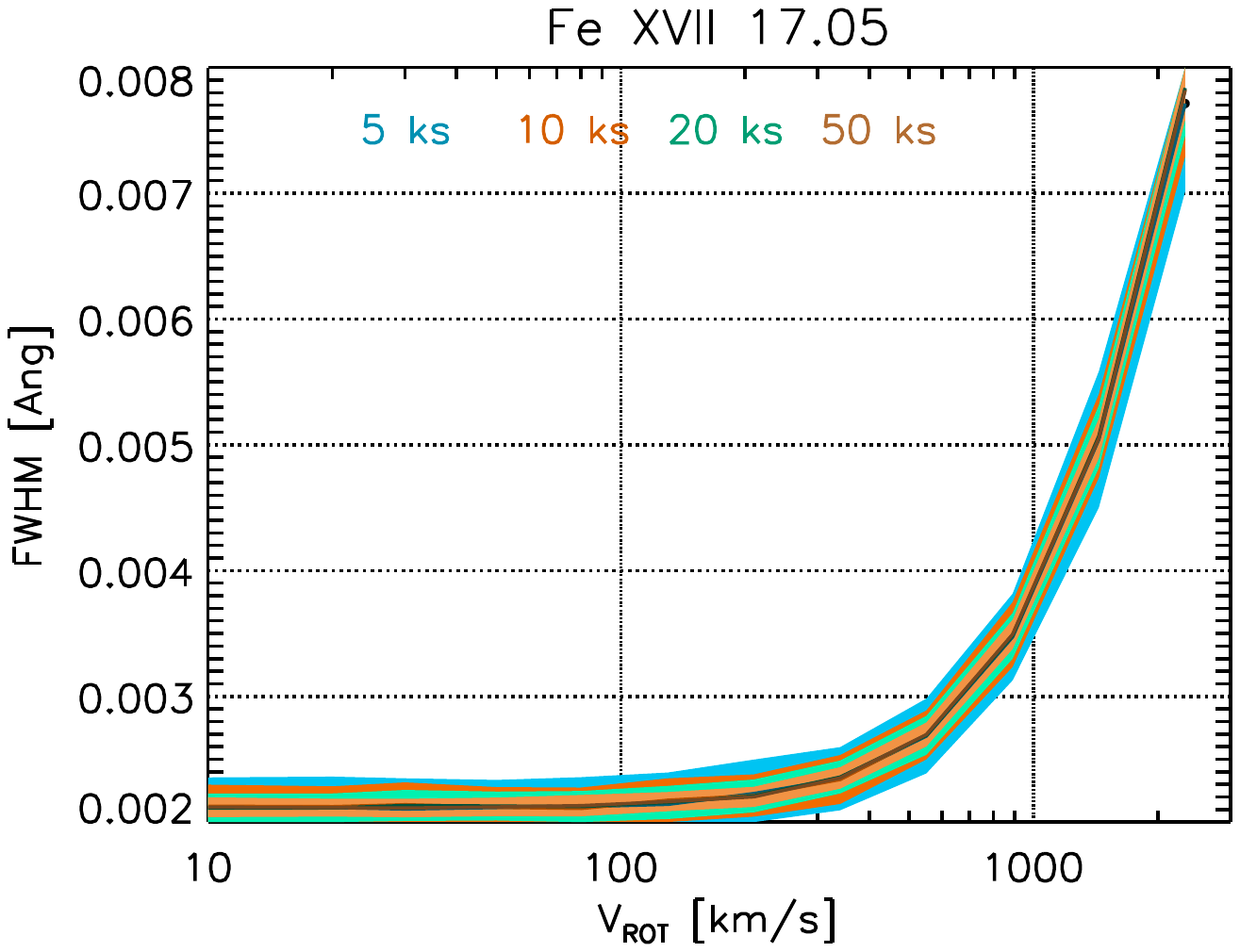}
    \includegraphics[width=0.48\linewidth]{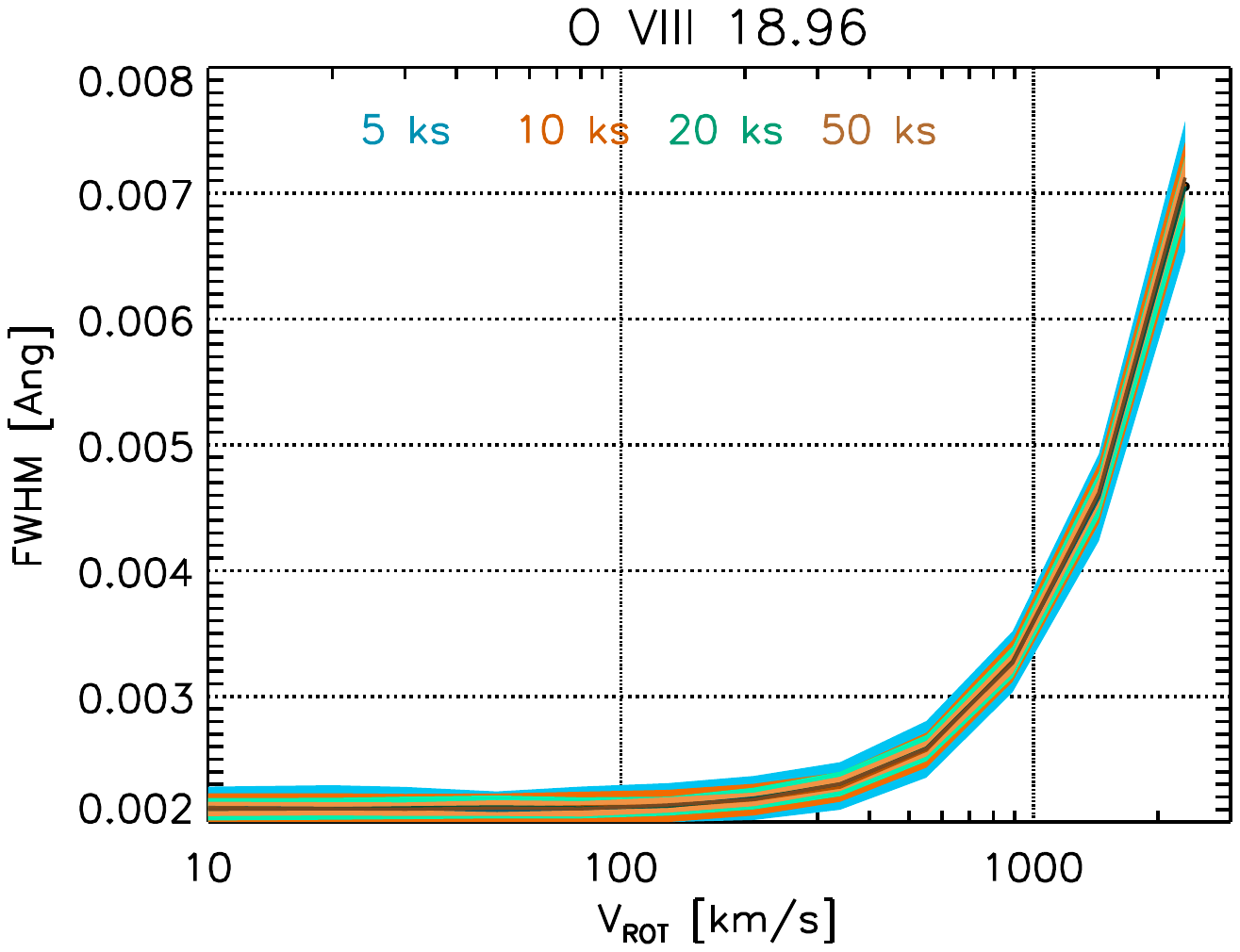}
    \caption{Demonstrating {\it LEM}'s ability to detect Doppler shifts ({\sl top}) and Doppler broadening ({\sl bottom}) in symbiotic spectral lines Fe\,XVII at 17.05~\AA\ ({\sl left}) and O\,VIII at 18.96~\AA\ ({\sl right}).  Line fluxes are as in CH\,Cyg \citep{2023MNRAS.522.6102T}.  The bands are uncertainty ranges based on 200 simulations, with the different colors representing different exposure times as listed in the legend.}
    \label{fig:symb_doppler}
\end{figure*}

A major limitation of current studies of symbiotics is to understand the origin of the emission components.  It is unclear where precisely the observed soft emission arises, whether it is detected due to patchy absorption in and around the accretion disk, or occurs due to periodic outbursts of jet-like emission, or both.  The observed spectrum is typically a combination of emission from the central source, an accretion disk, and a jet.  Hence, component separation must be carried out spectroscopically.  We expect that emission from the accretion disk will be Doppler broadened by $>$500~km~s$^{-1}$, while jets should exhibit Doppler shifts $>$500~km~s$^{-1}$.  A simple measurement of Doppler shifts in lines or line broadening attributable to rotational smearing can pin down the source of the emission, and {\it LEM} is capable of distinguishing these scenarios.  We show in Figure~\ref{fig:symb_doppler}, simulations made for two representative lines (with fluxes based on a 35~ks XMM/RGS observation of 
CH\,Cyg\cite{2023MNRAS.522.6102T}).
Even measurements based on individual lines (Fe\,XVII~17.05~\AA\ and O\,VIII~18.96~\AA\ are shown) are capable of detecting Doppler shifts of as small as $V_{orb}\approx$200-300~km~s$^{-1}$ and line broadening $V_{rot}>$300~km~s$^{-1}$ in observations with durations as small as 5~ks.  When data from several lines from similar species are modeled together, the uncertainties can be reduced further.

In addition, the spectra of symbiotics is likely to also include reflection/photoionization components which have been hard to pin down because of large background and low effective areas in current high-resolution instruments \citep[][]{2023MNRAS.521..969Z}.  With its high effective area and high spectral resolution, detailed sifting of the emission components becomes feasible in {\it LEM} spectra.

\section{CATACLYSMIC VARIABLES}
\label{s:cvs}

Cataclysmic variables (CVs) are interacting binaries, which consists of a white dwarf (WD) and a Roche lobe filling low mass companion star (usually late type main sequence star). Depending on the strength of the magnetic fields of the WD, they are classified into non-magnetic CVs where the accretion disk extends up to the primary WD, and magnetic CVs where the magnetic fields are strong enough to prevent the formation of an accretion disk, at least close to the WD. An intermediary subclass of magnetic CVs are Intermediate Polars (IPs) where a partial accretion disk exists and accretion onto the white dwarf is magnetically funneled onto its poles from the inner edge of the truncated accretion disk.

While the general broad brush picture of CV behaviour resulting from Roche lobe overflow and accretion has been accepted for a number of years, as is often the case the details of this picture come into question and still remain to be fleshed out. 

In IPs, the X-ray emission originates from the accreted matter shock heated up to high temperatures (kT$\sim$10--50 keV), which must cool before settling onto the surface of the WD. Therefore the X-ray spectrum is expected to be the resultant emission from a multi-temperature plasma, which has a continuous temperature distribution from the shock temperature (of kT$\sim$10--50 keV) to the WD photospheric temperature (kT$\sim$0.02 keV). The post-shock plasma in many IPs can be well represented by a cooling flow model, which contains a bremsstrahlung continuum and collisionally-excited emission lines from a range of temperatures \citep{done1995, mukai2003}. The emergent X-ray emission from this  shock-heated, multi-temperature plasma is modified by the presence of a complex distribution of absorbers in the pre-shock region. Given the X-ray fluxes from the post-shock plasma region, we expect these complex absorbers to be ionized, which is observationally motivated by the presence of photo-ionized emission lines and warm absorber features in the soft X-ray spectra of IPs \citep{islam2021}.

In Ref.~\citenum{islam2021}, an ionized absorber model {\tt zxipab} was used to model the Chandra/HETG spectrum of two IPs: NY Lup and V1223 Sgr. The ionized absorber model along with a cooling flow model describes well many of the H- and He-like emission lines from medium-Z elements, which arise from a collisionally-excited plasma. However the model fails to account for some of the He-like triplets from the medium Z elements. Using a combination of collisionally ionized plasma with a cooling flow model and photoionized models, \citep{chakraborty2022} suggested a hybrid origin of O VII, O VIII, Ne IX, Ne X, and Mg XI lines in the {\it Chandra}/HETG spectrum. These issues indicate that we are still some way from understanding the accretion scenario in IPs and the factors that go into shaping the emerging spectra.

Other aspects of the behaviour of IPs that remain to be understood are orbital and spin-dependent variationsin X-ray emission. For a magnetic pole closely aligned with the WD rotation axis, spin modulation is expected to be small. However, variations in the height of the accretion shock are thought to be likely candidates explaining spin modulation.\cite{Mukai2015} Spectroscopic diagnostics of the emitting regions could help with this, although existing instrumentation is insufficiently sensitive and requires of the order of 100~ks exposure times for a single spectrum of the brighter sources\cite{Mukai2017}. Moreover, He-like diagnostics of plasma density are prone to contamination by FUV radiative excitation effects---an aspect that has been put to good use in the study of X-rays from the winds of massive stars.\cite{Kahn.etal:01} Other diagnostics immune to such effects are available, in the "L shell" lines of Fe for example, but these are too faint to be of practical use with current generation spectrometers.

There are at least two breakthroughs that {\it LEM} will make in the study of CVs. In the example above, the high sensitivity of {\it LEM} would be crucial in understanding the origin of He-like emission lines in IPs, and get an accurate picture of the accretion, reflection and absorption components, and being able to utilise density diagnostics of weaker lines.  The second is pushing studies toward time-dependent spectroscopic probes of the accretion physics, in which {\it LEM} spectra can provide significant signal-to-noise ratio in short exposures, or enable phase-folding into sufficiently finely resolve bins to unveil the underlying behaviour.

\section{NOVAE}
\label{s:novae}

Nova explosions arise from the interaction between a white dwarf and a close companion star from which hydrogen-rich matter is accreted onto the white dwarf surface that is genuinely hydrogen-poor. When the accreted gas density and temperature exceeds a critical threshold, runaway nuclear fusion occurs in the form of CNO burning (fusing hydrogen to helium). The resulting explosion expels the outer layers into space, emitting intense radiation across all wavelengths, depending on the opacity of the ejecta. Directly after the initial detonation, a nova is brightest in the optical/UV because the high-energy radiation produced via nuclear burning on the white dwarf surface undergoes complex radiation transport processes through the large optically thick ejecta. Nuclear burning continues for several weeks to months powering continuous expansion via radiation pressure, and the ejecta become gradually optically thinner. Radiation transport through an atmosphere with a smaller radius allows the radiation to escape after fewer interactions with dense matter, and what escapes from the photosphere will thus be of successively higher energy. The gradual shift of the peak of the Spectral Energy Distribution (SED) from low to higher energies has been observed to stop in the soft X-ray regime peaking at 0.5-1\,keV, observable as a Super-Soft-Source (SSS) spectrum. The brightness and effective temperature of these spectral yield radii consistent with typical white dwarf radii, and this emission can thus be considered the closest we can get to observe nuclear burning in vivo. Typical SSS spectra fit perfectly into the sensitive energy range of {\it LEM}.

Until the dynamic hot photosphere from steady-state surface nuclear burning emits an SSS spectrum, the white dwarf itself is X-ray dark. However, there are multiple other X-ray emission processes from novae including accretion onto the white dwarf, shock heating of the ejected material and circumbinary medium, the interaction between the expelled gas and intense magnetic fields. Observations of this X-ray emission provide crucial insights into the dynamics of the explosion, as well as the properties of the accretion disk and circumstellar material. Consequently, nova explosions are recognized as key events for our understanding of binary stellar evolution and the behavior of matter and radiation in extreme environments, including  high-energy astrophysics of non-equilibrium nuclear fusion, shock waves, particle acceleration and radiatively-driven outflows. 

Similarly to the advances {\it LEM} can bring to other areas of stellar physics, advances in our understanding of novae will come from both the ability to obtain high-resolution spectra for sources up to two orders of magnitude fainter than existing instrumentation, and the opening up of the time domain to high-resolution study.


\begin{figure*}
\begin{center}
\includegraphics[width=0.95\textwidth]{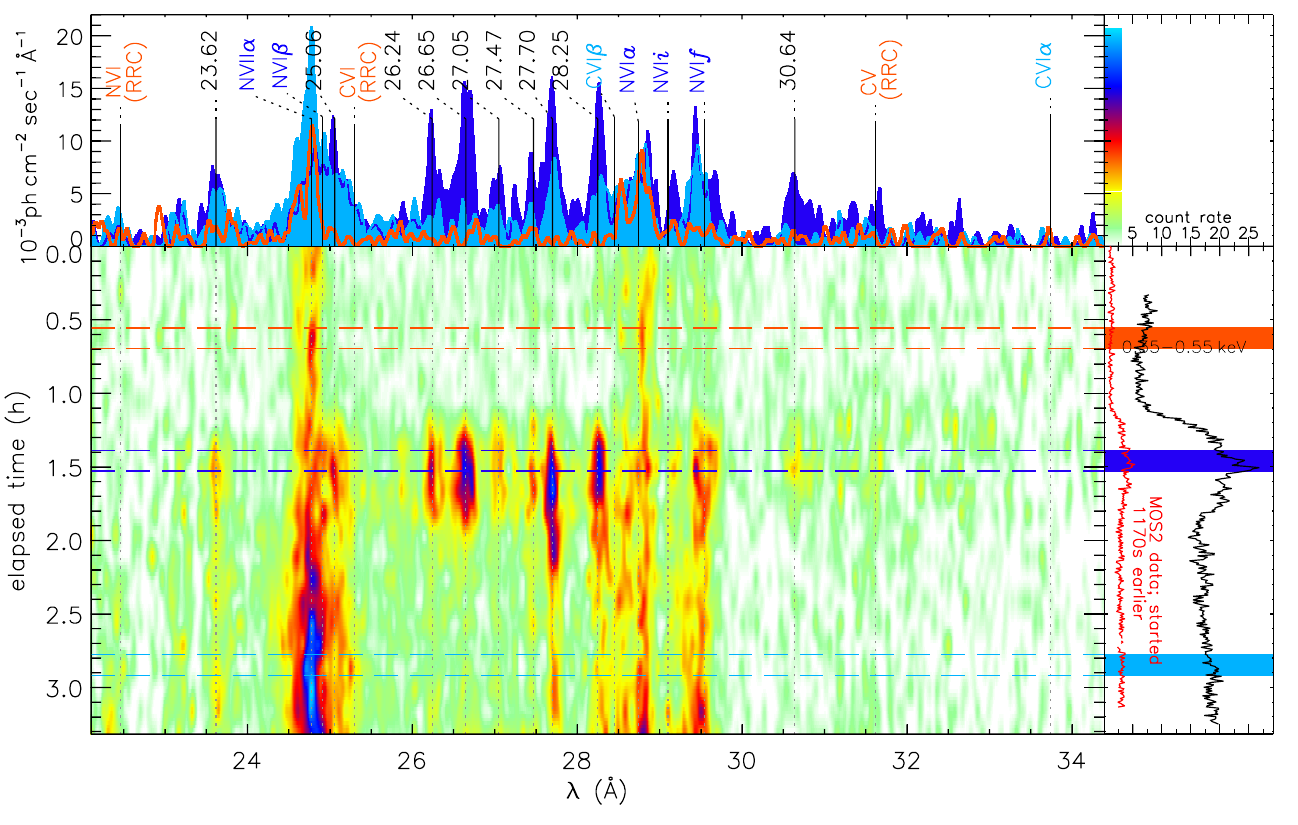}
\end{center}
\caption{The spectral time-map of the {\it XMM-Newton} observation of the recurrent nova RS\,Oph on day 26.1 after its 2006 outburst. Top left: spectra extracted over selected time intervals showing line identifications and also noting the wavelengths of unidentified features. The color coding of the different spectra corresponds to the intervals indicated in the lower right panel. Top right: a color bar assigning a color to each intensity. Bottom right: EPIC/pn
and MOS2 light curves showing the brightness evolution in count rate with time running downward, and intervals of spectrum extraction noted that correspond to the spectra in the upper left panel. Bottom left (main panel): the time evolution of the spectrum using the colors defined in the top right.
Taken from Ref.~\citenum{ness+2015}.}
\label{f:timemap_spec}
\end{figure*}

\begin{figure*}
\begin{center}
\includegraphics[width=0.95\textwidth]{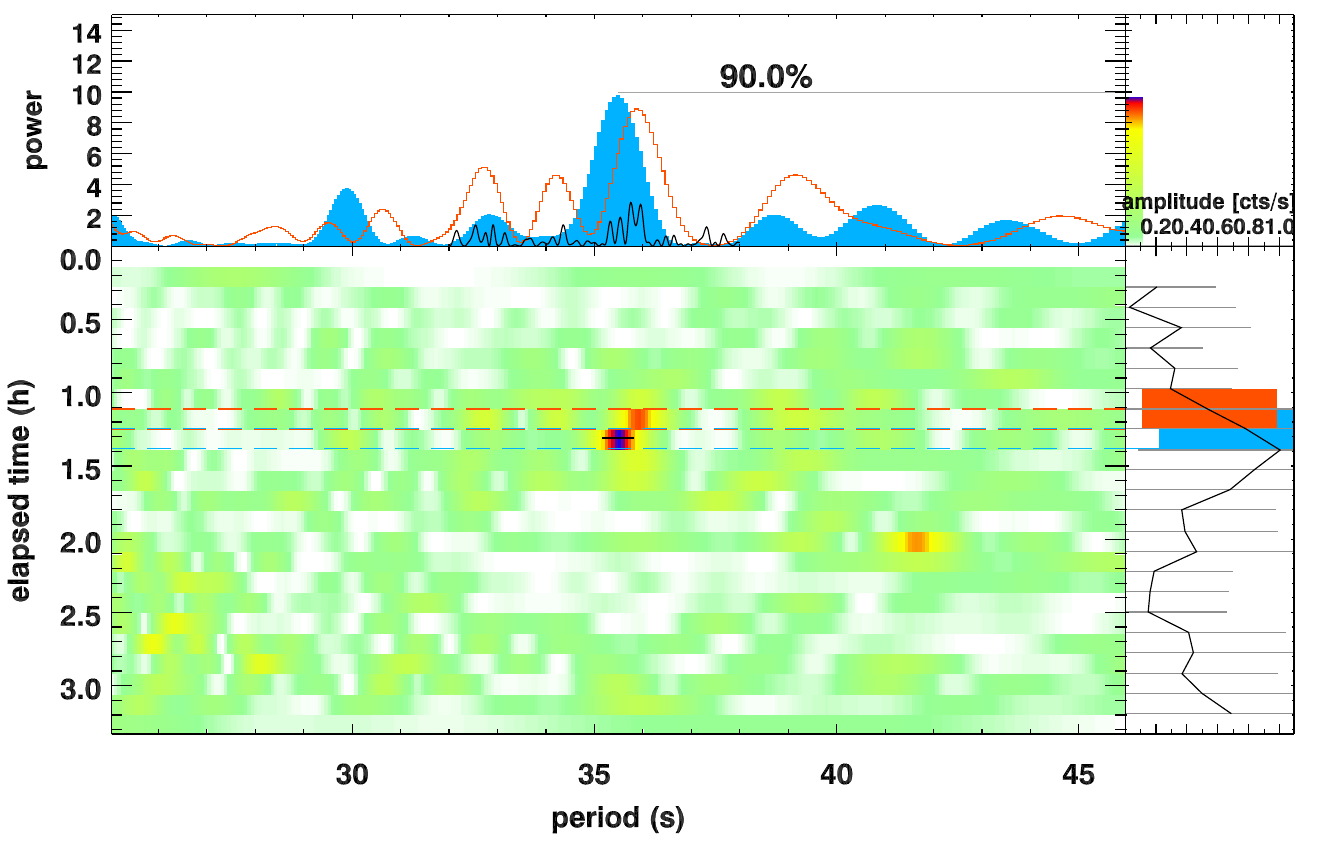}
\end{center}
\caption{Similar to Figure~\ref{f:timemap_spec}, showing the data on the recurrent nova RS\,Oph observed with {\it XMM-Newton} on day 26.1 after its 2006 outburst. This figure concentrates on the periodogram and its variations. Top left: periodogram extracted over selected time intervals indicated in the lower right panel.  Color coding of the periodograms corresonds to the extraction interval illustrations. In addition to the two power spectra from the colored partial
light-curve intervals, the power spectrum from the full light-curve
is also shown in black, containing no significant signal.  Top right: a color bar assigning a color to each Fourier power. Bottom right: modulation amplitude of the signal with time running downwards. Bottom left (main panel): the time evolution of the periodogram computed from 24 overlapping 1000s time segments from the pn light curve. The colors indicate the power (corresponding to detection likelihood, see scale in the top right):  blue colors mark
periods with $>$90\% confidence, red marks 30--80\%, and yellow/light green reflect noise.
Taken from Ref.~\citenum{ness+2015}}
\label{f:timemap_qpo}
\end{figure*}

\subsection{Early Onset X-rays and Blast Waves}
\label{s:blast}

As CNO burning produces intense high-energy radiation, it is expected that the initial blast should be observable in the X-ray regime for a short time until the outer layers start to lift off the surface via radiation pressure. Observations of the early blast wave are only possible in serendipitous observations as this phase is too short to point an X-ray telescope via a ToO observation. Such a serendipitous observation was achieved with eROSITA during one of the systematic scans of the sky, and \cite{koenig+2022} described a SSS spectrum similar to the SSS spectra observed later during the outburst.


\subsection{Probing early shock emission}
\label{s:earlyshock}

Before the bright SSS emission dominates X-ray observations, X-ray observers are entertained with faint emission originating from shock emission. Such emission consists of bremsstrahlung continuum and emission lines, and there are several ways of producing such emission. Shocks can occur between the ejecta and pre-existing circumstellar material, e.g. a stellar wind of the companion star, ejecta from a previous outburst, or even material of a planetary nebula. There can also be shocks between shock fronts moving at different velocities. This emission is faint and requires deep X-ray observations.

\subsection{Probing the Supersoft Source evolution}
\label{s:sss}

Models of the opacity evolution of the ejecta assume a homogeneous evolution, however, observations reveal extreme variability by more than an order of magnitude on short time scales of $<$hours. Further, quasi-periodic transient modulations of <100 seconds have been detected in some novae which might be related to nuclear-burning instabilities. High-spectral resolution timing studies would constitute an important contribution to understanding the origin of these periods.

Understanding the origin of these short-term variations requires not only high spectral resolution. High-resolution timing requires high sensitivity allowing the observations to be spliced into short time segments while still yielding well-exposed high-resolution spectra. The shorter the sub-time intervals, the lower the number of counts in each spectrum, and even for these very bright sources, the time resolution with which spectral variations can be studied require the high sensitivity of {\it LEM}.

The diagnostic power of high time and spectral resolution is demonstrated with Figures~\ref{f:timemap_spec} and \ref{f:timemap_qpo} for the example of the recurrent nova RS\,Oph. An {\it XMM-Newton} observation taken on Mach 11, 2006 (26.1 days after the 2006 outburst) contains simultaneous high-resolution spectra from the RGS and high-sensitivity data from the EPIC. The X-ray light curve contained an
increase in count rate by a factor $\sim 3$ starting $\sim 1.2$ hours after the start of the observation, and the origin of this brightness increase is of interest for understanding the opacity evolution of the nova ejecta. Figure~\ref{f:timemap_spec} demonstrates that the brightness increase resolves into the appearance of a number of soft emission lines between 26 and 30\,\AA\ \citep{ness+2015, jness_basi}. They lived for only half an hour, after which the N\,{\sc vii} line at 24.8\,\AA\ became stronger, thus the overall flux hardly changed while important changes took place. The simultaneous EPIC spectra yield perfect blackbody spectra, and without high spectral resolution, these important spectral changes completely escape our attention.

In Figure~\ref{f:timemap_qpo}, it can seen that during the rise towards the higher count rate level, a 35-second period appeared for a short time. Such a transient period was seen at much higher significance during later X-ray observations and is thus considered real. This short appearance of the 35s-periodic signal can only be seen in the EPIC light curve. The RGS data are too noisy to resolve it at such high time resolution, let alone allowing extraction of high-resolution spectra that could probe the spectral changes with the 35s-period.

We have thus demonstrated that slicing high spectral-resolution observations into very short time slices yields important new physical insights, and the high sensitivity of {\it LEM} is can thus making transformative progress in this field by increasing the time resolution in which we can investigate brighter sources, and in pushing the high-resolution observational capability to much fainter novae.

\section{SERENDIPITOUS SCIENCE}
\label{s:serendip}

It is worth noting that studies of stellar coronal transients like the flares and the CMEs will likely be possible also with the analysis of interesting events that serendipitously will occur on late type stars falling in the {\it LEM} fov when observing a variety of non-stellar targets. The archive of {\it LEM} observations will be a valuable resource of interesting events. An estimate of the number of  "serendipitous coronal sources" on which interesting events could occurs can be computed from the RASS source list derived on a band-pass that is quite similar to the {\it LEM} one. The latest RASS2 catalogue contains about 135000 distinct X-ray sources and about 39700 of them have as a counterpart a normal star \citep{Boller+2016AA}. 

\begin{figure}
\begin{center}
\includegraphics[width=0.5\textwidth]{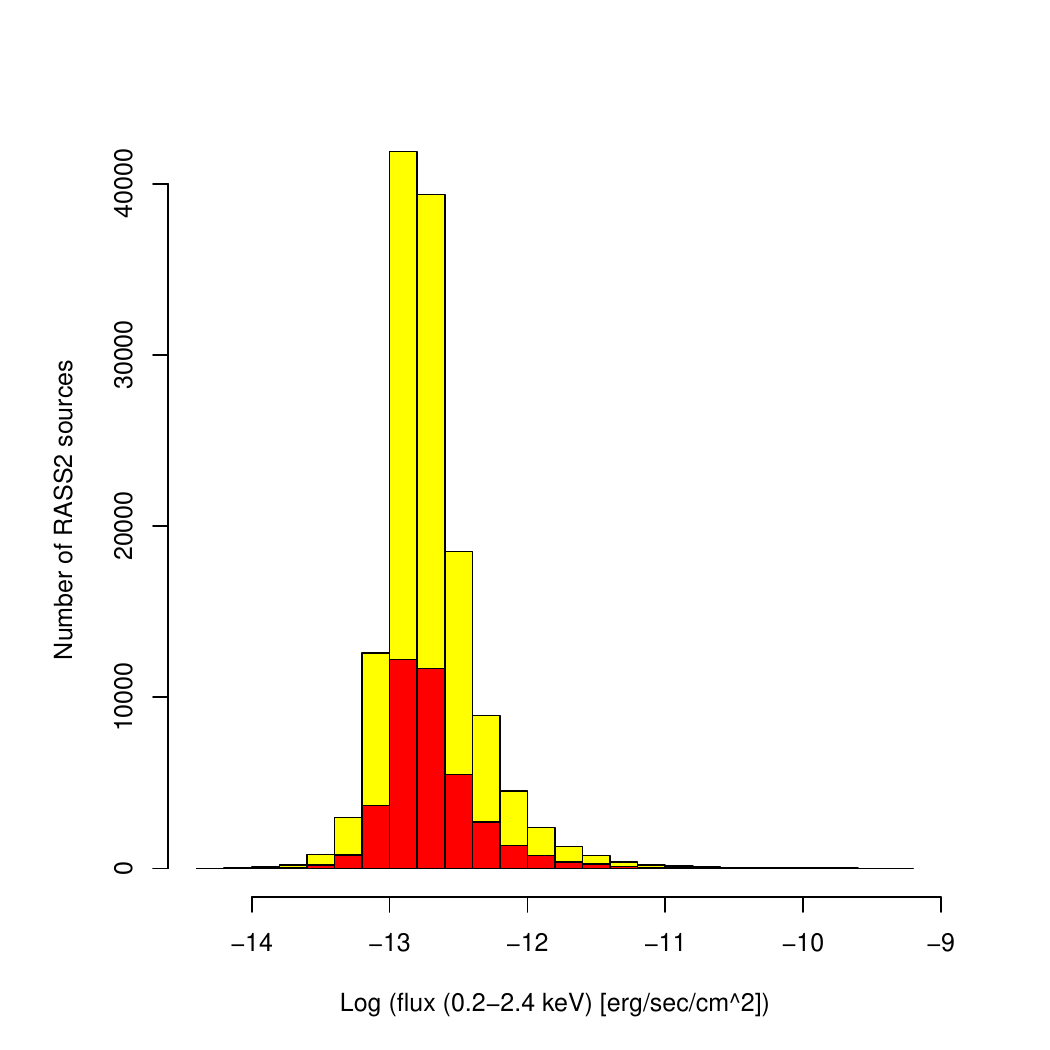}
\end{center}
\caption{The yellow histogram shows the flux distribution of the RASS2 sources derived assuming a conversion factor from rate to flux of 6 10$^{-12}$ cnt/erg/cm$^2$, while the red histogram shows the same flux distribution for the sources having as a counterpart a normal star as listed in available catalogues.} 
\label{fig.rass_hist}
\end{figure}

The RASS limiting flux is about 2 10$^{-13}$ erg/sec/cm$^2$ (cf. Fig. \ref{fig.rass_hist}); one can, probably conservatively, assume that the (stellar) sources for which it will be valuable to perform time resolved high resolution spectroscopy should have an X-ray flux higher than about 10$^{-12}$ erg/sec/cm$^2$. The number of RASS2 sources above this flux limit is about 5270 and 430 of them have a normal star as counterpart.

\begin{figure}
\begin{center}
\includegraphics[width=0.5\textwidth]{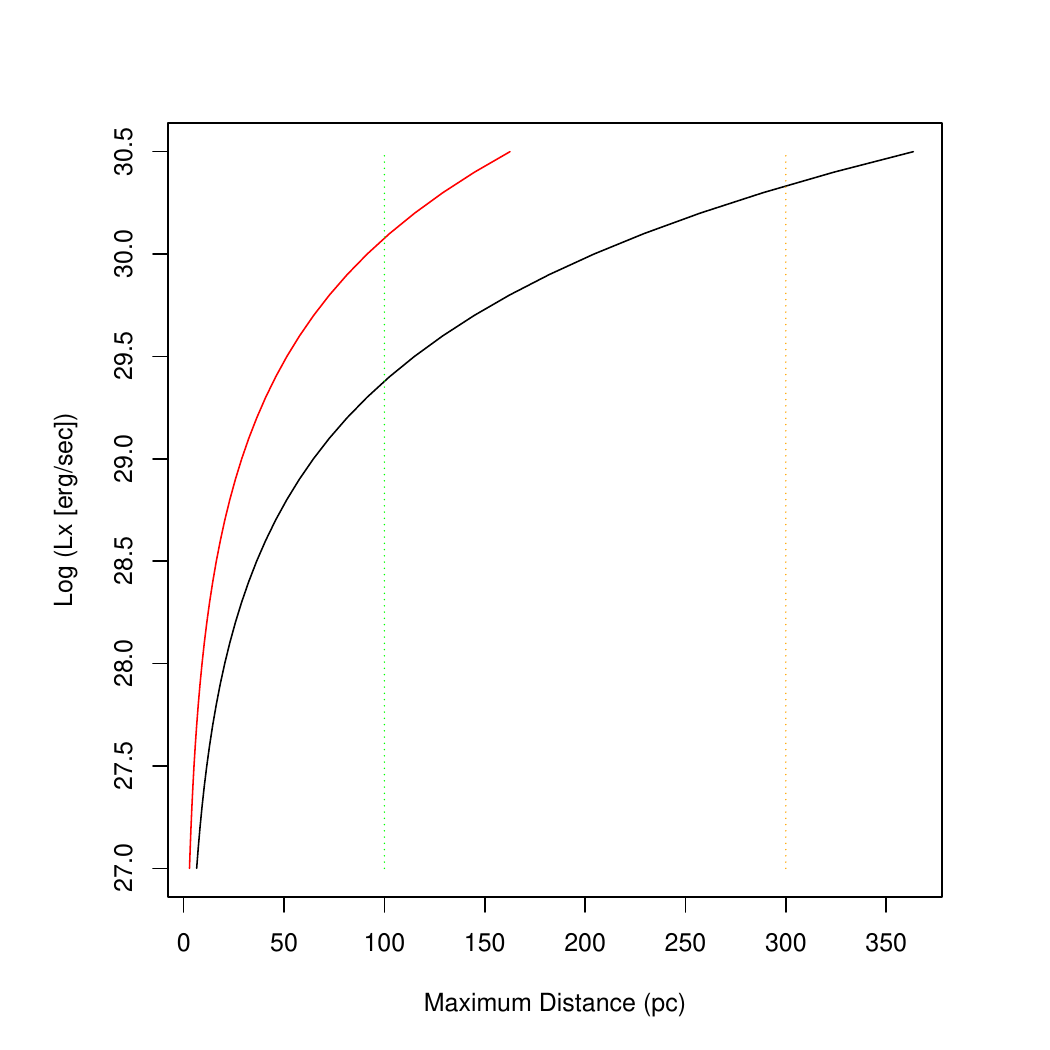}
\end{center}
\caption{The curves show the predicted maximum accessible distance for a (stellar) source as a function of its X-ray luminosity for the RASS typical limiting flux (black. f$_X$ = 2 10$^{-13}$ erg/sec/cm$^2$) and for f$_X$ = 10$^{-12}$ erg/sec/cm$2$ (red), respectively. The vertical dashed line are the typical scale height values for the young (green) and old (orange) stellar populations. }
\label{fig.distance}
\end{figure}

We show in Fig. \ref{fig.distance} the accessible distance (assuming negligible absorption) as a function of the logarithm of stellar X-ray luminosity. Log (L$_X$) ranges between about 26.5 and 29.3 for the old (age > 300 Myr) stellar population and between about 31.5 and 29.5 for the young (age < 50 Myr) stellar population (eg. \cite{Preibisch.and.Feigelson2005}), with median values of about 27.2 and 30.0, respectively. 

Given the size of the {\it LEM} fov and the planned duration of its operative life one can estimates that {\it LEM} will cover about 10\% of the entire sky with the vast majority of observations performed at high galactic latitude. It is worth to note that at the flux f$_X$ = 1e$^{-12}$ erg/sec/cm$^2$ the spatial distribution of old population stars can be considered as almost uniform, since the accessible distance is smaller than the scale height of 300 pc of this population, the situation is different for the young stellar population whose scale height is about 100 pc, much smaller than the predicted accessible distance. and whose spatial distribution, as GAIA has clearly shown, can hardly be considered as uniform (e.g. \cite{Prisinzano+2022AA}). As a consequence one can predict that the {\it LEM} archive could sample transient phenomena in "old" stars, while a proper coverage of "young" stars will require properly selected observations.

In conclusion we expect about 40-50 serendipitous stellar sources in the {\it LEM} archive for which time-resolved high resolution spectroscopy can reliably be performed, assuming 20-30\% of them will host transient phenomena the number of expected interesting events will likely range between 8 and 15.

The {\it LEM} archive will also be very valuable for all the high resolution studies based on stellar X-ray spectra serendipitously collected during the program of GO observations. For all the stellar sources that have been detected with the RASS good quality spectra could be collected if the source serendipitously fall in a {\it LEM} fov. Overall several thousands of stellar spectra will be available as can easily be extimatedd with a simple extrapolation based on the RASS2 source list. Moreover in the case of deeper GO observations one can expect to collect high quality spectra for numerous stellar sources fainter that those detected with the RASS. 

\section{Stacking Analyses}
\label{s:stacking}

While the effective area of {\it LEM} is unprecedented for a high-resolution X-ray spectrometer, there are still key parts of high energy stellar physics that will remain out of its reach within reasonable pointed observation exposure times. In some of these cases, stacking analysis based on the known positions of stars from deep surveys such as {\it Gaia} can provide key insights into the X-ray characteristics of groups of similar types of stars, such as plasma temperature distributions, densities, and chemical compositions. We cite three examples here, each of which are relevant for exoplanet science as well as intrinsic astrophysical interest. 

High-mass stars with masses $M >6 M_\odot$ and spectral types O to early B generate copious X-rays through instabilities and resulting shocks in their radiatively-driven winds. Low-mass stars with masses $M <1.5 M_\odot$ generate X-rays through convection and rotation-driven magnetic activity. The intermediate mass later B- and A-type stars are X-ray dark. Little is known about the transitions between these regimes. Stacking analysis of intermediate mass stars will probe the physics and plasma conditions of these boundaries. 

Little is known about magnetic activity at the end of the main-sequence. The very cool, neutral photospheres of late M dwarfs and brown dwarfs should be largely incapable of sustaining magnetic stresses, yet flaring on these objects is frequent. Generally too X-ray faint for detailed study,  these objects are sufficiently numerous in the solar vicinity that stacking of {\it LEM} spectra can be key for revealing the details of their plasma physics. 

Radiative losses from stellar coronae are dominated by metal lines. One outstanding problem in coronal physics is whether metal-poor stars sustain X-ray emitting coronae, and if so what is the effect of metal paucity on the coronal energy balance. Population II stars are metal-poor, but they are also old and expected to be of low magnetic activity and faint in X-rays. Stacked {\it LEM} spectra of Population II stars will provide a window into this astrophysical regime that will be generally inaccessible through pointed observations.  

\section{CONCLUDING REMARKS}
\label{s:conc}

We have outlined some of the major aspects of the scientific case for next generation soft X-ray spectroscopy of cool stars. The great strength of {\it LEM} in the realm of cool star physics is its substantial effective area, coupled with the inherent energy resolution of its microcalorimeter, which easily achieves resolving powers of 1000. The resulting high sensitivity enables the observation of much fainter and more distant stars, greatly broadening the range and number of potential targets for high-resolution study. 

For brighter sources, this next generation sensitivity level facilitates time-domain studies, where enough signal can be collected in short exposures to enable the study of source variability on short timescales. 

Finally, the large 30'$\times$30' field of view of {\it LEM} will produce a spectacular and rich array of spectra of stars in star forming regions and in open clusters that will be invaluable for probing stellar physics and its evolution in time.


\small
\vspace{-6mm}
\parindent=0cm
\baselineskip=12pt

\bibliography{coolstarsLEMwhitepaper,corabun}
\bibliographystyle{aasjournal}

\end{document}

%% file: headstuff.tex
\usepackage[super,comma]{natbib} 
\usepackage{apjfonts} 
\usepackage{graphicx} 
\usepackage{amssymb,amsmath} 
\usepackage{color} 
\usepackage{enumitem} 
\usepackage{wrapfig} 
\usepackage{tikz} 
\usepackage{authblk}

\usepackage{multirow} 
\usepackage{tabularx} 
\usepackage{array} 

\usepackage{hyperref} 
\hypersetup{
    colorlinks,
    linkcolor={blue!80!black},
    citecolor={blue!80!black},
    urlcolor={blue!80!black}
}

\usepackage{sidecap}
\sidecaptionvpos{figure}{c} 

\newcolumntype{L}[1]{>{\raggedright\let\newline\\\arraybackslash\hspace{0pt}}p{#1}}
\newcolumntype{C}[1]{>{\centering\let\newline\\\arraybackslash\hspace{0pt}}p{#1}}
\newcolumntype{R}[1]{>{\raggedleft\let\newline\\\arraybackslash\hspace{0pt}}p{#1}}
\renewcommand\arraystretch{1.3}

\definecolor{headcolor}{rgb}{0.65,0.65,0.65}
\usepackage{fancyhdr}
\renewcommand{\topmargin}{-9mm}

\renewcommand{\textfloatsep}{4mm}

\newcommand{\bsf}{\sffamily\bfseries}

\definecolor{callout}{rgb}{0.25,0.40,0.85}
\definecolor{synergies}{rgb}{0.20,0.45,0.99}
\definecolor{methods}{rgb}{0.20,0.70,0.45}

\definecolor{calllem}{rgb}{0.20,0.45,0.99}
\definecolor{tabledef}{rgb}{0.95,0.95,0.95}
\definecolor{tablealt}{rgb}{0.77,0.80,1.0}
\definecolor{tablelem}{rgb}{0.80,0.85,1.0}
\definecolor{whitelem}{rgb}{1.0,1.0,1.0}
\definecolor{greenlem}{rgb}{0.7,1.0,0.7}

%% file: main.bbl
\begin{thebibliography}{}
\expandafter\ifx\csname natexlab\endcsname\relax\def\natexlab#1{#1}\fi
\providecommand{\url}[1]{\href{#1}{#1}}
\providecommand{\dodoi}[1]{doi:~\href{http://doi.org/#1}{\nolinkurl{#1}}}
\providecommand{\doeprint}[1]{\href{http://ascl.net/#1}{\nolinkurl{http://ascl.net/#1}}}
\providecommand{\doarXiv}[1]{\href{https://arxiv.org/abs/#1}{\nolinkurl{https://arxiv.org/abs/#1}}}

\bibitem[{{Acharya} {et~al.}(2023){Acharya}, {Kashyap}, {Saar}, {Singh}, \&
  {Cuntz}}]{2023ApJ...951..152A}
{Acharya}, A., {Kashyap}, V.~L., {Saar}, S.~H., {Singh}, K.~P., \& {Cuntz}, M.
  2023, \apj, 951, 152, \dodoi{10.3847/1538-4357/acd054}

\bibitem[{{Andrews} {et~al.}(2016){Andrews}, {Wilner}, {Zhu}, {Birnstiel},
  {Carpenter}, {P{\'e}rez}, {Bai}, {{\"O}berg}, {Hughes}, {Isella}, \&
  {Ricci}}]{Andrews2016}
{Andrews}, S.~M., {Wilner}, D.~J., {Zhu}, Z., {et~al.} 2016, \apjl, 820, L40,
  \dodoi{10.3847/2041-8205/820/2/L40}

\bibitem[{{Argiroffi} {et~al.}(2017){Argiroffi}, {Drake}, {Bonito}, {Orlando},
  {Peres}, \& {Miceli}}]{argiroffi17}
{Argiroffi}, C., {Drake}, J.~J., {Bonito}, R., {et~al.} 2017, \aap, 607, A14,
  \dodoi{10.1051/0004-6361/201731342}

\bibitem[{{Argiroffi} {et~al.}(2019){Argiroffi}, {Reale}, {Drake},
  {Ciaravella}, {Testa}, {Bonito}, {Miceli}, {Orlando}, \&
  {Peres}}]{Argiroffi2019}
{Argiroffi}, C., {Reale}, F., {Drake}, J.~J., {et~al.} 2019, Nature Astronomy,
  3, 742, \dodoi{10.1038/s41550-019-0781-4}

\bibitem[{{Audard} {et~al.}(2014){Audard}, {{\'A}brah{\'a}m}, {Dunham},
  {Green}, { Grosso}, {Hamaguchi}, {Kastner}, {K{\'o}sp{\'a}l}, {Lodato},
  {Romanova}, {Skinner}, {Vorobyov}, \& {Zhu}}]{audard14}
{Audard}, M., {{\'A}brah{\'a}m}, P., {Dunham}, M.~M., {et~al.} 2014, Protostars
  \& Planets~VI, ed. H.~{Beuther}, R.~{Klessen}, C.~{Dullemond}, \&
  T.~{Henning} (Univ.\ of Arizona Space Sci.\ Series), 387--410.
\newblock \doarXiv{1401.3368}

\bibitem[{{Bacciotti} {et~al.}(2000){Bacciotti}, {Mundt}, {Ray},
  {Eisl{\"o}ffel}, {Solf}, \& {Camezind}}]{bacciotti00}
{Bacciotti}, F., {Mundt}, R., {Ray}, T.~P., {et~al.} 2000, \apjl, 537, L49,
  \dodoi{10.1086/312745}

\bibitem[{{Boller} {et~al.}(2016){Boller}, {Freyberg}, {Tr{\"u}mper}, {Haberl},
  {Voges}, \& {Nandra}}]{Boller+2016AA}
{Boller}, T., {Freyberg}, M.~J., {Tr{\"u}mper}, J., {et~al.} 2016, \aap, 588,
  A103, \dodoi{10.1051/0004-6361/201525648}

\bibitem[{{Brickhouse} {et~al.}(2012){Brickhouse}, {Cranmer}, {Dupree},
  {G{\"u}nther}, {Luna}, \& {Wolk}}]{brickhouse12}
{Brickhouse}, N.~S., {Cranmer}, S.~R., {Dupree}, A.~K., {et~al.} 2012, \apjl,
  760, L21, \dodoi{10.1088/2041-8205/760/2/L21}

\bibitem[{{Brickhouse} {et~al.}(2010){Brickhouse}, {Cranmer}, {Dupree}, {Luna},
  \& {Wolk}}]{brickhouse10}
{Brickhouse}, N.~S., {Cranmer}, S.~R., {Dupree}, A.~K., {Luna}, G.~J.~M., \&
  {Wolk}, S. 2010, \apj, 710, 1835, \dodoi{10.1088/0004-637X/710/2/1835}

\bibitem[{{Brinkman} {et~al.}(2001){Brinkman}, {Behar}, {G{\" u}del}, {Audard},
  {den Boggende}, {Branduardi-Raymont}, {Cottam}, {Erd}, {den Herder},
  {Jansen}, {Kaastra}, {Kahn}, {Mewe}, {Paerels}, {Peterson}, {Rasmussen},
  {Sakelliou}, \& {de Vries}}]{Brinkman.etal:01}
{Brinkman}, A.~C., {Behar}, E., {G{\" u}del}, M., {et~al.} 2001, \aap, 365,
  L324.
\newblock
  \url{http://adsabs.harvard.edu/cgi-bin/nph-bib_query?bibcode=2001A%26A...365L.324B&db_key=AST}

\bibitem[{{Chabrier} {et~al.}(2023){Chabrier}, {Baraffe}, {Phillips}, \&
  {Debras}}]{chabrier23}
{Chabrier}, G., {Baraffe}, I., {Phillips}, M., \& {Debras}, F. 2023, \aap, 671,
  A119, \dodoi{10.1051/0004-6361/202243832}

\bibitem[{{Chakraborty} {et~al.}(2022){Chakraborty}, {Ferland}, {Chatzikos},
  {Fabian}, {Bianchi}, {Guzm{\'a}n}, \& {Su}}]{chakraborty2022}
{Chakraborty}, P., {Ferland}, G.~J., {Chatzikos}, M., {et~al.} 2022, \apj, 935,
  70, \dodoi{10.3847/1538-4357/ac7eb9}

\bibitem[{{Coffaro} {et~al.}(2020){Coffaro}, {Stelzer}, {Orlando}, {Hall},
  {Metcalfe}, {Wolter}, {Mittag}, {Sanz-Forcada}, {Schneider}, \&
  {Ducci}}]{Coffaro2020}
{Coffaro}, M., {Stelzer}, B., {Orlando}, S., {et~al.} 2020, \aap, 636, A49,
  \dodoi{10.1051/0004-6361/201936479}

\bibitem[{{Dempsey} {et~al.}(1997){Dempsey}, {Linsky}, {Fleming}, \&
  {Schmitt}}]{Dempsey1997}
{Dempsey}, R.~C., {Linsky}, J.~L., {Fleming}, T.~A., \& {Schmitt}, J.~H.~M.~M.
  1997, \apj, 478, 358, \dodoi{10.1086/303786}

\bibitem[{{Done} {et~al.}(1995){Done}, {Osborne}, \& {Beardmore}}]{done1995}
{Done}, C., {Osborne}, J.~P., \& {Beardmore}, A.~P. 1995, \mnras, 276, 483,
  \dodoi{10.1093/mnras/276.2.483}

\bibitem[{{Dougados} {et~al.}(2000){Dougados}, {Cabrit}, {Lavalley}, \&
  {Ménard}}]{dougados00}
{Dougados}, C., {Cabrit}, S., {Lavalley}, C., \& {Ménard}, F. 2000, A\&A

\bibitem[{{Drake}(2002)}]{Drake:02}
{Drake}, J. 2002, NASA STI/Recon Technical Report N, 2, 42360

\bibitem[{{Drake}(1996{\natexlab{a}})}]{Drake:96}
{Drake}, J.~J. 1996{\natexlab{a}}, in Astronomical Society of the Pacific
  Conference Series, Vol. 109, Cool Stars, Stellar Systems, and the Sun, ed.
  {R.~Pallavicini \& A.~K.~Dupree}, 203--+

\bibitem[{{Drake} {et~al.}(2001){Drake}, {Brickhouse}, {Kashyap}, {Laming},
  {Huenemoerder}, {Smith}, \& {Wargelin}}]{Drake.etal:01}
{Drake}, J.~J., {Brickhouse}, N.~S., {Kashyap}, V., {et~al.} 2001, \apjl, 548,
  L81.
\newblock
  \url{http://adsabs.harvard.edu/cgi-bin/nph-bib_query?bibcode=2001ApJ...548L..81D&db_key=AST}

\bibitem[{{Drake} {et~al.}(2013){Drake}, {Cohen}, {Yashiro}, \&
  {Gopalswamy}}]{drake2013}
{Drake}, J.~J., {Cohen}, O., {Yashiro}, S., \& {Gopalswamy}, N. 2013, \apj,
  764, 170, \dodoi{10.1088/0004-637X/764/2/170}

\bibitem[{{Drake} {et~al.}(2009){Drake}, {Ratzlaff}, {Laming}, \&
  {Raymond}}]{Drake2009}
{Drake}, J.~J., {Ratzlaff}, P.~W., {Laming}, J.~M., \& {Raymond}, J. 2009,
  \apj, 703, 1224, \dodoi{10.1088/0004-637X/703/2/1224}

\bibitem[{{Drake}(1996{\natexlab{b}})}]{Drake:96b}
{Drake}, S.~A. 1996{\natexlab{b}}, in Astronomical Society of the Pacific
  Conference Series, Vol.~99, Cosmic Abundances, ed. {S.~S.~Holt \&
  G.~Sonneborn}, 215--+

\bibitem[{{Dupree} {et~al.}(2012){Dupree}, {Brickhouse}, {Cranmer}, {Luna},
  {Schneider}, {Bessell}, {Bonanos}, {Crause}, {Lawson}, {Mallik}, \&
  {Schuler}}]{dupree12}
{Dupree}, A.~K., {Brickhouse}, N.~S., {Cranmer}, S.~R., {et~al.} 2012, \apj,
  750, 73, \dodoi{10.1088/0004-637X/750/1/73}

\bibitem[{{Eisl{\"o}ffel} \& {Mundt}(1998)}]{eisloffel98}
{Eisl{\"o}ffel}, J., \& {Mundt}, R. 1998, \aj, 115, 1554,
  \dodoi{10.1086/300282}

\bibitem[{{Favata} {et~al.}(2005){Favata}, {Flaccomio}, {Reale}, {Micela},
  {Sciortino}, {Shang}, {Stassun}, \& {Feigelson}}]{Favata2005}
{Favata}, F., {Flaccomio}, E., {Reale}, F., {et~al.} 2005, \apjs, 160, 469,
  \dodoi{10.1086/432542}

\bibitem[{{Feigelson} \& {Montmerle}(1999)}]{feigelson99}
{Feigelson}, E.~D., \& {Montmerle}, T. 1999, \araa, 37, 363,
  \dodoi{10.1146/annurev.astro.37.1.363}

\bibitem[{{Feigelson} {et~al.}(2005){Feigelson}, {Getman}, {Townsley},
  {Garmire}, {Preibisch}, {Grosso}, {Montmerle}, {Muench}, \&
  {McCaughrean}}]{Feigelson2005}
{Feigelson}, E.~D., {Getman}, K., {Townsley}, L., {et~al.} 2005, \apjs, 160,
  379, \dodoi{10.1086/432512}

\bibitem[{{Fleming} \& {Tagliaferri}(1996)}]{Fleming1996}
{Fleming}, T.~A., \& {Tagliaferri}, G. 1996, \apjl, 472, L101,
  \dodoi{10.1086/310361}

\bibitem[{{G{\" u}del} {et~al.}(2005){G{\" u}del}, {Skinner}, {Briggs},
  {Audard}, {Arzner}, \& {Telleschi}}]{guedel05}
{G{\" u}del}, M., {Skinner}, S.~L., {Briggs}, K.~R., {et~al.} 2005, \apjl, 626,
  L53, \dodoi{10.1086/431666}

\bibitem[{{Galeev} {et~al.}(1979){Galeev}, {Rosner}, \& {Vaiana}}]{Galeev1979}
{Galeev}, A.~A., {Rosner}, R., \& {Vaiana}, G.~S. 1979, \apj, 229, 318,
  \dodoi{10.1086/156957}

\bibitem[{{Getman} \& {Feigelson}(2021)}]{getman21}
{Getman}, K.~V., \& {Feigelson}, E.~D. 2021, \apj, 916, 32,
  \dodoi{10.3847/1538-4357/ac00be}

\bibitem[{{Getman} {et~al.}(2005){Getman}, {Flaccomio}, {Broos}, {Grosso},
  {Tsujimoto}, {Townsley}, {Garmire}, {Kastner}, {Li}, {Harnden}, {Wolk},
  {Murray}, {Lada}, {Muench}, {McCaughrean}, {Meeus}, {Damiani}, {Micela},
  {Sciortino}, {Bally}, {Hillenbrand}, {Herbst}, {Preibisch}, \&
  {Feigelson}}]{Getman2005}
{Getman}, K.~V., {Flaccomio}, E., {Broos}, P.~S., {et~al.} 2005, \apjs, 160,
  319, \dodoi{10.1086/432092}

\bibitem[{{Glassgold} {et~al.}(2004){Glassgold}, {Najita}, \&
  {Igea}}]{glassgold04}
{Glassgold}, A.~E., {Najita}, J., \& {Igea}, J. 2004, \apj, 615, 972,
  \dodoi{10.1086/424509}

\bibitem[{{Grosso} {et~al.}(2010){Grosso}, {Hamaguchi}, {Kastner}, {Richmond},
  \& {Weintraub}}]{grosso10}
{Grosso}, N., {Hamaguchi}, K., {Kastner}, J.~H., {Richmond}, M.~W., \&
  {Weintraub}, D.~A. 2010, \aap, 522, A56, \dodoi{10.1051/0004-6361/200913850}

\bibitem[{{Grosso} {et~al.}(2007){Grosso}, {Briggs}, {G{\"u}del}, {Guieu},
  {Franciosini}, {Palla}, {Dougados}, {Monin}, {M{\'e}nard}, {Bouvier},
  {Audard}, \& {Telleschi}}]{grosso07}
{Grosso}, N., {Briggs}, K.~R., {G{\"u}del}, M., {et~al.} 2007, \aap, 468, 391,
  \dodoi{10.1051/0004-6361:20065559}

\bibitem[{{G{\"u}del} {et~al.}(2004){G{\"u}del}, {Audard}, {Reale}, {Skinner},
  \& {Linsky}}]{guedel2004}
{G{\"u}del}, M., {Audard}, M., {Reale}, F., {Skinner}, S.~L., \& {Linsky},
  J.~L. 2004, \aap, 416, 713, \dodoi{10.1051/0004-6361:20031471}

\bibitem[{{G{\"u}del} {et~al.}(2008){G{\"u}del}, {Skinner}, {Audard}, {Briggs},
  \& {Cabrit}}]{guedel08b}
{G{\"u}del}, M., {Skinner}, S.~L., {Audard}, M., {Briggs}, K.~R., \& {Cabrit},
  S. 2008, \aap, 478, 797, \dodoi{10.1051/0004-6361:20078141}

\bibitem[{{G{\"u}del} {et~al.}(2007{\natexlab{a}}){G{\"u}del}, {Telleschi},
  {Audard}, {L.~Skinner}, {Briggs}, {Palla}, \& {Dougados}}]{guedel07d}
{G{\"u}del}, M., {Telleschi}, A., {Audard}, M., {et~al.} 2007{\natexlab{a}},
  \aap, 468, 515, \dodoi{10.1051/0004-6361:20065736}

\bibitem[{{G{\"u}del} {et~al.}(2001){G{\"u}del}, {Audard}, {Briggs}, {Haberl},
  {Magee}, {Maggio}, {Mewe}, {Pallavicini}, \& {Pye}}]{Gudel2001}
{G{\"u}del}, M., {Audard}, M., {Briggs}, K., {et~al.} 2001, \aap, 365, L336,
  \dodoi{10.1051/0004-6361:20000220}

\bibitem[{{G{\"u}del} {et~al.}(2007{\natexlab{b}}){G{\"u}del}, {Briggs},
  {Arzner}, {Audard}, {Bouvier}, {Feigelson}, {Franciosini}, {Glauser},
  {Grosso}, {Micela}, {Monin}, {Montmerle}, {Padgett}, {Palla}, {Pillitteri},
  {Rebull}, {Scelsi}, {Silva}, {Skinner}, {Stelzer}, \& {Telleschi}}]{guedel07}
{G{\"u}del}, M., {Briggs}, K.~R., {Arzner}, K., {et~al.} 2007{\natexlab{b}},
  \aap, 468, 353, \dodoi{10.1051/0004-6361:20065724}

\bibitem[{{G{\"u}nther} {et~al.}(2009){G{\"u}nther}, {Matt}, \&
  {Li}}]{guenther09b}
{G{\"u}nther}, H.~M., {Matt}, S.~P., \& {Li}, Z.~Y. 2009, \aap, 493, 579,
  \dodoi{10.1051/0004-6361:200810886}

\bibitem[{{Hall} {et~al.}(2022){Hall}, {Kashyap}, {Drake}, {Pillitteri},
  {Poppenhaeger}, \& {Wolk}}]{2022HEAD...1911063H}
{Hall}, J., {Kashyap}, V., {Drake}, J., {et~al.} 2022, in AAS/High Energy
  Astrophysics Division, Vol.~54, AAS/High Energy Astrophysics Division, 110.63

\bibitem[{{He} {et~al.}(2022){He}, {Wang}, {Luo}, {Li}, {Liu}, \&
  {Jiang}}]{He2022}
{He}, Z., {Wang}, K., {Luo}, Y., {et~al.} 2022, \apjs, 262, 7,
  \dodoi{10.3847/1538-4365/ac7c17}

\bibitem[{{Herbig}(1989)}]{Herbig1989}
{Herbig}, G.~H. 1989, in European Southern Observatory Conference and Workshop
  Proceedings, Vol.~33, European Southern Observatory Conference and Workshop
  Proceedings, 233--246

\bibitem[{{Houck} \& {Denicola}(2000)}]{houck00}
{Houck}, J.~C., \& {Denicola}, L.~A. 2000, 216, 591.
\newblock \url{https://www.adass.org/adass/proceedings/adass99/P2-39/}

\bibitem[{{Huenemoerder} {et~al.}(2011){Huenemoerder}, {Houck}, {Nowak},
  {Schulz}, \& {Davis}}]{huenemoerder11b}
{Huenemoerder}, D.~P., {Houck}, J.~C., {Nowak}, M.~A., {Schulz}, N.~S., \&
  {Davis}, J.~E. 2011, Atomic Data Unleashed: A Compact Database of Helium
  Triplet X-Ray Line Emissivities
  (http://space.mit.edu/cxc/analysis/he\_modifier/index.html)

\bibitem[{{Irving} {et~al.}(2023){Irving}, {Saar}, {Wargelin}, \& {do
  Nascimento}}]{Irving2023}
{Irving}, Z.~A., {Saar}, S.~H., {Wargelin}, B.~J., \& {do Nascimento}, J.-D.
  2023, \apj, 949, 51, \dodoi{10.3847/1538-4357/acc468}

\bibitem[{{Islam} \& {Mukai}(2021)}]{islam2021}
{Islam}, N., \& {Mukai}, K. 2021, \apj, 919, 90,
  \dodoi{10.3847/1538-4357/ac134e}

\bibitem[{{Kastner} {et~al.}(2002){Kastner}, {Huenemoerder}, {Schulz},
  {Canizares}, \& {Weintraub}}]{kastner02}
{Kastner}, J.~H., {Huenemoerder}, D.~P., {Schulz}, N.~S., {Canizares}, C.~R.,
  \& {Weintraub}, D.~A. 2002, \apj, 567, 434, \dodoi{10.1086/338419}

\bibitem[{{K{\"o}nig} {et~al.}(2022){K{\"o}nig}, {Wilms}, {Arcodia}, {Dauser},
  {Dennerl}, {Doroshenko}, {Haberl}, {H{\"a}mmerich}, {Kirsch}, {Kreykenbohm},
  {Lorenz}, {Malyali}, {Merloni}, {Rau}, {Rauch}, {Sala}, {Schwope},
  {Suleimanov}, {Weber}, \& {Werner}}]{koenig+2022}
{K{\"o}nig}, O., {Wilms}, J., {Arcodia}, R., {et~al.} 2022, Nature, 605, 248,
  \dodoi{10.1038/s41586-022-04635-y}

\bibitem[{{Laming}(2015)}]{Laming:15}
{Laming}, J.~M. 2015, Living Reviews in Solar Physics, 12, 2,
  \dodoi{10.1007/lrsp-2015-2}

\bibitem[{{Llama} \& {Shkolnik}(2015)}]{2015ApJ...802...41L}
{Llama}, J., \& {Shkolnik}, E.~L. 2015, \apj, 802, 41,
  \dodoi{10.1088/0004-637X/802/1/41}

\bibitem[{{Luna} {et~al.}(2013){Luna}, {Sokoloski}, {Mukai}, \&
  {Nelson}}]{2013A&A...559A...6L}
{Luna}, G.~J.~M., {Sokoloski}, J.~L., {Mukai}, K., \& {Nelson}, T. 2013, \aap,
  559, A6, \dodoi{10.1051/0004-6361/201220792}

\bibitem[{{Mittag} {et~al.}(2019){Mittag}, {Schmitt}, {Hempelmann}, \&
  {Schr{\"o}der}}]{Mittag2019}
{Mittag}, M., {Schmitt}, J.~H.~M.~M., {Hempelmann}, A., \& {Schr{\"o}der},
  K.~P. 2019, \aap, 621, A136, \dodoi{10.1051/0004-6361/201834319}

\bibitem[{{Mokler} \& {Stelzer}(2002)}]{mokler02}
{Mokler}, F., \& {Stelzer}, B. 2002, \aap, 391, 1025,
  \dodoi{10.1051/0004-6361:20020887}

\bibitem[{{Montmerle} \& {Townsley}(2012)}]{montmerle12}
{Montmerle}, T., \& {Townsley}, L.~K. 2012, Astronomische Nachrichten, 333,
  355, \dodoi{10.1002/asna.201211678}

\bibitem[{{Moschou} {et~al.}(2019){Moschou}, {Drake}, {Cohen},
  {Alvarado-G{\'o}mez}, {Garraffo}, \& {Fraschetti}}]{moschou2019}
{Moschou}, S.-P., {Drake}, J.~J., {Cohen}, O., {et~al.} 2019, \apj, 877, 105,
  \dodoi{10.3847/1538-4357/ab1b37}

\bibitem[{{Mukai} {et~al.}(2003){Mukai}, {Kinkhabwala}, {Peterson}, {Kahn}, \&
  {Paerels}}]{mukai2003}
{Mukai}, K., {Kinkhabwala}, A., {Peterson}, J.~R., {Kahn}, S.~M., \& {Paerels},
  F. 2003, \apjl, 586, L77, \dodoi{10.1086/374583}

\bibitem[{{Nakajima} {et~al.}(1995){Nakajima}, {Oppenheimer}, {Kulkarni},
  {Golimowski}, {Matthews}, \& {Durrance}}]{nakajima95}
{Nakajima}, T., {Oppenheimer}, B.~R., {Kulkarni}, S.~R., {et~al.} 1995, \nat,
  378, 463, \dodoi{10.1038/378463a0}

\bibitem[{{Ness}(2012)}]{jness_basi}
{Ness}, J.~U. 2012, Bulletin of the Astronomical Society of India, 40, 353.
\newblock \doarXiv{1209.2153}

\bibitem[{{Ness} {et~al.}(2015){Ness}, {Beardmore}, {Osborne}, {Kuulkers},
  {Henze}, {Piro}, {Drake}, {Dobrotka}, {Schwarz}, {Starrfield}, {Kretschmar},
  {Hirsch}, \& {Wilms}}]{ness+2015}
{Ness}, J.-U., {Beardmore}, A.~P., {Osborne}, J.~P., {et~al.} 2015, A\&A, 578,
  A39, \dodoi{10.1051/0004-6361/201425178}

\bibitem[{{Neuh{\"a}user} \& {Comer{\'o}n}(1998)}]{neuhaeuser98}
{Neuh{\"a}user}, R., \& {Comer{\'o}n}, F. 1998, Science, 282, 83

\bibitem[{{Orlando} {et~al.}(2017){Orlando}, {Favata}, {Micela}, {Sciortino},
  {Maggio}, {Schmitt}, {Robrade}, \& {Mittag}}]{Orlando2017}
{Orlando}, S., {Favata}, F., {Micela}, G., {et~al.} 2017, \aap, 605, A19,
  \dodoi{10.1051/0004-6361/201731301}

\bibitem[{{Osten} \& {Wolk}(2015)}]{osten2015}
{Osten}, R.~A., \& {Wolk}, S.~J. 2015, \apj, 809, 79,
  \dodoi{10.1088/0004-637X/809/1/79}

\bibitem[{{Pabst} {et~al.}(2019){Pabst}, {Higgins}, {Goicoechea}, {Teyssier},
  {Berne}, {Chambers}, {Wolfire}, {Suri}, {Guesten}, {Stutzki}, {Graf},
  {Risacher}, \& {Tielens}}]{pabst19}
{Pabst}, C., {Higgins}, R., {Goicoechea}, J.~R., {et~al.} 2019, \nat, 565, 618,
  \dodoi{10.1038/s41586-018-0844-1}

\bibitem[{{Pabst} {et~al.}(2020){Pabst}, {Goicoechea}, {Teyssier}, {Bern{\'e}},
  {Higgins}, {Chambers}, {Kabanovic}, {G{\"u}sten}, {Stutzki}, \&
  {Tielens}}]{pabst20}
{Pabst}, C.~H.~M., {Goicoechea}, J.~R., {Teyssier}, D., {et~al.} 2020, \aap,
  639, A2, \dodoi{10.1051/0004-6361/202037560}

\bibitem[{{Pabst} {et~al.}(2022){Pabst}, {Goicoechea}, {Hacar}, {Teyssier},
  {Bern{\'e}}, {Wolfire}, {Higgins}, {Chambers}, {Kabanovic}, {G{\"u}sten},
  {Stutzki}, {Kramer}, \& {Tielens}}]{pabst22}
{Pabst}, C.~H.~M., {Goicoechea}, J.~R., {Hacar}, A., {et~al.} 2022, \aap, 658,
  A98, \dodoi{10.1051/0004-6361/202140805}

\bibitem[{{Poppenhaeger} {et~al.}(2013){Poppenhaeger}, {Schmitt}, \&
  {Wolk}}]{2013ApJ...773...62P}
{Poppenhaeger}, K., {Schmitt}, J.~H.~M.~M., \& {Wolk}, S.~J. 2013, \apj, 773,
  62, \dodoi{10.1088/0004-637X/773/1/62}

\bibitem[{{Porquet} {et~al.}(2010){Porquet}, {Dubau}, \& {Grosso}}]{porquet10}
{Porquet}, D., {Dubau}, J., \& {Grosso}, N. 2010, \ssr, 157, 103,
  \dodoi{10.1007/s11214-010-9731-2}

\bibitem[{{Preibisch} \& {Feigelson}(2005)}]{Preibisch.and.Feigelson2005}
{Preibisch}, T., \& {Feigelson}, E.~D. 2005, \apjs, 160, 390,
  \dodoi{10.1086/432094}

\bibitem[{{Preibisch} {et~al.}(2005){Preibisch}, {McCaughrean}, {Grosso},
  {Feigelson}, {Flaccomio}, {Getman}, {Hillenbrand}, {Meeus}, {Micela},
  {Sciortino}, \& {Stelzer}}]{preibisch05}
{Preibisch}, T., {McCaughrean}, M.~J., {Grosso}, N., {et~al.} 2005, \apjs, 160,
  582, \dodoi{10.1086/432098}

\bibitem[{{Prisinzano} {et~al.}(2022){Prisinzano}, {Damiani}, {Sciortino},
  {Flaccomio}, {Guarcello}, {Micela}, {Tognelli}, {Jeffries}, \&
  {Alcal{\'a}}}]{Prisinzano+2022AA}
{Prisinzano}, L., {Damiani}, F., {Sciortino}, S., {et~al.} 2022, \aap, 664,
  A175, \dodoi{10.1051/0004-6361/202243580}

\bibitem[{{Reale} {et~al.}(2004){Reale}, {G{\"u}del}, {Peres}, \&
  {Audard}}]{reale2004}
{Reale}, F., {G{\"u}del}, M., {Peres}, G., \& {Audard}, M. 2004, \aap, 416,
  733, \dodoi{10.1051/0004-6361:20034027}

\bibitem[{{Reale} {et~al.}(2018){Reale}, {Lopez-Santiago}, {Flaccomio},
  {Petralia}, \& {Sciortino}}]{Reale2018}
{Reale}, F., {Lopez-Santiago}, J., {Flaccomio}, E., {Petralia}, A., \&
  {Sciortino}, S. 2018, \apj, 856, 51, \dodoi{10.3847/1538-4357/aaaf1f}

\bibitem[{{Rebolo} {et~al.}(1995){Rebolo}, {Zapatero-Osorio}, \&
  {Martin}}]{rebolo95}
{Rebolo}, R., {Zapatero-Osorio}, M.~R., \& {Martin}, E.~L. 1995, \nat, 377,
  129, \dodoi{10.1038/377129a0}

\bibitem[{{Rivera} {et~al.}(2019){Rivera}, {Landi}, {Lepri}, \&
  {Gilbert}}]{rivera2019}
{Rivera}, Y.~J., {Landi}, E., {Lepri}, S.~T., \& {Gilbert}, J.~A. 2019, \apj,
  874, 164, \dodoi{10.3847/1538-4357/ab0e11}

\bibitem[{{Rutledge} {et~al.}(2000){Rutledge}, {Basri}, {Mart{\'{\i}}n}, \&
  {Bildsten}}]{rutledge00}
{Rutledge}, R.~E., {Basri}, G., {Mart{\'{\i}}n}, E.~L., \& {Bildsten}, L. 2000,
  \apjl, 538, L141, \dodoi{10.1086/312817}

\bibitem[{{Sanz-Forcada} {et~al.}(2019){Sanz-Forcada}, {Stelzer}, {Coffaro},
  {Raetz}, \& {Alvarado-G{\'o}mez}}]{SanzForcada2019}
{Sanz-Forcada}, J., {Stelzer}, B., {Coffaro}, M., {Raetz}, S., \&
  {Alvarado-G{\'o}mez}, J.~D. 2019, \aap, 631, A45,
  \dodoi{10.1051/0004-6361/201935703}

\bibitem[{{Schneider} \& {Schmitt}(2008)}]{schneider08}
{Schneider}, P.~C., \& {Schmitt}, J.~H.~M.~M. 2008, \aap, 488, L13,
  \dodoi{10.1051/0004-6361:200810261}

\bibitem[{{Stelzer}(2004)}]{stelzer04}
{Stelzer}, B. 2004, \apjl, 615, L153, \dodoi{10.1086/426121}

\bibitem[{{Stelzer} \& {Schmitt}(2004)}]{stelzer2004b}
{Stelzer}, B., \& {Schmitt}, J.~H.~M.~M. 2004, \aap, 418, 687,
  \dodoi{10.1051/0004-6361:20040041}

\bibitem[{{Su{\'a}rez Mascare{\~n}o} {et~al.}(2016){Su{\'a}rez Mascare{\~n}o},
  {Rebolo}, \& {Gonz{\'a}lez Hern{\'a}ndez}}]{SuarezMascareno2016}
{Su{\'a}rez Mascare{\~n}o}, A., {Rebolo}, R., \& {Gonz{\'a}lez Hern{\'a}ndez},
  J.~I. 2016, \aap, 595, A12, \dodoi{10.1051/0004-6361/201628586}

\bibitem[{{Teets} {et~al.}(2012){Teets}, {Weintraub}, {Kastner}, {Grosso},
  {Hamaguchi}, \& {Richmond}}]{teets12}
{Teets}, W.~K., {Weintraub}, D.~A., {Kastner}, J.~H., {et~al.} 2012, \apj, 760,
  89, \dodoi{10.1088/0004-637X/760/1/89}

\bibitem[{{Toal{\'a}} {et~al.}(2023){Toal{\'a}}, {Gonz{\'a}lez-Mart{\'\i}n},
  {Karovska}, {Montez}, {Botello}, \& {Sabin}}]{2023MNRAS.522.6102T}
{Toal{\'a}}, J.~A., {Gonz{\'a}lez-Mart{\'\i}n}, O., {Karovska}, M., {et~al.}
  2023, \mnras, 522, 6102, \dodoi{10.1093/mnras/stad1401}

\bibitem[{{Townsley} {et~al.}(2011){Townsley}, {Broos}, {Chu}, {Gagn{\'e}},
  {Garmire}, {Gruendl}, {Hamaguchi}, {Mac Low}, {Montmerle}, {Naz{\'e}}, {Oey},
  {Park}, {Petre}, \& {Pittard}}]{townsley11b}
{Townsley}, L.~K., {Broos}, P.~S., {Chu}, Y.-H., {et~al.} 2011, \apjs, 194, 15,
  \dodoi{10.1088/0067-0049/194/1/15}

\bibitem[{{Wheatley} \& {Kallman}(2006)}]{2006MNRAS.372.1602W}
{Wheatley}, P.~J., \& {Kallman}, T.~R. 2006, \mnras, 372, 1602,
  \dodoi{10.1111/j.1365-2966.2006.10959.x}

\bibitem[{{White}(1996)}]{White:96}
{White}, N.~E. 1996, in Astronomical Society of the Pacific Conference Series,
  Vol. 109, Cool Stars, Stellar Systems, and the Sun, ed. {R.~Pallavicini \&
  A.~K.~Dupree}, 193--+

\bibitem[{{Wood} {et~al.}(2018){Wood}, {Laming}, {Warren}, \&
  {Poppenhaeger}}]{Wood.etal:18}
{Wood}, B.~E., {Laming}, J.~M., {Warren}, H.~P., \& {Poppenhaeger}, K. 2018,
  \apj, 862, 66, \dodoi{10.3847/1538-4357/aaccf6}

\bibitem[{{Yang} {et~al.}(2020){Yang}, {Zhang}, \& {Ji}}]{yang20b}
{Yang}, H., {Zhang}, S., \& {Ji}, L. 2020, \apj, 894, 22,
  \dodoi{10.3847/1538-4357/ab80c9}

\bibitem[{{Zhang} {et~al.}(2023){Zhang}, {Algeri}, {Kashyap}, \&
  {Karovska}}]{2023MNRAS.521..969Z}
{Zhang}, X., {Algeri}, S., {Kashyap}, V., \& {Karovska}, M. 2023, \mnras, 521,
  969, \dodoi{10.1093/mnras/stad398}

\end{thebibliography}
